\documentclass[final,5p,times,twocolumn]{elsarticle}

\usepackage{amssymb}
\usepackage{amsmath}
\usepackage{amsmath}
\usepackage{mathrsfs}
\usepackage{subfigure}
\usepackage{soul,color}
\usepackage{siunitx}
\usepackage[percent]{overpic}
\usepackage{placeins}

\begin{document}

\begin{frontmatter}



\title{Resolving the inverse problem in pulse response analysis of TAP reactors
}


\author[label1]{Anjali Aleria}
\ead{anjali2@iitb.ac.in}
\author[labelR]{Evgeniy Redekop}
\author[labelS]{A. K. Suresh}
\author[label1]{Jason R. Picardo\corref{cor1}}
\ead{picardo@iitb.ac.in}
\cortext[cor1]{corresponding author}
\affiliation[label1]{organization={Department of Chemical Engineering, Indian Institute of Technology Bombay},
            addressline={Powai},
            city={Mumbai},
            postcode={400076},
            country={India}}

\affiliation[labelR]{
organization={Department of Chemistry, University of Oslo},
addressline={ },
city={ Oslo},
postcode={ },
country={Norway}}

\affiliation[labelS]{
organization={Indian Institute of Technology Madras Zanzibar},
addressline={Bweleo},
city={Zanzibar },
postcode={71215 },
country={Tanzania}}

%

\begin{abstract}

Pulse experiments in the temporal analysis of products (TAP) reactor are one of the most important methods for studying transient kinetics of gas-solid catalytic reactions. The Y-procedure (\textit{Yablonsky et al., Chem. Eng. Sci. 62, 6754, 2007}) is a model-free analysis framework for inferring the relationship between the reaction-rate $R$ and the reactant concentration $C$ from measurements of the outlet flux of gas. 
While elegant in conception, its application is hindered by the amplification of measurement noise that results from having to backtrack diffusive transport from the outlet to the reaction zone. Here, we explicitly recognize the inverse problem inherent in the Y-procedure and treat it using well-developed tools from the field of inverse problems. While previous implementations of the Y-procedure used Fourier-based filtering, we do not pre-process the measurements with an \textit{ad hoc} noise-filter. Instead, we use a basis of localized square pulses to formulate a discrete inverse problem, whose regularized solution is obtained via the 
truncated singular value decomposition (TSVD) method. This method requires one to select a cutoff mode number; while we show how the choice of this regularization parameter can be guided by a Picard plot, we also develop an objective selection strategy for state defining experiments, for which $R(C)$ is a single-valued function, i.e., for every value of $C$, there is a single value of $R$. We apply our proposed inverse-problem approach to synthetic measurement data corresponding to linear and nonlinear reactions and compare the results with the Fourier-filtration method. The former produces better reconstructions of the $R$ vs $C$ relationship, especially for nonlinear reactions. 
Our work facilitates the automation of pulse response analyses and enables the application of other discrete inverse-problem techniques, such as Tikhonov regularization or machine-learning methods.


\end{abstract}




\end{frontmatter}




\section{Introduction}


Catalysts underpin a wide range of industrial processes and are vital to achieving a sustainable and energy secure future \citep{GARCIASERNA2022237}. The promise of new cutting-edge catalysts \citep{Lim2016} will be realized, however, only when they are implemented in industrial operations. But translating advances in catalysis research, driven by the increasingly powerful methods of experimental surface science and computational chemistry, remains challenging because of the materials-pressure gap~\citep{Friend,Morgan2017}. Transient kinetic studies have an important role to play in this regard~\citep{REDEKOP2023}. Unlike steady-state tests, which only resolve the slow steps of a complex catalytic-reaction mechanism, transient studies can yield insight into fast elementary steps, the lifetimes of unstable surface intermediates, and the distribution of active sites~\citep{GLEAVES2010,Kim24,Yonge24}.

Arguably, the most advanced research instrument for studying time-resolved kinetics of heterogeneous gas-phase catalytic reactions is the temporal analysis of products (TAP) reactor \citep{GLEAVES2010,Morgan2017}. In this automated system, a solid catalyst sample is contained in a packed-bed microreactor and exposed to a controlled, small gas pulse under conditions of high vacuum. The pulse is injected at the reactor inlet and travels through the bed by Knudsen diffusion. An online mass spectrometer logs the efflux of gas at the outlet. The TAP reactor effectively bridges the material and pressure gaps: it allows the characterization of a range of catalyst material types, from ideal single crystals to practical particles, and can also be operated under a range of pressures that are intermediate to those encountered in surface science experiments and industrial reactors~\citep{Gleaves1988,BRANDAO2023147489,Morgan2017}. 

\begin{figure*}
	\centering
	\includegraphics[width=.7\textwidth]{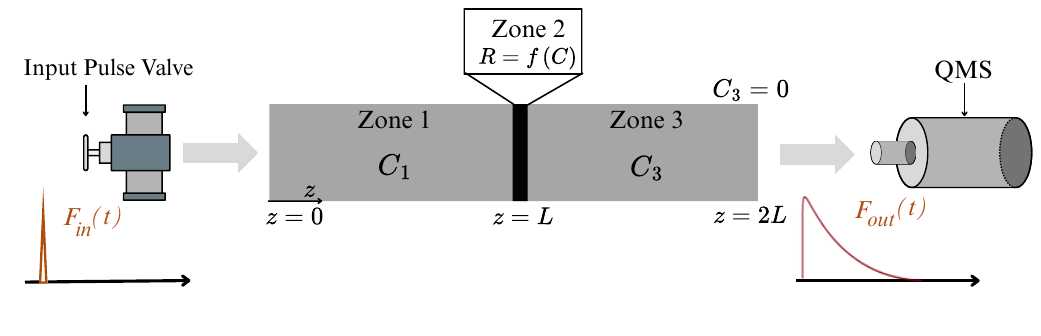}
	\caption{Schematic of a thin-zone TAP reactor. A pulse of gas molecules diffuses through the bed of inert particles in zone 1, reacts in the thin zone of catalyst-loaded particles in zone 2, and diffuses again through the inert zone 3, to finally emerge as the outlet pulse. $F_{in}(t)$ and $F_{out}(t)$ are the inlet and outlet flux profiles; the former is known while the latter is measured by the Quadrupole Mass Spectrometer (QMS). 
    }
	\label{fig:TAP}
\end{figure*}

A common configuration of the TAP reactor involves three zones, with the catalyst-bearing zone sandwiched between two inert zones~\citep{CONSTALES2001, CONSTALES2006} (see Fig.~\ref{fig:TAP}). In standard TAP experiments, the pressures are sufficiently low for collisions and reactions in the gas phase to be negligible. The pulse of gas molecules therefore undergoes Knudsen diffusion in the inert zones and is fully exposed to the catalytic surfaces in the middle zone. These well-defined conditions, particularly the simple mode of diffusive mass transport, enable one to infer qualitative and quantitative kinetic information from the differences between the known injected input pulse and the measured output pulse~\citep{Marinbook,Morgan2017}. Such a pulse response analysis may be carried out using the method of moments \cite{Yablonsky1998, Shekhtman2003} or model-fitting \cite{Kondratenk02006,Prasad2009,Kumar2011}. However, while the former averages over time, the latter requires an \textit{a priori} assumption on the form of the kinetic model. A powerful, model-free method, which retains transient kinetic information, is the Y-procedure that was introduced by \citet{YABLONSKY20076754} in the context of the thin-zone TAP reactor (TZTR) \citep{SHEKHTMAN19994371}, wherein an excellent approximation to a uniform concentration is realized in the middle reaction zone by making it sufficiently thin. 

The Y-procedure takes advantage of the separation between reaction and transport in the thin-zone TAP reactor, to reconstruct the temporal variation of reaction rates and concentrations in the thin reaction zone, thereby yielding a reaction rate vs. concentration curve without having to assume a kinetic model \citep{YABLONSKY20076754}. 
Since its introduction, the Y-procedure has been extended to include irreversible adsorption and to compute the surface concentration~\citep{REDEKOP20116441}; it has also been implemented within a data-science workflow \citep{ROSSKUNZ201846}. The Y-procedure is based on three key steps: (i) The measured outlet flux profile (leaving zone 3) is used to reconstruct the flux leaving the thin reaction zone (entering zone 3), as well as the concentration $C(t)$ in the reaction zone (see Fig.~\ref{fig:TAP}); (ii) The known temporal variation of the input flux to zone 1, i.e, the input pulse profile, is used to obtain the flux entering the thin reaction zone (zone 2); (iii) The difference between the fluxes entering and leaving the reaction zone is used to compute the reaction rate $R(t)$. Now, while the second step involves the straightforward simulation of diffusion in zone~1, the first step requires one to go backward in time and infer the flux that must have entered zone~3 so as to yield the measured outlet flux at the exit of zone~3. Solving the diffusion equation backwards in time is Hadamard-ill-posed~\citep{Hadamard1902,MuellerBook2012}: since diffusion dampens high-frequency fluctuations, going backward in time will amplify them. As a consequence, even tiny measurement noise in the outlet flux will get amplified by a naive backward solution so that the reconstructed flux entering zone-3 is entirely corrupted by noise. Recognizing this issue, \citet{YABLONSKY20076754} used Fourier-filtering to smooth the measured outlet flux. This pre-processing step has henceforth been the typical approach to dealing with the illposedness of the first step \citep{ROSSKUNZ201846} (though an alternate G-procedure has also been developed based on probability distribution functions~\citep{KUNZ2020}). 

This amplification of measurement noise is a characteristic feature of inverse problems, which is a well-developed branch of applied mathematics \citep{MuellerBook2012,Hansenbook2010}. 
Ad hoc Fourier-filtration of the measured signal is not the recommended strategy for solving such problems, for the following reasons: (a) in the absence of an objective basis for selecting the extent of filtering, one must rely on human judgement, which not only is subjective but also prevents automation of the analysis; 
(b) oversmoothing can produce artifacts like a non-zero reconstructed flux at times prior to the start of the experiment; 
(c) the use of a generic Fourier basis requires one to treat the signal as a portion of a \textit{continuous} periodic function; this also contributes to artefacts in the result while preventing consideration of inputs like a step function. Implementations of the Y-procedure that use Fourier-filtering naturally suffer from these drawbacks \citep{YABLONSKY20076754,ROSSKUNZ201846}. Our aim in this paper is to overcome these difficulties in implementing the Y procedure, by formulating its first step as a discrete inverse problem and then obtaining a regularized solution via the truncated singular value decomposition (TSVD). 

Like all regularization techniques, the TSVD method requires one to specify the value of a regularization parameter, which in this case is the cutoff SVD mode number. While a Picard plot can be used to guide this choice (as demonstrated here), we would ideally like to have an objective way to select the regularization parameter, so that the pulse response analysis can be automated. Here, we devise such a strategy for the case of a \textit{state-defining} experiment, in which the surface of the catalyst remains effectively unaltered during the course of a pulse. In this scenario, which is realized for a sufficiently small pulse of gas, the reaction rate is a single-valued function of the gas-phase concentration~\citep{REDEKOP20116441}. We introduce a positive-semi-definite measure of the multi-valuedness of the reconstructed rate-vs-concentration curve and show that minimizing this measure is an effective way to select the regularization parameter. 

The paper unfolds as follows. The mathematical model for the thin zone TAP reactor is presented in \S \ref{sec:TAPmaths}, along with the numerical methods used to simulate the reactor and generate synthetic measurement data. For simplicity, we consider the case of a single gas species undergoing an irreversible reaction, as in \citet{YABLONSKY20076754}. The rationale and the key steps of the the Y-procedure are recalled in \S~\ref{sec:Y_procedure}. The Fourier-filtration method that has been used thus far to implement the procure is outlined in \S~\ref{sec:filtration}, along with an example of a reconstruction of a linear reaction rate function.  The new inverse-problem methodology is presented in \S \ref{sec:inverse}. Representing the unknown flux entering zone~3 with a
basis of localized functions, specifically square pulses, we formulate in \S~\ref{sec:discrete-inverse} a discrete inverse problem $\mathbf{A} \boldsymbol{w} = \boldsymbol{b}$ that relates the discretized inlet flux $\boldsymbol{w}$ to the measured outlet flux $\boldsymbol{b}$.  The ill-posed nature of the problem manifests in a very large condition number for the matrix $\mathbf{A}$. The regularized solution of this problem is then obtained using the TSVD, in \S~\ref{sec:TSVD}, where we also show how the Picard plot helps to distinguish between information-bearing SVD modes and noise-dominated SVD modes, thereby informing the choice of the truncation mode number $m_\mathrm{cut}$. Results for the reconstructed flux and rate function are presented here, for the case of a linear reaction. \S~\ref{sec:single-valued} introduces a metric $\Psi$ that quantifies the extent to which the reconstructed reaction rate function departs from being a single-valued function of concentration. $\Psi$ is then shown to vary nonmonotonically with the regularization parameter $m_\mathrm{cut}$, so that the minimum of $\Psi$ yields the optimal value of $m_\mathrm{cut}$; this approach to selecting the regularization parameter is shown to work even for the Fourier filtration parameter $\lambda$. The ability of the new implementation of the Y-procedure to reconstruct nonlinear rate functions is then demonstrated using synthetic data based on quadratic and cubic reaction rate expressions, in \S~\ref{sec:nonlinear}. The reconstructions from the proposed inverse-problem regularization approach are found to be better than those obtained from Fourier-filtration, especially in the case of nonlinear kinetics. We conclude in \S \ref{sec:conclusion} with a summary of our work and a discussion of the avenues for future work that are opened up by the explicit recognition and formulation of the inverse problem at the heart of the Y-procedure.

\section{Simulation of the thin-zone TAP reactor}
\label{sec:TAPmaths}

\subsection{Mathematical model}\label{sec:model}

For this study, we adopt the model of the TZTR that was used to develop the Y-procedure ~\citep{YABLONSKY20076754}. A schematic of the TZTR is presented in Fig.~\ref{fig:TAP}. Here, zones $1$ and $3$ are inert packed beds, both of length $L$. The Knudsen diffusion of the gas species, of concentration $C$ (\si{mol.m^{-3}}), in these zones is governed by 
\begin{equation}
	\epsilon \frac{\partial C_i}{\partial t}= \mathcal{D} \frac{\partial ^2 C_i}{\partial z^2},
	\label{eqn:diffusion}
\end{equation}
where $\mathcal{D}$ is the effective Knudsen diffusivity, $\epsilon$ is the void fraction, and the subscript $i = 1,3$ denotes the zone number.

 The thin zone 2 is the catalyst-bearing reaction zone; it is idealized as a surface across which the flux of gas molecules has a discontinuity due to the reaction, which occurs with a rate $R$ (\si{mol.m^{-2}.s^{-1}}). The concentration, of course, is continuous across this zone. Denoting the diffusive flux by $F_i(z,t) \equiv - \mathcal{D} {\partial C_i}/{\partial z}$, we have
 \begin{equation}\label{eq:zone2}
    F_3 (L,t) -  F_1(L,t) = R(t), \quad C_3 (L,t) =  C_1(L,t) = C(t)
 \end{equation}
where $C$ without the subscript represents the concentration at the thin-zone reaction surface. We restrict ourselves to state defining experiments in this work, for which $R$ is a single-valued function of~$C$. We assume a rate form of the type
\begin{equation}\label{eq:R}
    R(C) =  k C^n.
\end{equation}
We mainly focus on the first-order case of $n = 1$, for which $k$ is an apparent rate constant with units \si{m.s^{-1}}. 
However, we also consider quadratic and cubic cases ($n = 2,3$) in the penultimate section, to illustrate the ability of the inverse-method to reconstruct nonlinear kinetics. 

\begin{figure}
    \centering    
    {\includegraphics[width=0.4\textwidth]{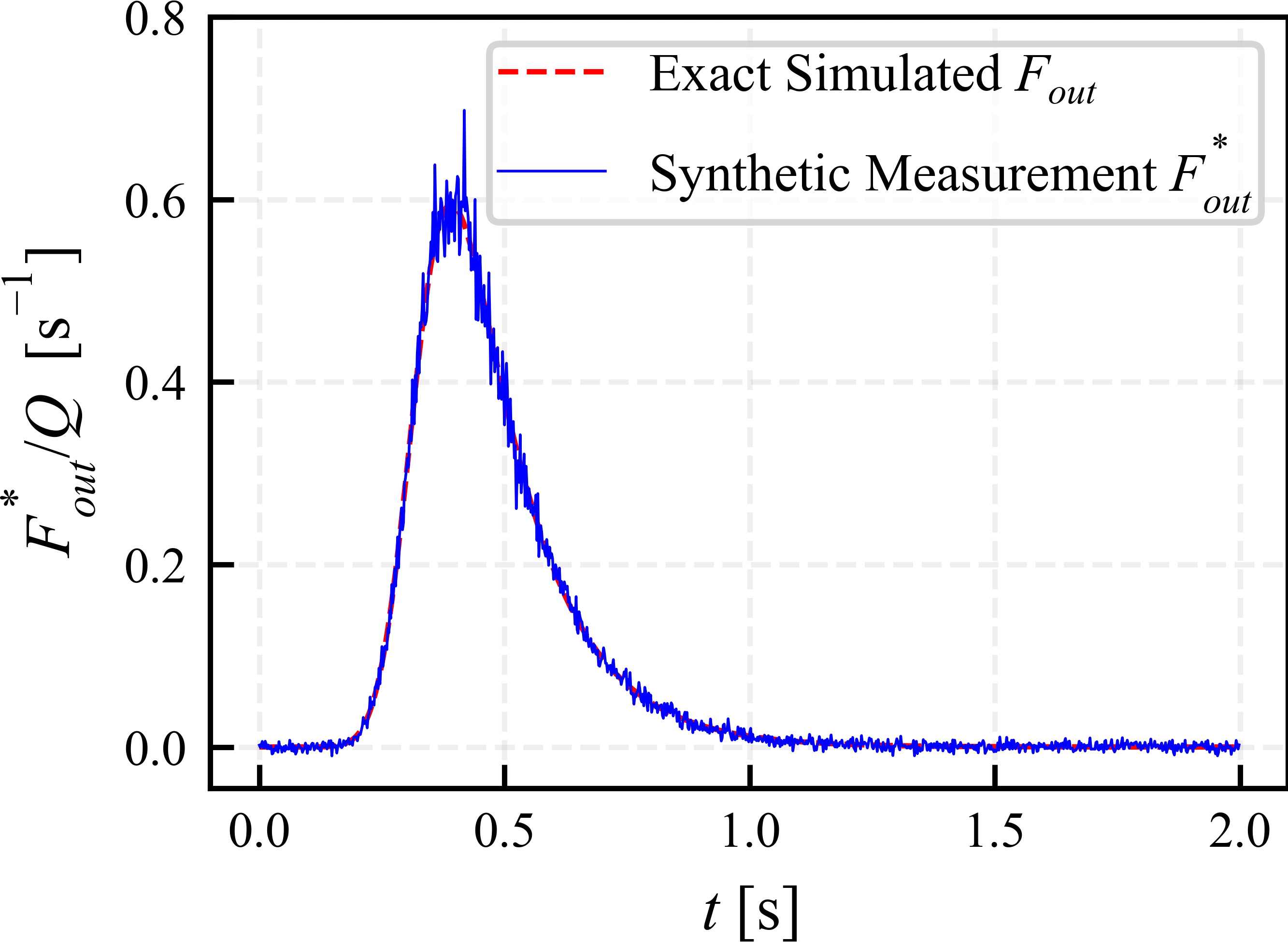}} 
	\caption{
    Outlet flux profile emerging at the exit of zone 3 after undergoing a linear reaction in zone 2. Noise is added to the exact outlet flux from the forward simulation (red-dashed line) to obtain the synthetic measured outlet flux (blue line).}
	\label{fig:OF}
\end{figure}

Equation~\eqref{eqn:diffusion} is complemented by the initial condition $C_i(z,0)=0 $ and the boundary conditions at the inlet and outlet of the TZTR:
\begin{align}
	F_{in}(t) &= -\mathcal{D}\frac{\partial C_i}{\partial z} \quad \mathrm{at}\quad z = 0, \label{eqn:F_inlet} \\
    C_3 &= 0  \quad \mathrm{at}\quad z = 2L \label{eqn:c_outlet}
\end{align}
The latter condition is a simple way to account for the applied vacuum that continuously draws gas out of the reactor and into the mass spectrometer. Indeed, 
this approximate exit condition has been shown to yield predictions in close agreement with experiments~\citep{Marinbook}. The inlet pulse is typically modelled as a delta function, when the equations are solved analytically. A numerical solution, which is required in case of nonlinear reactions, necessitates the use of an approximation of the delta function. Here we use a Gaussian pulse of strength $Q$ (\si{mol.m^{-2}.s^{-1}}), with mean $\mu$ and standard deviation $\sigma$:
\begin{equation}\label{eq:Fin}
	F_{in}(t)=\frac{Q}{\sigma \sqrt{2 \pi}} \exp{\frac{-1}{2}\frac{(t-\mu)^2}{ \sigma^2}}
\end{equation}
We set $\mu = 1/4$ \si{s} and $\sigma = 1/20$ \si{s} throughout the main text. 
Fig.~\ref{fig:OF} depicts the corresponding outlet flux profile, after the gas has undergone a linear reaction in zone 2 (this plot shows the predicted profile for $F_{out}$, to which noise is added to obtain $F_{out}^*$, as discussed below). Here, we have intentionally used a relatively broad pulse and shifted the initial time ($t=0$) a little before the start of the pulse in order to more clearly visualize the effectiveness of the reconstruction procedure, for both the rising and falling parts of the pulse. In the Appendix, we repeat our key calculations for a narrower inlet pulse ($\mu = 1/12.5 \si{s}$ and $\sigma = 1/40$ $\si s$) that starts just after $t=0$. 

The precise value of the pulse strength $Q$ does not affect the results presented here (provided it is small enough for realizing a state defining experiment in the Knudsen regime) because we present the results after normalizing with $Q$, as done in Ref.~\citep{YABLONSKY20076754}. A typical value, corresponding to a pulse of intensity $10^{-10}$ \si{mol} and width $250$~\si{\mu s} passing through a reactor of diameter $5$ \si{mm}, is $Q = 0.02$ \si{mol.m^{-2}s^{-1}}~\citep{Marinbook,Fushimi2018}. 

The model parameters are set to the representative values used in \citet{YABLONSKY20076754}: $L=10^{-2}$ m, $\epsilon=1/2$, $\mathcal{D}=2 \times 10^{-4}\,$ \si{m^2.s^{-1}}. These choices result in a diffusive time-scale $\tau=\epsilon \frac{L^2}{\mathcal{D}}=1$ \si{s} and transport velocity-scale $\gamma={\mathcal{D}}/{L} = 2$ \si{m.s^{-1}}. When considering a linear reaction, we use $k = 0.1$ \si{m.s^{-1}} so that $k/\gamma = kL/\mathcal{D} = 0.05$. For the second and third order reactions, we choose the rate constants so that $k_2 Q = k_3 Q^2 = k$ (thus, with $Q = 0.02$ \si{mol.m^{-2}.s^{-1}}, we have $k_2 = 5$ \si{m.s^{-1}.mol^{-1}} and $k_3 = 250$ \si{m.s^{-1}.mol^{-2}}); this choice results in a rate of reaction that is of the same order of magnitude in all three cases, thereby facilitating a comparison among them.  


\subsection{Forward simulation and synthetic measurement data}

In the forward simulation, the reaction kinetics is specified via $R(C)$ in Eq.~\eqref{eq:R} and one calculates the outlet flux $F_{out}$ by solving Eq.~\eqref{eqn:diffusion} subject to the boundary conditions~\eqref{eq:zone2}-\eqref{eqn:c_outlet}. When the reaction is linear, an analytical solution may be obtained via Laplace/Fourier transforms, as discussed in \citep{YABLONSKY20076754}. For the case of nonlinear reactions, we solve the equations numerically using the method of lines. Specifically, we discretize space using the second-order finite difference method, with central differences; the boundary conditions are implemented by applying backward/forward first-order finite-differences at the start/end of the zones~\cite{gdsmith, VANDERLINDE199727,schiesser2012numerical}. The resulting ordinary differential equations are integrated in time using
 the LSODA solver, provided by the solve{\textunderscore}ivp function from the SciPy library in Python \cite{virtanen2020scipy}. We have validated the simulations against analytical results for the case of a linear reaction~\citep{YABLONSKY20076754}. A spatial grid of 500 points is sufficient for grid independence, in the case of linear and quadratic reactions, while 800 points are required for the cubic reaction. 

In pulse response analysis, the goal is to solve the inverse problem, i.e, to determine $R(C)$ from measurements of $F_{out}$. To test and illustrate our inverse problem methodology, we use synthetic measurement data for $F_{out}$, generated by solving the forward problem and then adding noise to the outlet flux to mimic experimental noise (which arises in practice from a variety of sources, including power and heater coil fluctuations~\cite{ROSSKUNZ201846}). Following \citet{YABLONSKY20076754}, a zero-mean unit-variance Gaussian noise ($\mathcal{N}(0,1)$) is added to the outlet flux from the forward simulation $F_{out}$ to obtain the synthetic measurement $F_{out}^*$ (where $*$ signifies a measured function that bears experimental noise):
\begin{equation}\label{eq:noise}
    F_{out}^* = F_{out} +(0.004+0.05\,F_{out})\,\mathcal{N}(0,1)
\end{equation}

The exact outlet flux from the forward calculation and the corresponding perturbed function, which serves as the measured outlet flux, are plotted in Fig.~\ref{fig:OF} for the case of a linear reaction.

\section{Key steps of the Y procedure}
\label{sec:Y_procedure}
We now briefly recall the rationale of the Y-procedure~\citep{YABLONSKY20076754}, which enables the determination of $R(C)$ given an input pulse $F_{in}=F_1(0,t)$ and a measured outlet pulse $F_{out}^*=F_3(2L,t)$ (see Fig.~\ref{fig:TAP}).  The procedure is based on the fact that Eq.~\eqref{eqn:diffusion}, being a linear second-order partial differential equation (PDE), may be solved by specifying two boundary conditions. These are available for zone 3, since we have the outlet Dirichlet condition Eq.~\eqref{eqn:c_outlet} in addition to the measured outlet flux condition $ -\mathcal{D}\partial C_3/\partial z\rvert_{z=2L} = F_{out}^*$. After solving Eq.~\eqref{eqn:diffusion} in zone 3, we can evaluate $F_3(L,t)$ and $C_3(L,t)$ (which equals $C$ and $C_1(L,t), $ see Eq.~\eqref{eq:zone2}). The latter then allows us to specify a Dirichlet boundary condition at the outlet of zone 1, which along with the inlet condition Eq.~\eqref{eqn:F_inlet} provides the two boundary conditions needed to solve Eq.~\eqref{eqn:diffusion} in zone 1. Thereby, we can evaluate $F_1(L,t)$, whose difference with $F_3(L,t)$ yields $R$ (see Eq.~\eqref{eq:zone2}). Thus, one can obtain the curve of $R$ vs $C$.

As discussed in the Introduction, the main difficulty in applying this procedure is the solution of Eq.~\eqref{eqn:diffusion} in zone 3, given only outlet boundary conditions. In a forward calculation, the temporal fluctuations in the flux entering zone 3, $F_3(L,t)$, will be damped by diffusion so that the flux leaving zone 3, $F_{out} = F_3(2L,t)$ will have a smoother temporal profile. When backtracking from the measured $F_{out}^*$ to obtain $F_3(L,t)$, i.e., when naively solving the inverse problem, the inevitable small fluctuations in $F_{out}^*$ will be amplified so that the reconstructed $F_3(L,t)$ is dominated by noise. This is a typical property of inverse problems. Here, we obtain a physically meaningful reconstruction of $F_3(L,t)$ by 
regularizing the inverse problem using the TSVD method.
But first, we outline the Fourier-filtration approach that has been used thus far in implementations of the Y-procedure.


\section{Fourier-based noise filtration}\label{sec:filtration}

\begin{figure*}
	\centering
    \includegraphics[width=0.33\textwidth]{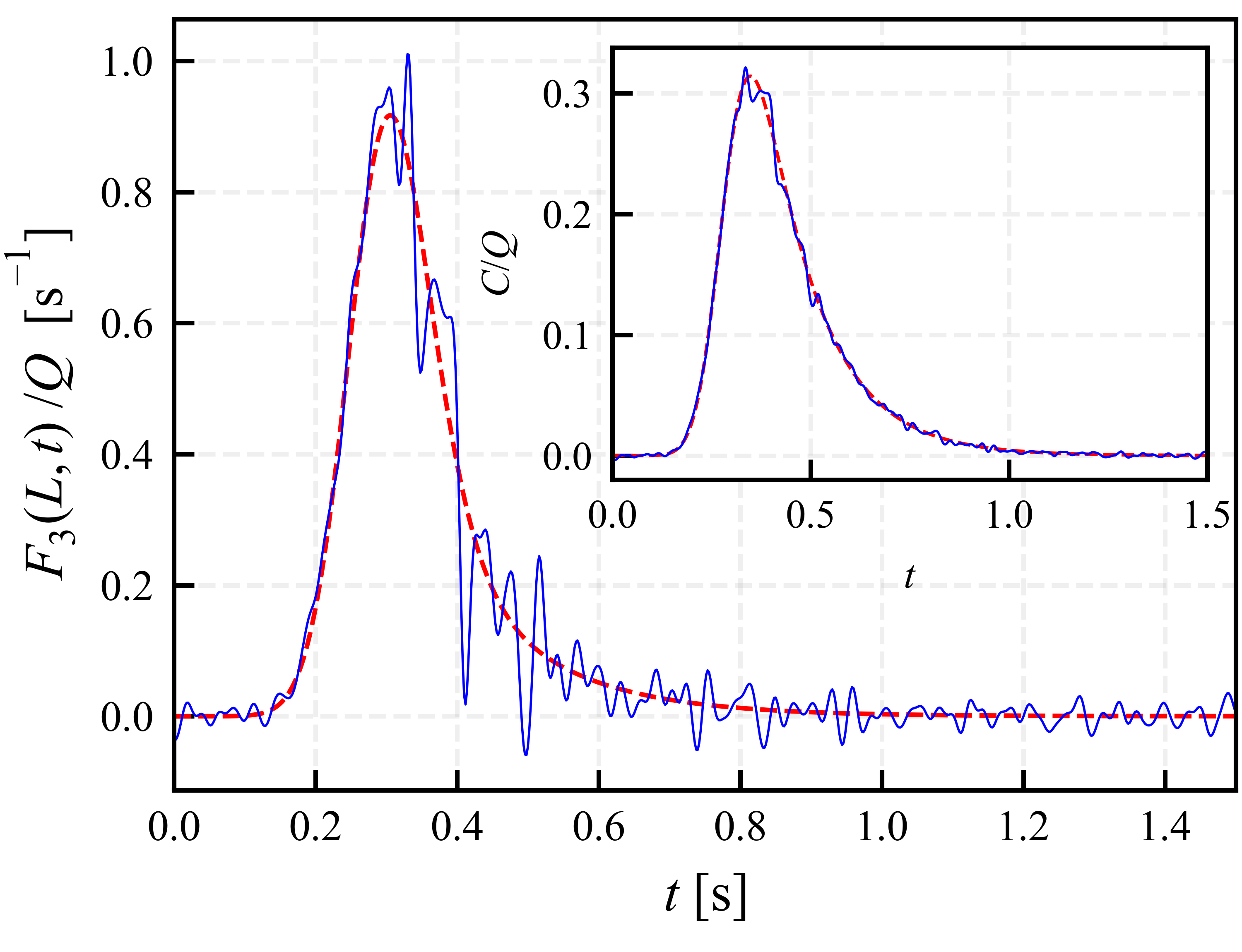}
    \put (-145,120){\footnotesize (a)}
	\includegraphics[width=0.33\textwidth]{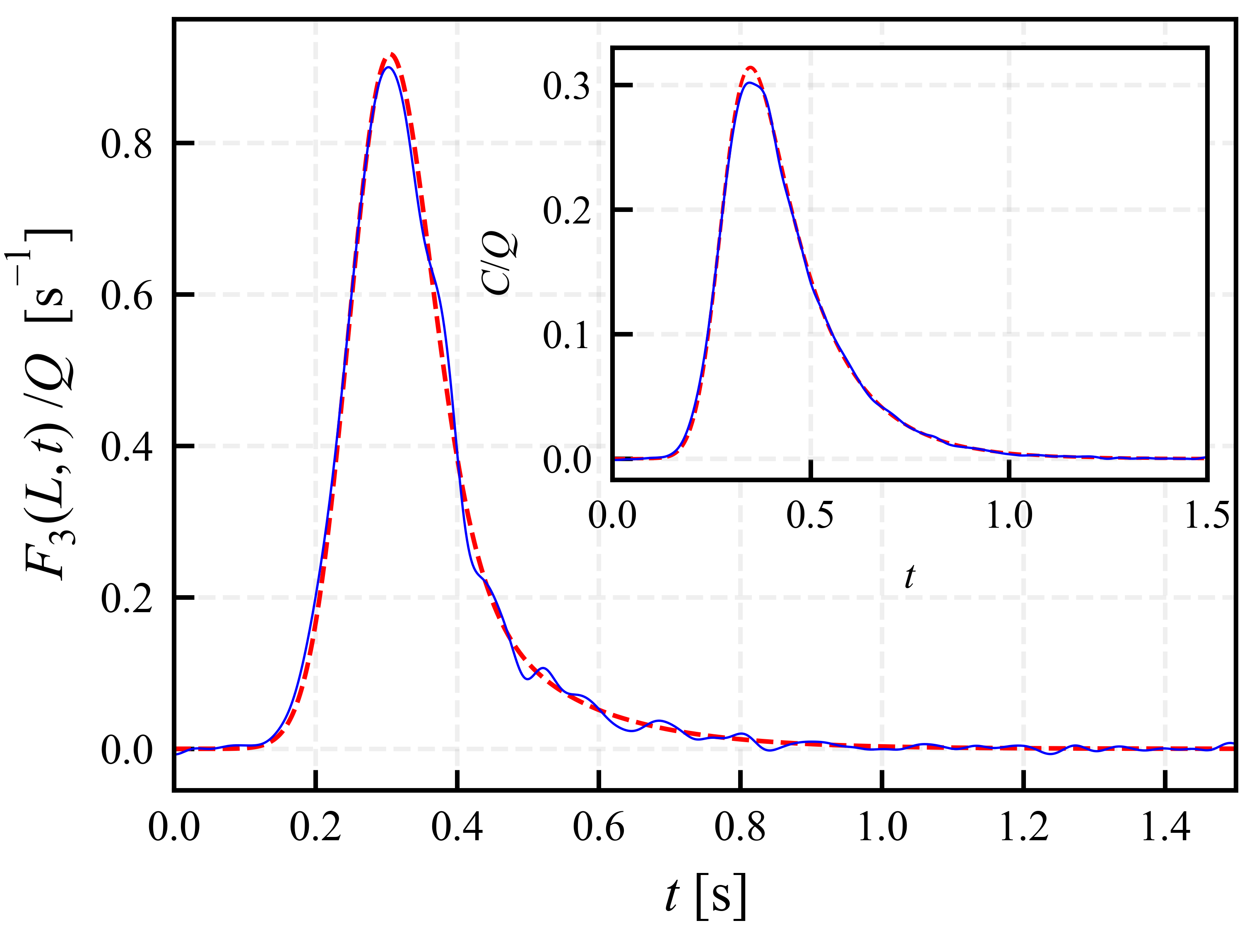}
    \put (-145,120){\footnotesize (b)}
	\includegraphics[width=0.33\textwidth]{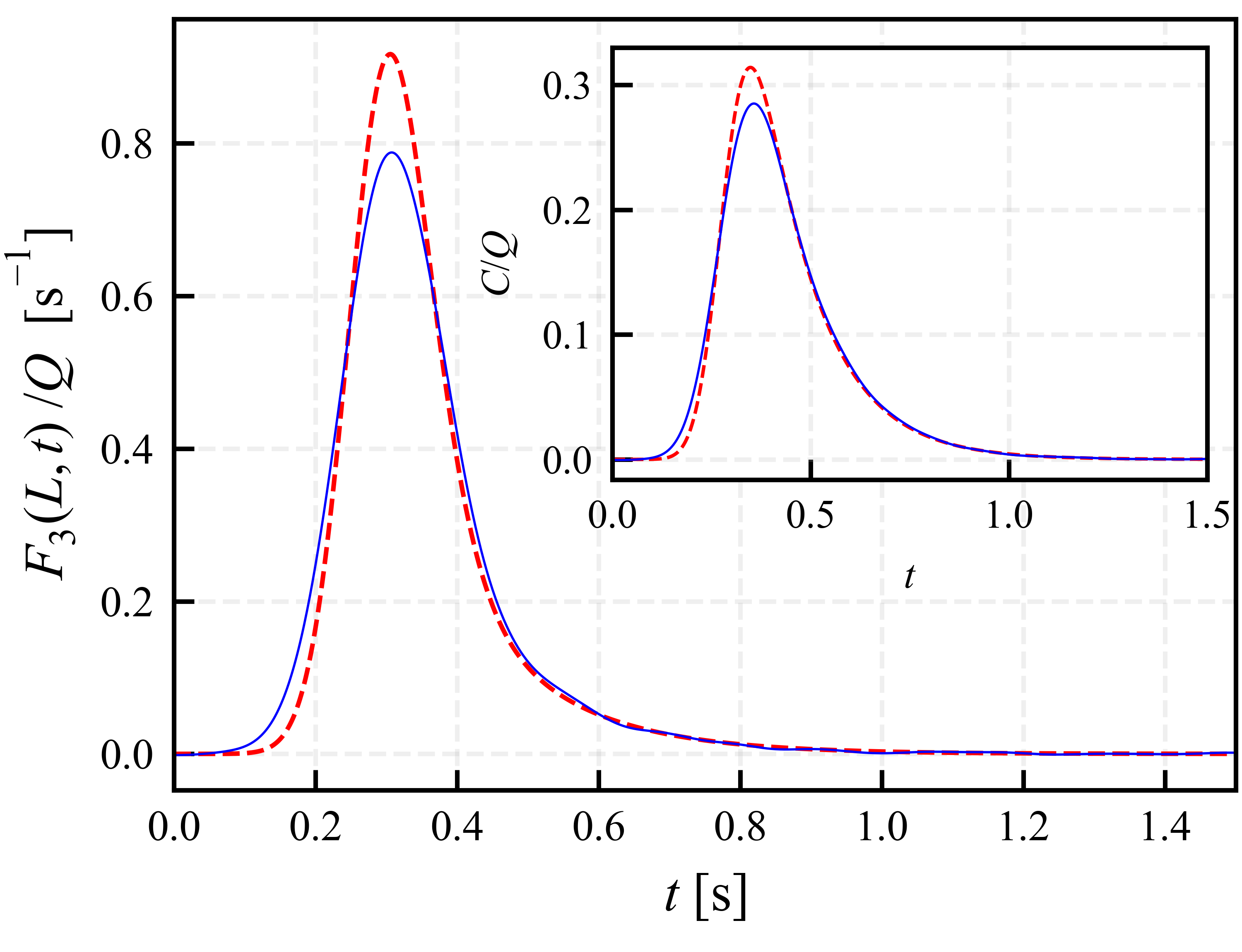}
    \put (-145,120){\footnotesize (c)}
    \hspace{0.02\textwidth}
	\includegraphics[width=0.33\textwidth]{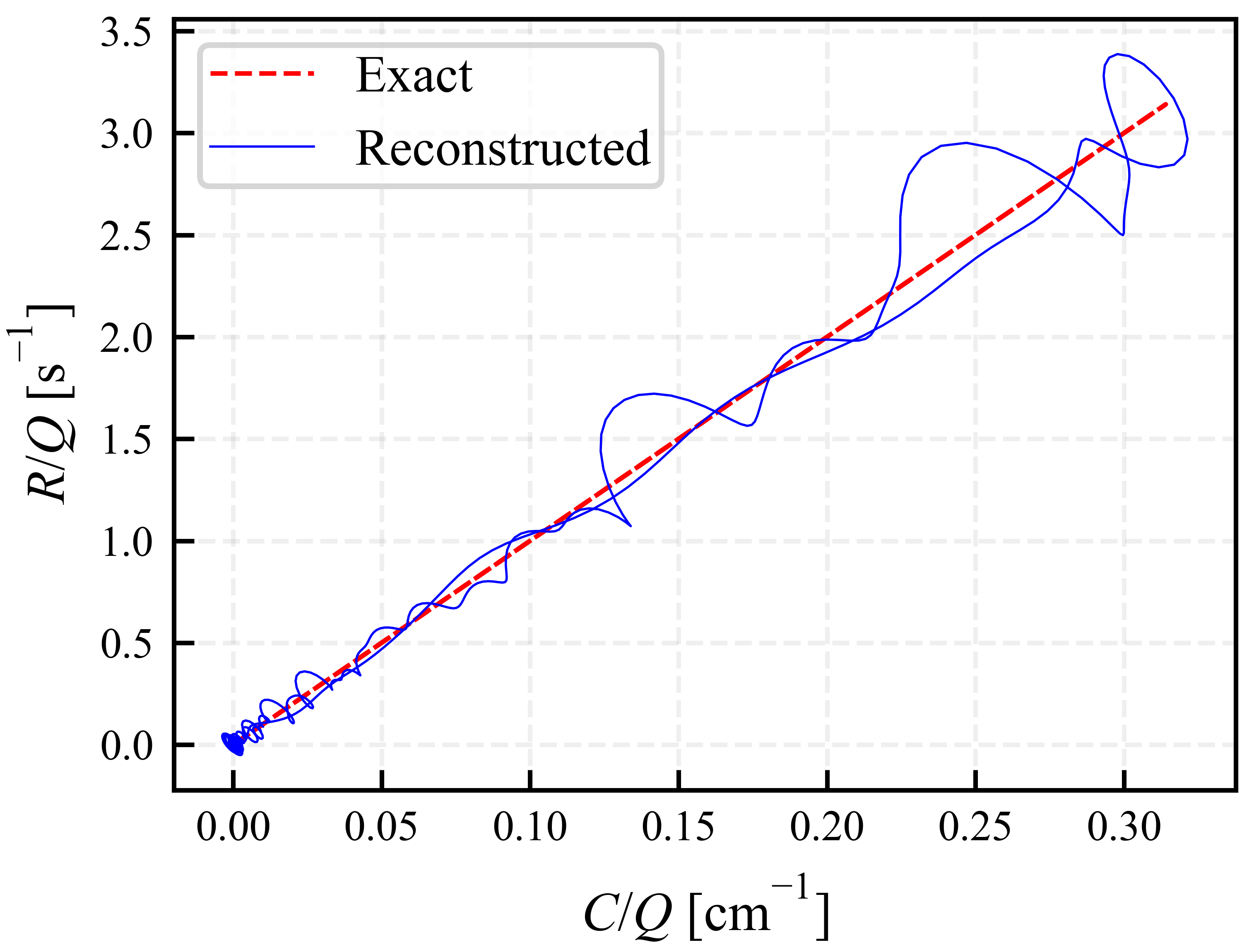}
    \put (-15,30){\footnotesize (d)}
	\includegraphics[width=0.33\textwidth]{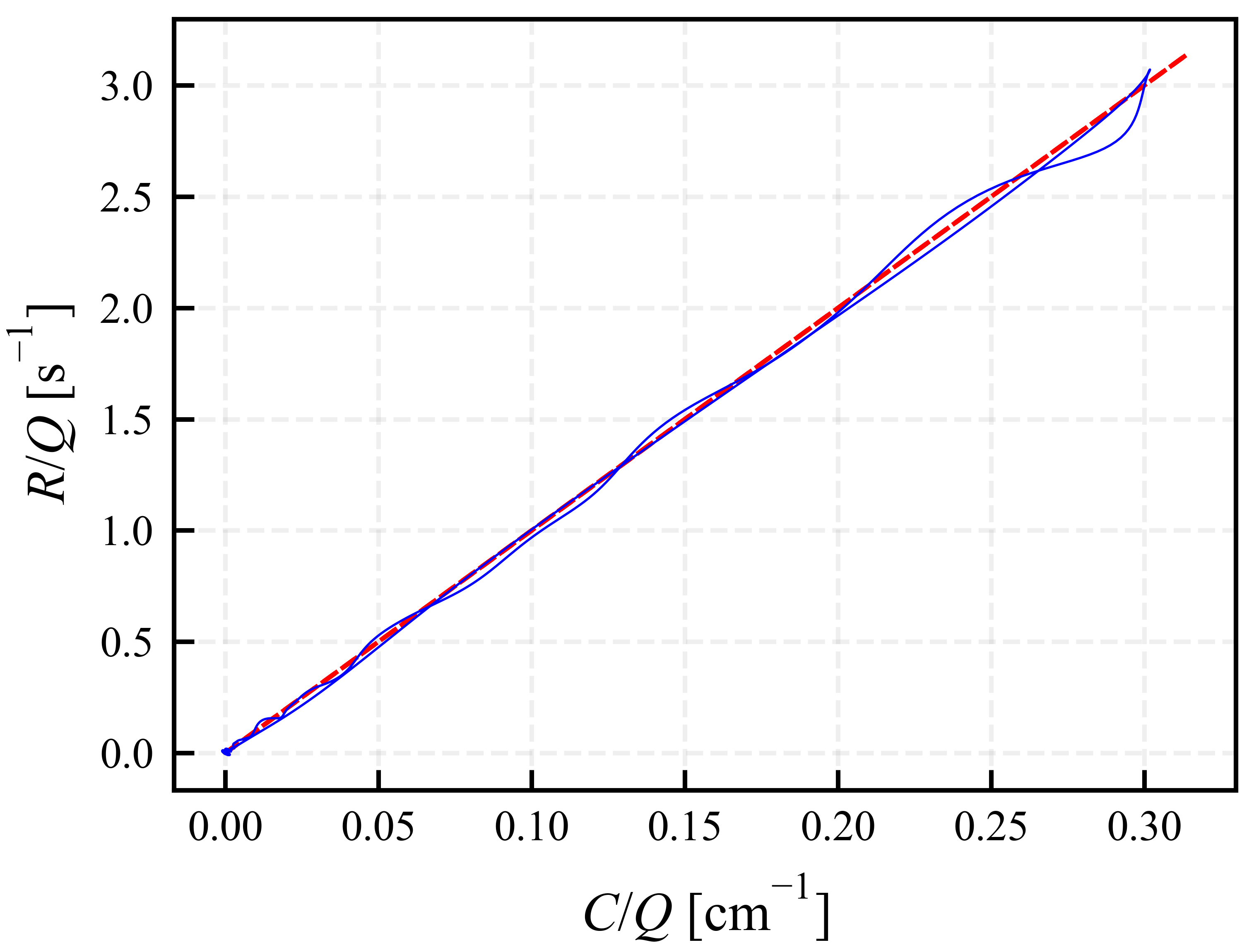}
    \put (-15,30){\footnotesize (e)}
	\includegraphics[width=0.33\textwidth]{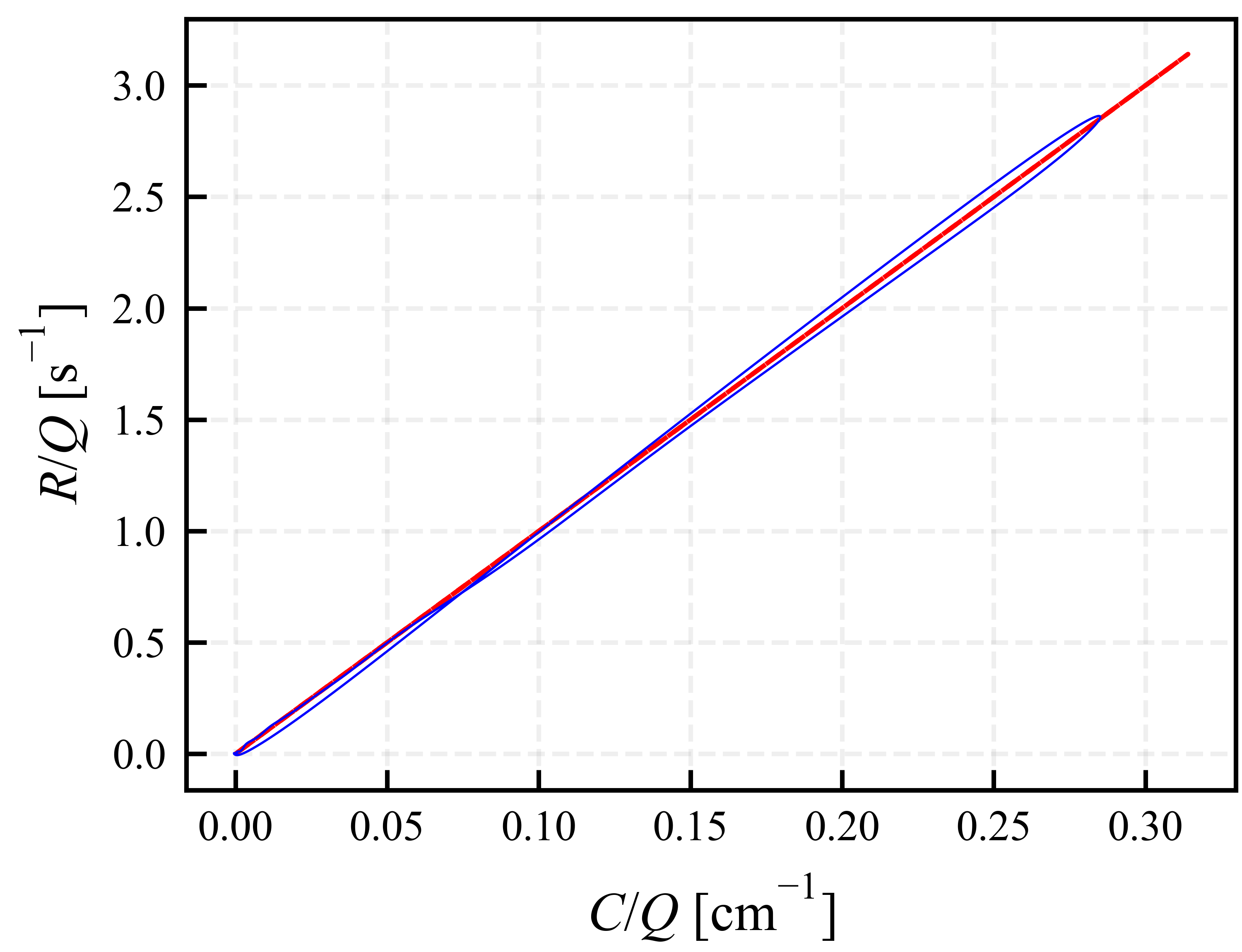}
    \put (-15,30){\footnotesize (f)}
	\caption{Y-procedure reconstruction using Fourier-filtration. The top row presents the inlet flux entering zone 3 (with the concentration in the thin reaction zone in insets), while the bottom row presents the reconstructed reaction-rate vs concentration curve. The three columns correspond to different values of the filtration parameter, $\lambda$: (a,d) 5, (b,e) 10 and (c,f) 20. The exact results from the forward simulation (red-dashed line) are shown for comparison along with the reconstructed results (blue line).}
	\label{fig:fourier}
\end{figure*}

In \citet{YABLONSKY20076754}, the issue of noise-amplification was dealt with by Fourier-filtering the measured signal $F_{out}^*(t)$. Since this operation requires the use of the Fourier transform, it is convenient to also solve Eq.~\eqref{eqn:diffusion} using the Fourier transform method, as done in \citep{YABLONSKY20076754}. Applying the Fourier transform in time to Eq.~\eqref{eqn:diffusion} in zone 3 yields a boundary value problem for $\tilde{C}_3(x,\omega)$ which is the Fourier transform of $C_3(x,t)$ (here $\sim$ denotes a Fourier transform and $\omega$ is the frequency). On solving this problem, one obtains an algebraic equation relating $\tilde{F}_3(L,\omega)$ and $\tilde{C}(\omega)=\tilde{C}_3(L,\omega)$ to $\tilde{F}_{out}^*(\omega)$. A similar treatment of zone 1 yields a relation between $\tilde{F}_1(L,\omega)$ and $\tilde{F}_{in}(\omega)$. In this manner, the Fourier transforms of $R$ and $C$ can be expressed in terms of the Fourier transform of $F_{out}^*(t)$:
\begin{subequations}
\begin{align}
    \tilde{C}(\omega)&= W(\omega)\tilde{F}_{out}^*(\omega)\\
     \tilde{R}(\omega)&= Y(\omega)\tilde{F}_{in}(\omega)-Z(\omega)\tilde{F}_{out}^*(\omega),
\end{align}
where the transfer functions are given by
\begin{align}\label{eq:transferfunc}
    W(\omega)&=\frac{\mbox{sinh} \sqrt{\mbox{i} \omega \tau}}{\gamma \sqrt{\mbox{i} \omega \tau}},\quad
    Y(\omega)= \frac{1}{\mbox{cosh}\sqrt{\mbox{i} \omega \tau}},\nonumber \\
  Z(\omega)&= \mbox{cosh}\sqrt{\mbox{i} \omega \tau} +  \frac{\mbox{sinh} \sqrt{\mbox{i} \omega \tau} ~ \mbox{sinh} \sqrt{\mbox{i} \omega \tau}}{\sqrt{{\gamma}}\,\cosh \sqrt{\mathrm{i} \omega \tau}},  
\end{align}
\end{subequations}
The Discrete Fourier Transform (DFT) is used to numerically compute the Fourier transforms. If the input and output fluxes and sampled at $N_t$ time instants at regular intervals $\Delta t$, over a total time window $T$, then the DFT yields the Fourier transform at discrete frequencies $\omega = 2 \pi n/T$ where $n=-N_t/2+1,...,N_t/2$. The Inverse DFT, applied to $\tilde{R}$ and $\tilde C$, then yields $R(t)$ and $C(t)$ on the uniform temporal grid~\citep{YABLONSKY20076754}.

The transfer functions $W$ and $Z$ in Eq.~\eqref{eq:transferfunc} may be shown to increase rapidly with $\omega$ and hence, as anticipated, small amplitude high-frequency noise in $F_{out}^*(t)$ is greatly amplified in the reconstructed $R(t)$ and $C(t)$. In order to reconstruct the underlying, true relationship between $R$ and $C$, \citet{YABLONSKY20076754} applied a Fourier filter to $F_{out}^*(t)$ by multiplying $\tilde{F}_{out}^*(\omega)$ with a smoothing filter $S(\omega)$ given by
\begin{equation}\label{eq:filter}
    S(\omega)=\exp({-\omega^2 \Delta t^2 \lambda^2/2}),
\end{equation}
where $\lambda$ is a smoothing parameter. The value of $\lambda$ is, of course, unknown a priori. It is typically chosen based on a qualitative examination of the results obtained using different values of $\lambda$. 

Figures~\ref{fig:fourier}(a-c) presents reconstructions of the flux entering zone 3, $F_3(L,t)$, obtained by using the Fourier-filtration method with $\lambda = 5$, 10, and 20, respectively. These reconstructions all correspond to the synthetic measurement of $F_{out}^*(t)$ in Fig.~\ref{fig:OF}. The reconstructed concentration in the thin reaction zone, $C(t)$, is presented in the insets of Figs.~\ref{fig:fourier}(a-c). The exact functions for $F_3(L,t)$ and $C(t)$ from the forward simulation that generated $F_{out}(t)$ is also shown for comparison (red-dashed lines). The reconstructions of the reaction rate function $R(C)$ are shown in Figs.~\ref{fig:fourier}(d-f), along with the exact kinetic curve ($R = k C$, with $k = 0.1$ \si{m.s^{-1}}). We see that a small value of $\lambda$ allows spurious fluctuations to persist in the reconstructed results (Figs.~\ref{fig:fourier}(a,d)), while a large value of $\lambda$ leads to oversmoothing---the peak of $F_3(L,t)$ is not fully captured (Fig.~\ref{fig:fourier}(c)) and so the curve of $R(C)$ does not extend over the entire range of $C$ values realized by the pulse (Fig.~\ref{fig:fourier}(f)). An intermediate value of $\lambda$ should therefore be selected; a value between 10 (Figs.~\ref{fig:fourier}(b,e)) and 20 (Figs.~\ref{fig:fourier}(c,f)) would likely be chosen for this case. Now, in applications, one of course does not have the exact results with which to compare reconstructions; so selecting $\lambda$ is precarious and oversmoothing is probable.

Indeed, the ad hoc nature of the filter in Eq.~\eqref{eq:filter} and the subjectivity associated with choosing $\lambda$ are both drawbacks of the Fourier-filtration procedure. Both these difficulties are alleviated by the inverse problem methodology presented in the next section. 

\begin{figure*}
    \centering    
	\includegraphics[width=0.325\textwidth]{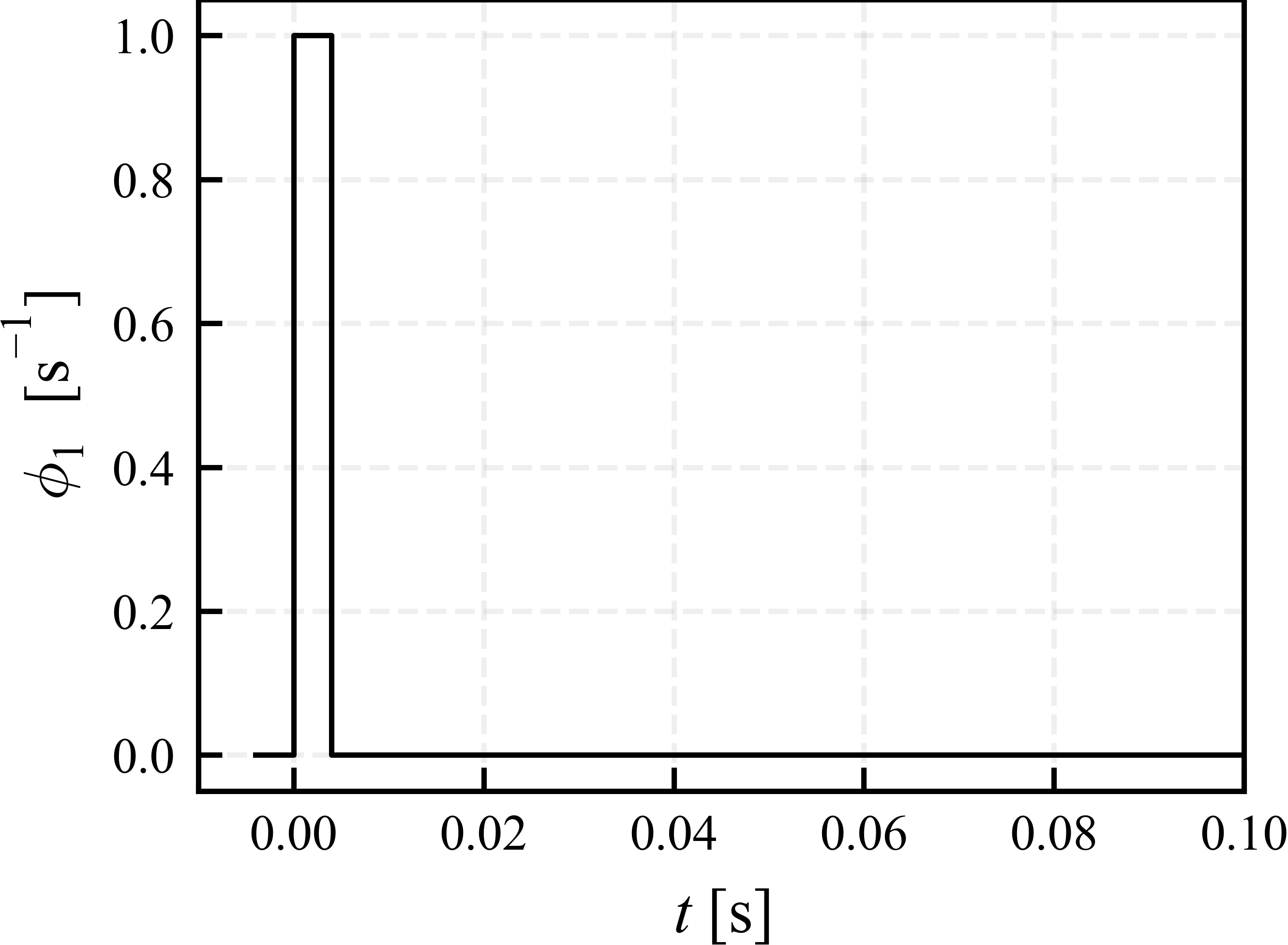}
     \put (-17,115){\footnotesize (a)}
    {\includegraphics[width=0.33\textwidth]{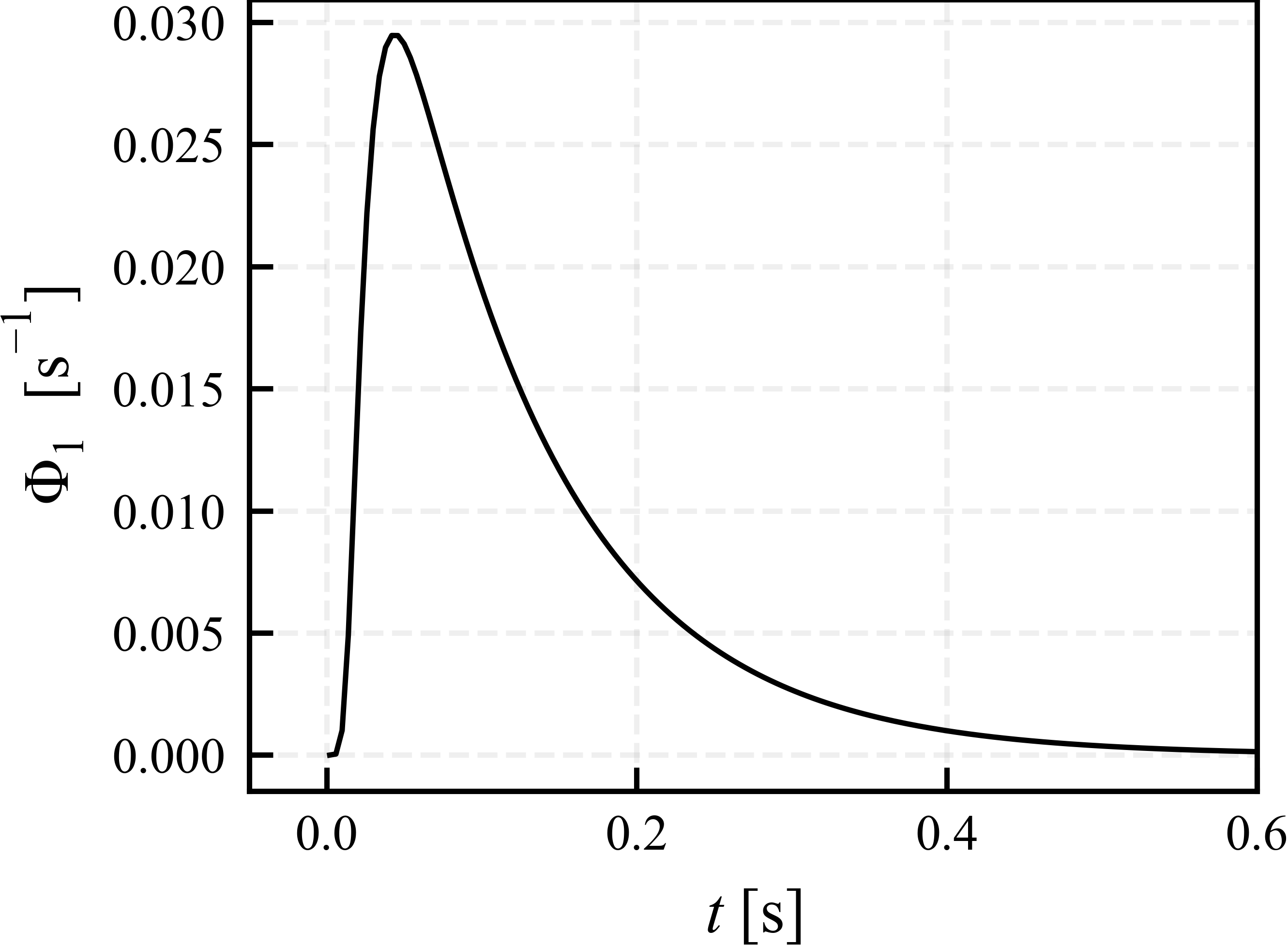}}  
    \put (-17,115){\footnotesize (b)}
    {\includegraphics[width=0.33\textwidth]{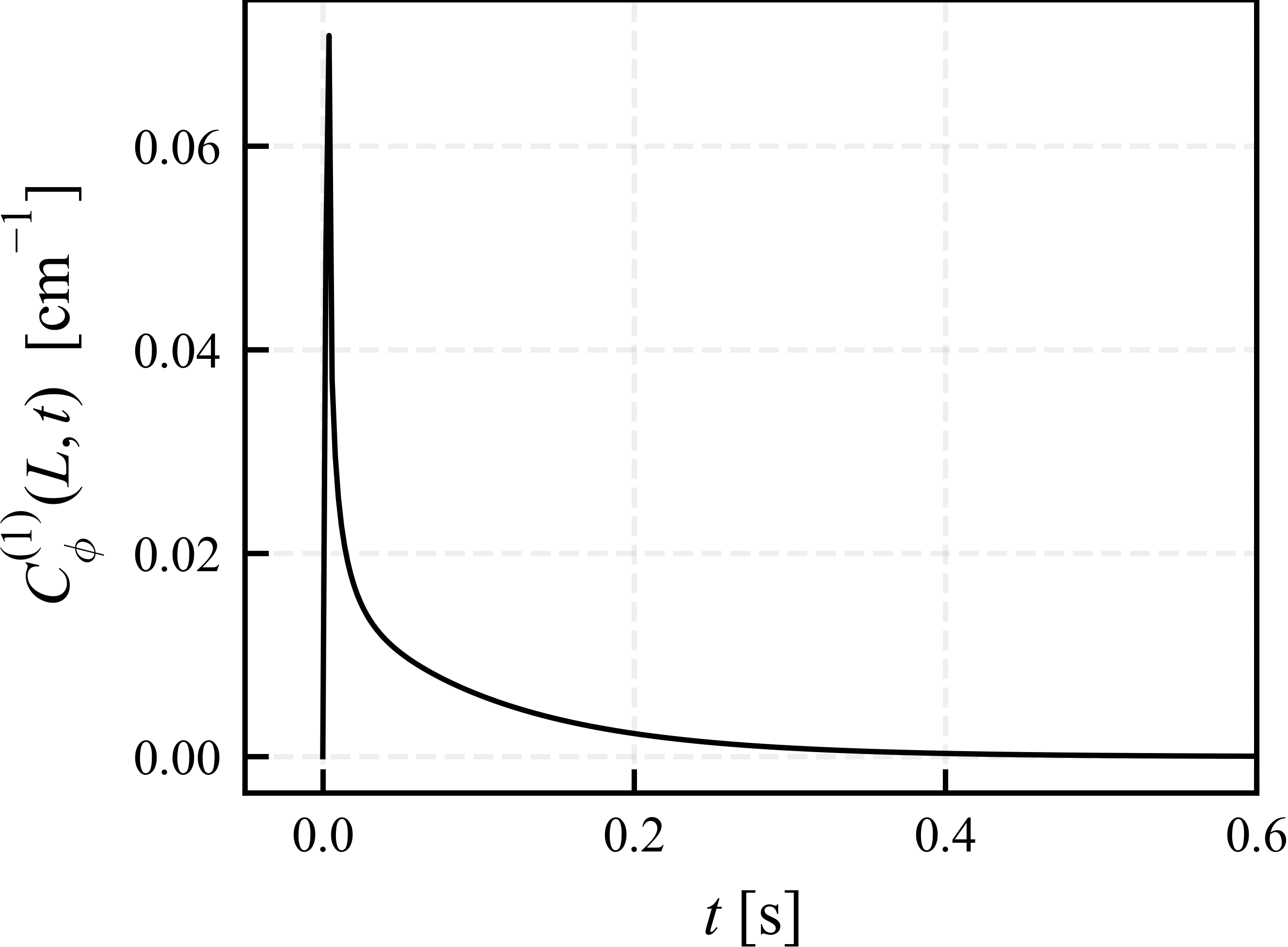}}  
    \put (-17,115){\footnotesize (c)}
	\caption{Illustration of (a) the first pulse function $\phi_1(t)$ of the basis used to represent the inlet flux to zone 3 (Eq.~\eqref{eq:basis}), along with (b) the corresponding outlet flux from zone 3, $\Phi_1(t)$, and (c) the corresponding temporal variation of the concentration at the inlet of zone 3, $C_\phi^{(1)}(L,t)$. The pulse has a width $\delta = T/M = 0.004$ \si{s} ($T = 2$ \si{s}, $M = 500$). All the basis functions $\phi_i$ ($i = 1, 2,..., M$) have the same profiles but are shifted in time (Eq.~\eqref{eq:phi-basis}). Note the difference in the scale of the time axis between panel (a) and panels (b), (c).
    }
	\label{fig:phi_Phi}
\end{figure*}

\section{The inverse problem methodology}
\label{sec:inverse}

\subsection{Formulation of a discrete inverse problem}\label{sec:discrete-inverse}

In this subsection, we formulate the problem of determining $F_3(L,t)$ and $C(t)=C_3(L,t)$ from $F_{out}^*(t)$, by using Eq.~\eqref{eqn:diffusion} in zone 3, as a discrete inverse problem of the form $\boldsymbol{A} \boldsymbol{w} = \boldsymbol{b}$, where $\boldsymbol{b}$ is the noisy measurement and $\boldsymbol{w}$ is the unknown vector that has to be reconstructed. The ill-posedness of the problem manifests in $\boldsymbol{A}$ having a very large condition number, due to which a naive direct solution (obtained by inverting $\boldsymbol{A}$) would yield a meaningless results for $\boldsymbol{w}$ that is dominated by amplified-noise. Such inverse problems have been well-studied and established methods exist for their regularized solution~\citep{Hansenbook2010}, including the TSVD method that will be applied in the subsequent subsection. 


The first step in formulating the inverse problem is to represent the flux entering zone 3, $F_3(L,t)$, as a sum of basis functions $\{\phi_i(t)\}_{i=1}^M$:
\begin{equation}\label{eq:basis}
    F_3(L,t)=\sum_{i=1}^{M} w_i \,\phi_i(t).
\end{equation}
Rather than using Fourier functions, which require periodicity and are ill-suited to describing localized sharp variations, we use a basis of square pulses. Specifically, the $i^\mathrm{th}$ basis function is a square pulse of width $\delta$ located at time $i\delta$:
\begin{equation}\label{eq:phi-basis}
    \phi_i(t)  = \begin{cases} 
          1 & \mathrm{if}\;\;(i-1) \delta \leq t \leq i\delta \\
          0 & \mathrm{otherwise}
       \end{cases}
\end{equation}
If the function $F_3(L,t)$ is defined over the time interval $[0,T]$ then $\delta = T/M$.

The next step is to determine the outlet flux that emerges from zone 3 when the input flux to zone 3 takes the form of any one basis function. Consider the first basis function $\phi_1(t)$, which is a unit-height square pulse lying between $t =0$ and $t = \delta$ (depicted in Fig.~\ref{fig:phi_Phi}(a)). 
Let us denote the concentration field inside zone 3 which would result from an input flux of $\phi_1(t)$ as $C^{(1)}_\phi(z,t)$ ($z \in [L,2L]$); let the corresponding outlet flux be $\Phi_1(t) = -\mathcal{D}\,\partial C^{(1)}_\phi/\partial z|_{z=2L}$. The concentration field $C^{(1)}_\phi$ is determined by solving the following diffusion equation:
\begin{equation}
	\epsilon \frac{\partial C^{(1)}_\phi}{\partial t}= \mathcal{D} \frac{\partial ^2 C^{(1)}_\phi}{\partial z^2},\quad z \in [L,2L], \;\; t>0,
\end{equation}
with boundary conditions
\begin{equation}
-\mathcal{D}\,\frac{\partial C^{(1)}_\phi}{\partial z}\bigg\rvert_{z=L} = \phi_1(t),\;\; C^{(1)}_\phi(2L,t) = 0.
\end{equation}
This forward problem can be solved numerically by the finite difference method or analytically using Laplace or Fourier transforms~\citep{YABLONSKY20076754}.
From the solution $C^{(1)}_{\phi}$, one can calculate the flux at the outlet, $\Phi_1(t)$, and the concentration at the inlet of zone-3, $C_\phi^{(1)}(L,t)$ (which will be used to reconstruct $C(t)$).
The analytical solutions for the Fourier transforms of $\Phi_1(t)$ and $C_\phi^{(1)}(L,t)$ are given below:
\begin{subequations}
\begin{align}
    \tilde{\Phi}_1(\omega)&=\frac{1}{\mbox{cosh}\sqrt{\mbox{i} \omega \tau}} \Bigg(\frac{1-\exp(-\mbox{i} \omega \delta)}{\mbox{i} \omega}\Bigg),\\
    \tilde{C}_\phi^{(1)}(L,\omega)&=\frac{\mbox{sinh} \sqrt{\mbox{i} \omega \tau}}{\gamma \sqrt{\mbox{i} \omega \tau}} \tilde{\Phi}_1(\omega).   
\end{align}
\end{subequations}

By evaluating these functions for a set of equispaced discrete frequencies $\omega = 2 \pi n/T$ ($n=-N_t/2+1,...,N_t/2$) and applying the IDFT one can obtain $\Phi_1(t)$ and $C_\phi^{(1)}(L,t)$ on a uniform temporal grid with resolution $T/N_t$. 
The corresponding solutions are presented in Figs.~\ref{fig:phi_Phi}(b,c).

Now, using the translational invariance of Eq.~\eqref{eqn:diffusion} and the fact that the different basis functions $\{\phi_i(t)\}_{i=1}^M$ have the same shape and are related to each other by temporal shifts, we can obtain the concentration field and the outlet flux corresponding to the $i^\mathrm{th}$ basis function by simply shifting $C^{(1)}_\phi(t)$ and $\Phi_1(t)$ in time by $i \delta$:
\begin{equation}
   C^{(i)}_\phi(t) = C^{(1)}_\phi(t+(i-1) \delta),\quad  \Phi_i(t) = \Phi_1(t+(i-1) \delta).
\end{equation}

Having solutions for all basis functions, we can use the principle of linear superposition to write the outlet flux from zone 3, $F_{out}^*(t)$, as a sum of $\{\Phi_i(t)\}_{i=1}^M$:
\begin{equation}\label{eq:Fout_Basis}
    F_{out}^*(t) = \sum_{i=1}^{M} w_i \Phi_i(t) = \sum_{i=1}^{M} w_i \Phi_1(t+(i-1) \delta)
\end{equation}
where the coefficients $w_i$ are the same as those in the basis expansion of the inlet flux to zone 3 (see Eq.~\eqref{eq:basis}). It remains to determine these coefficients given a measurement of $F_{out}^*(t)$.

\begin{figure*}
	\centering
    {\includegraphics[width=0.35\textwidth]{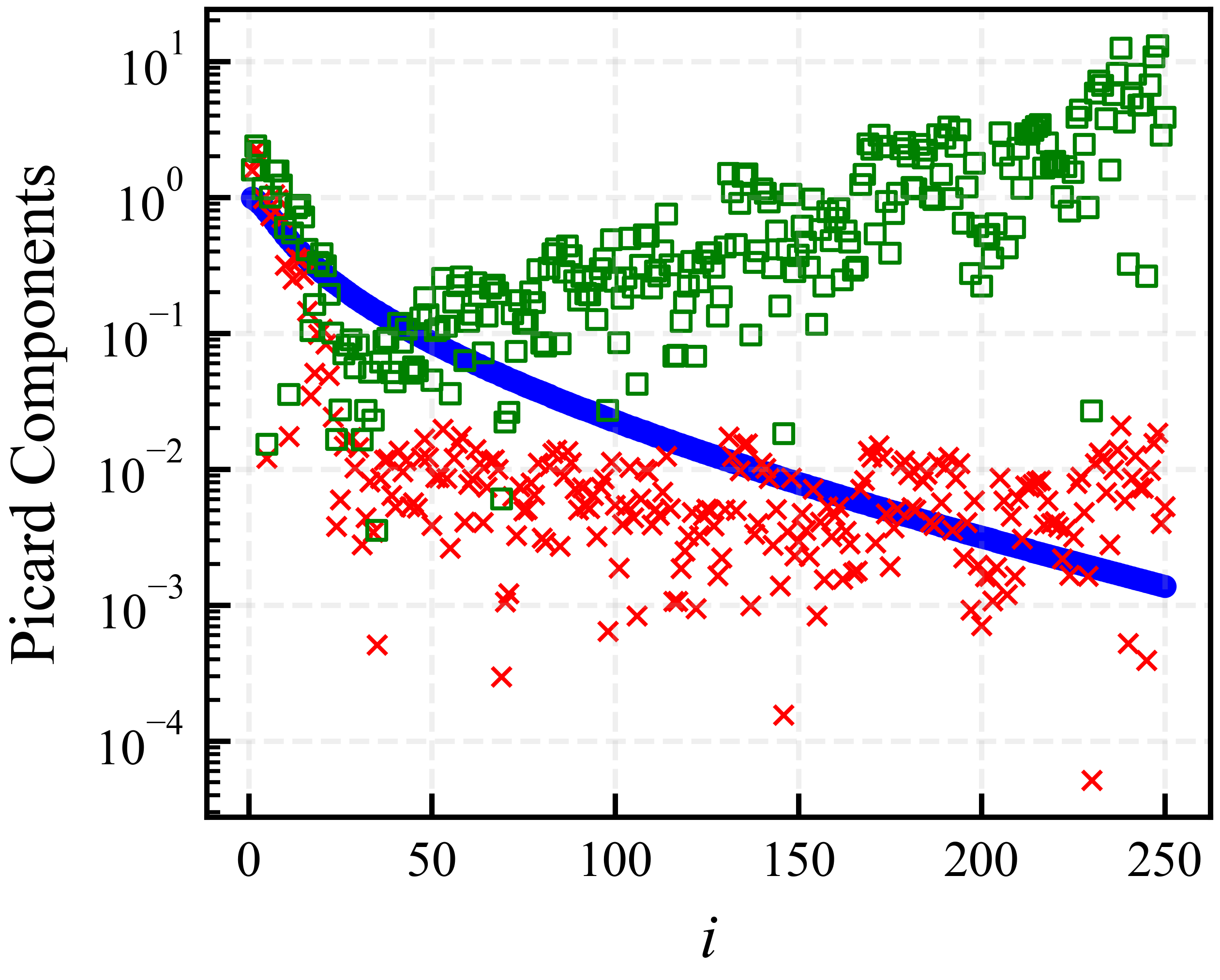}} 
    \put (-138,135){\footnotesize (a)}
\hspace{0.02\textwidth}
    {\includegraphics[width=0.35\textwidth]{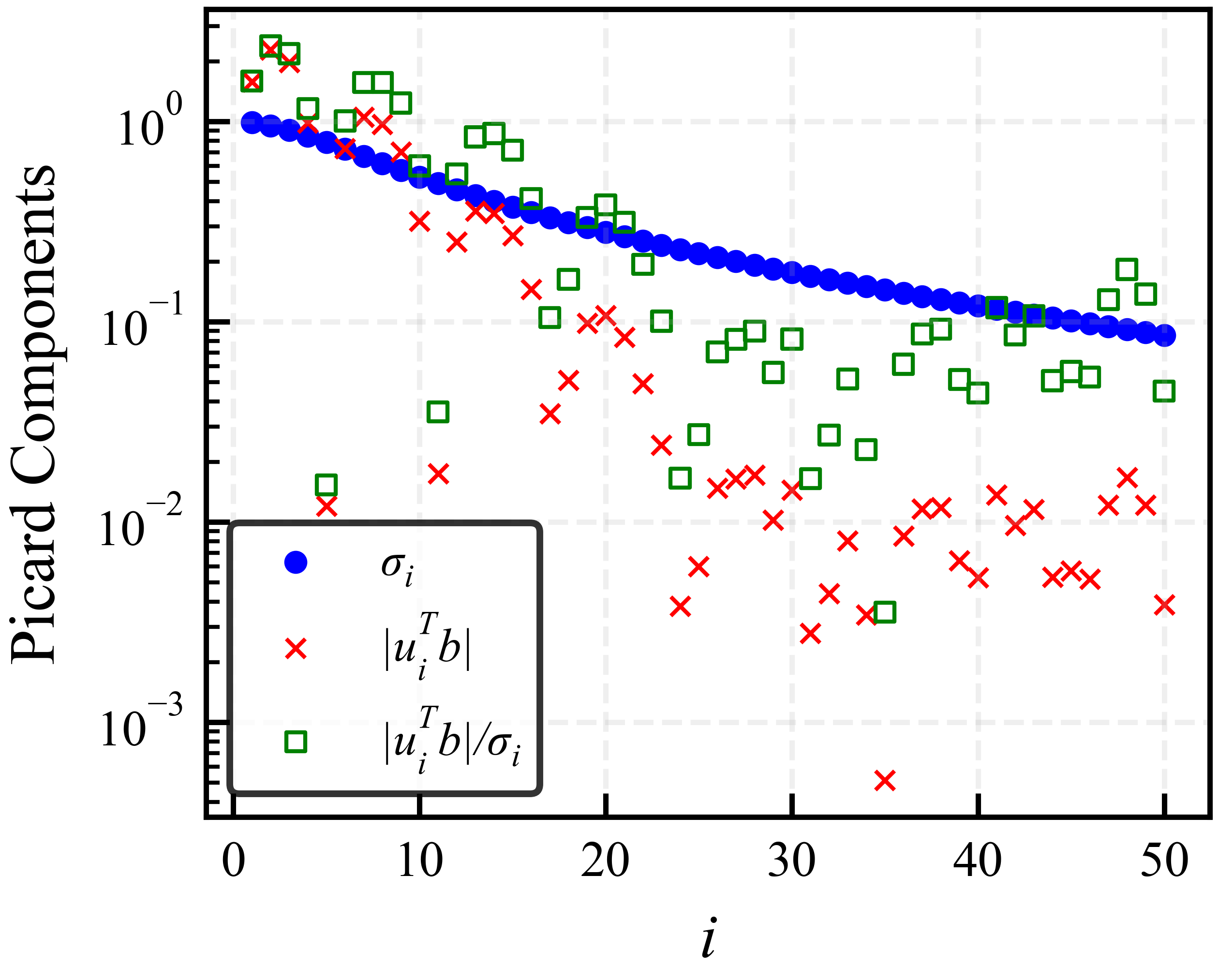}}  
    \put (-14,135){\footnotesize (b)}
	\caption{Picard plot showing the singular values $\sigma_i$ along with the magnitudes of the corresponding SVD components of the measured vector $\boldsymbol b$, $|u_i^Tb|$, and the solution vector $\boldsymbol{w}$, $|u_i^Tb|/\sigma_i$ (see the legend in panel (b)). The plot in (a) is restricted to the first 250 modes out of the total $M = 500$ modes, while the plot in (b) zooms into the transition region where noise first begins to dominate.} 
	\label{fig:Picard}
\end{figure*}

To obtain a system of $M$ linear algebraic equations for $\{w_i\}_{i=1}^M$, we use the Galerkin projection method. We multiply both sides of Eq.~\eqref{eq:Fout_Basis} by $\phi_j$ and then integrate over time:
\begin{equation}\label{eq:Fout_proj}
   \int_0^\infty \phi_j F_{out}^* \,dt = \sum_{i=1}^{M} w_i \int_0^\infty \Phi_i \,\phi_j \, dt,\quad j = 1,...,M,
\end{equation}
which on using the definition of $\phi_j$ simplifies to 
\begin{equation}\label{eq:Fout_proj_simple}
   \int_{(j-1)\delta}^{j\delta} F_{out}^* \,dt = \sum_{i=1}^{M} w_i \int_{(j-1)\delta}^{j\delta} \Phi_i \, dt.
\end{equation}
Given a measurement of $F_{out}^*(t)$, and knowing 
$\Phi_1(t)$, one may evaluate the integrals in the above equation numerically (using, e.g., the trapezoidal rule). Defining 
\begin{equation}\label{eq:A_b_def}
    A_{ji} \equiv 
    \int_{(j-1)\delta}^{j\delta} \Phi_1(t+(i-1) \delta) \, dt\;\; \mathrm{and} \;\; b_j = \int_{(j-1)\delta}^{j\delta} F_{out}^* \,dt,
\end{equation}
we obtain the following matrix equation for the vector of unknown coefficients $\boldsymbol{w}$:
\begin{equation} \label{eq:Aw_b}
    \boldsymbol{A} \boldsymbol{w} = \boldsymbol{b}.
\end{equation}
This is the discrete inverse problem whose regularized solution will allow us to reconstruct $F_3(L,t)$, via Eq.~\eqref{eq:basis}, as well as $C(t)$ via
\begin{equation}
    C(t) = \sum_{i=1}^M w_i C^{(i)}_\phi(L,t)= \sum_{i=1}^M w_i C^{(1)}_\phi(L,t+(i-1) \delta).
\end{equation} 
Recall that $C^{(1)}_\phi(L,t)$ is the concentration at the inlet of zone 3 corresponding to the first pulse basis function (see Fig.~\ref{fig:phi_Phi}(c)).

One can now proceed, in the usual manner, with the other steps of the Y-procedure, namely, the solution of the forward problem in zone 1 yielding $F_1(L,t)$, followed by the calculation of $R(t)$ (see Eq.~\eqref{eq:zone2}).

As noted at the beginning of this subsection, the ill-posed nature of calculating $F_3(L,t)$ and $C(t)$ by backtracking the diffusive transport in zone 3 manifests in $\boldsymbol{A}$ having a very large condition number. This implies that small errors in $\boldsymbol{b}$ will be greatly amplified so that a physically meaningful result for $\boldsymbol{w}$ cannot be obtained by inverting $\boldsymbol{A}$. Instead one must regularize the problem by seeking a vector $\boldsymbol{w}$ which minimizes the residual $\boldsymbol{A} \boldsymbol{w} - \boldsymbol{b}$ while also minimizing the noise in $\boldsymbol{w}$. One standard method is the TSVD, which is outlined below.

\subsection{Regularized solution via TSVD}\label{sec:TSVD}

\begin{figure*}
	\centering
    \includegraphics[width=0.33\textwidth]{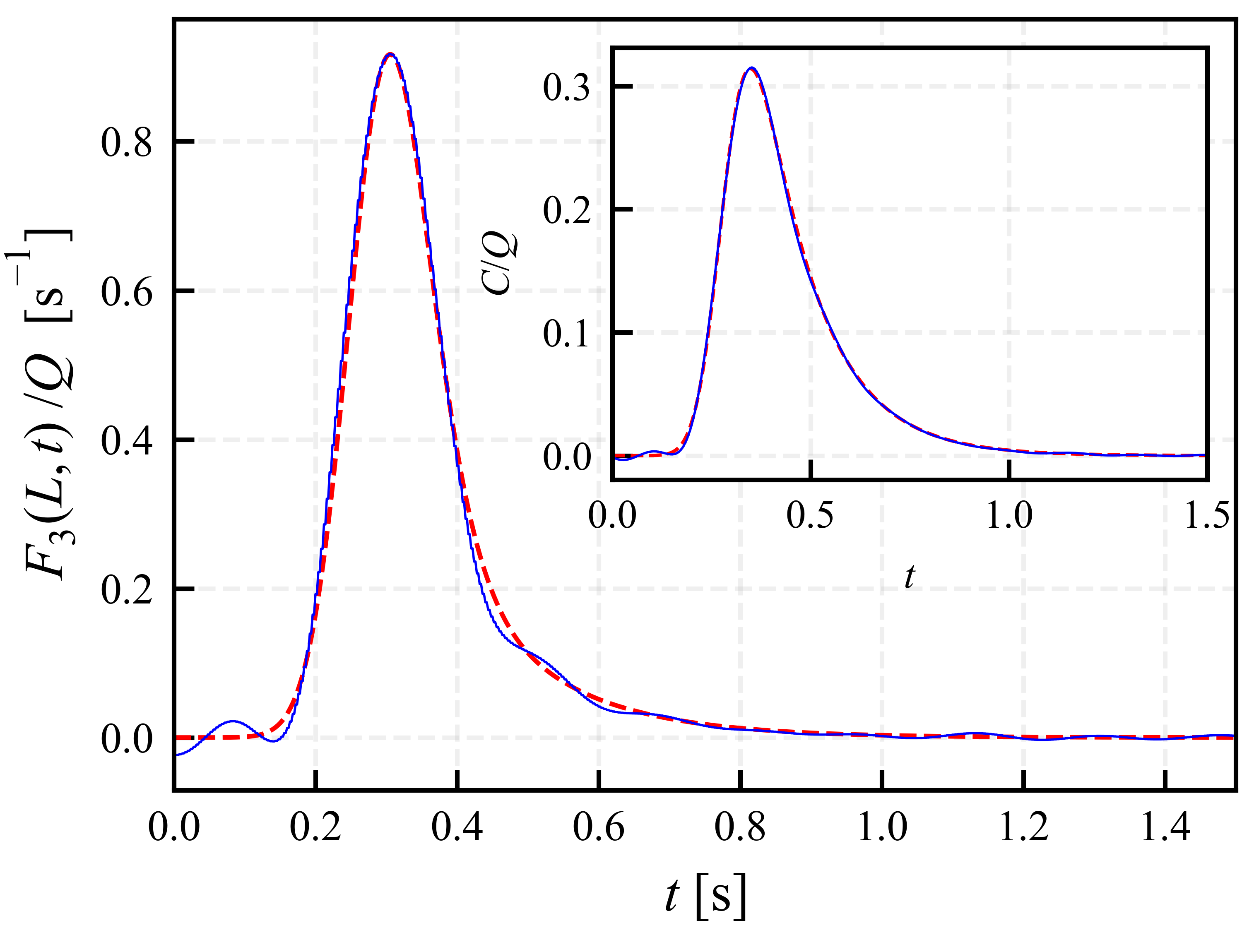}
    \put (-145,120){\footnotesize (a)}
	\includegraphics[width=0.33\textwidth]{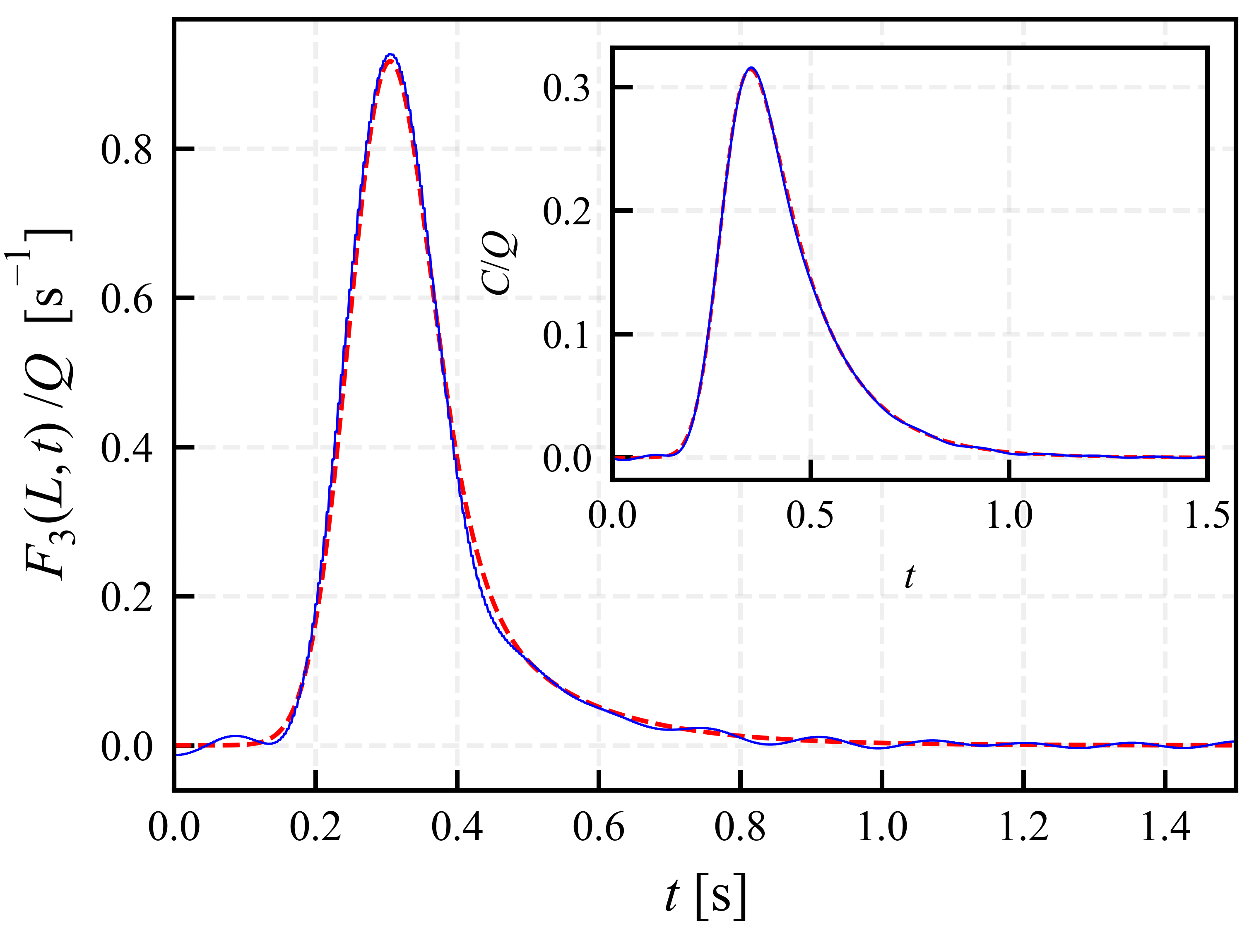}
    \put (-145,120){\footnotesize (b)}
	\includegraphics[width=0.33\textwidth]{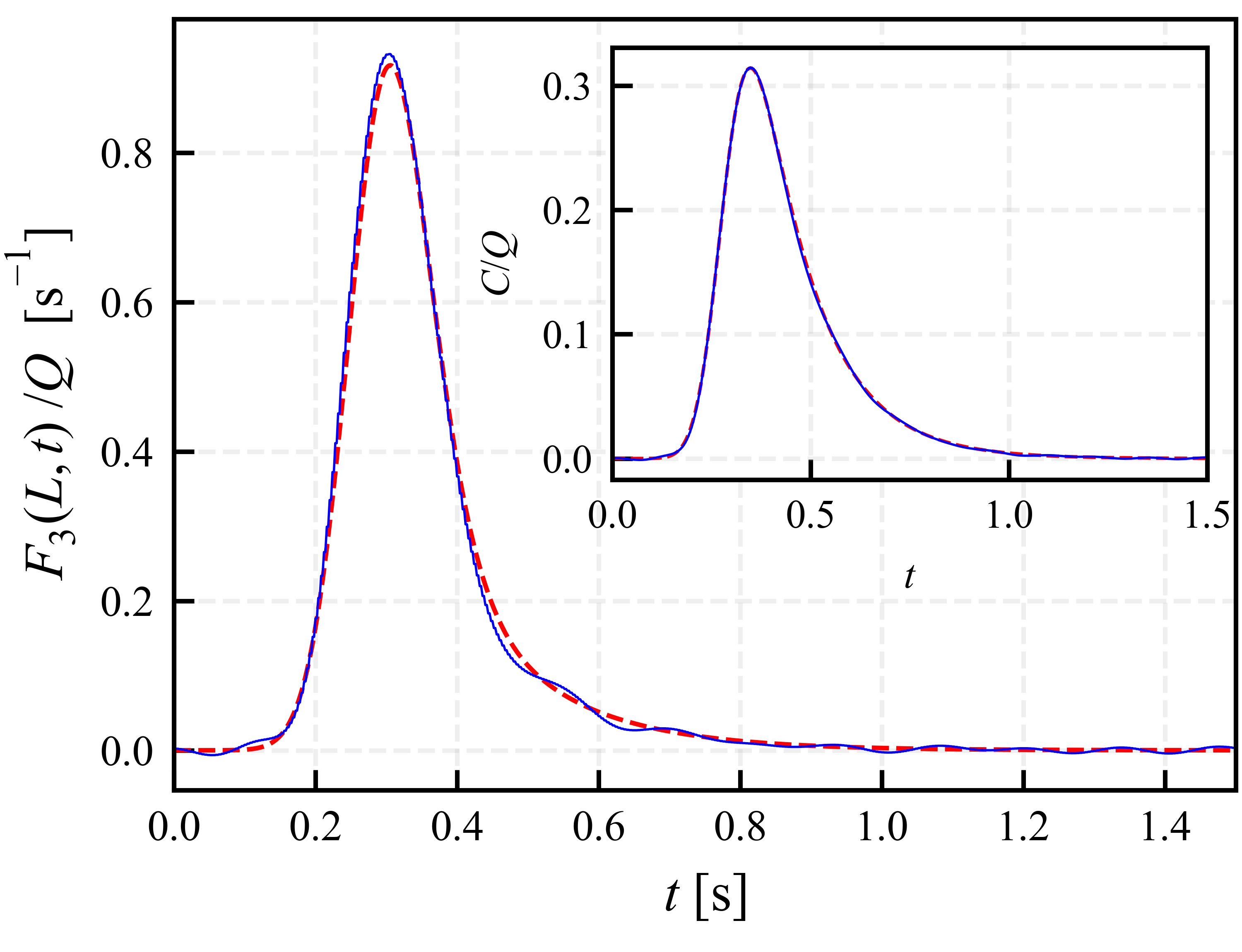}
    \put (-145,120){\footnotesize (c)}
    \hspace{0.02\textwidth}
	\includegraphics[width=0.33\textwidth]{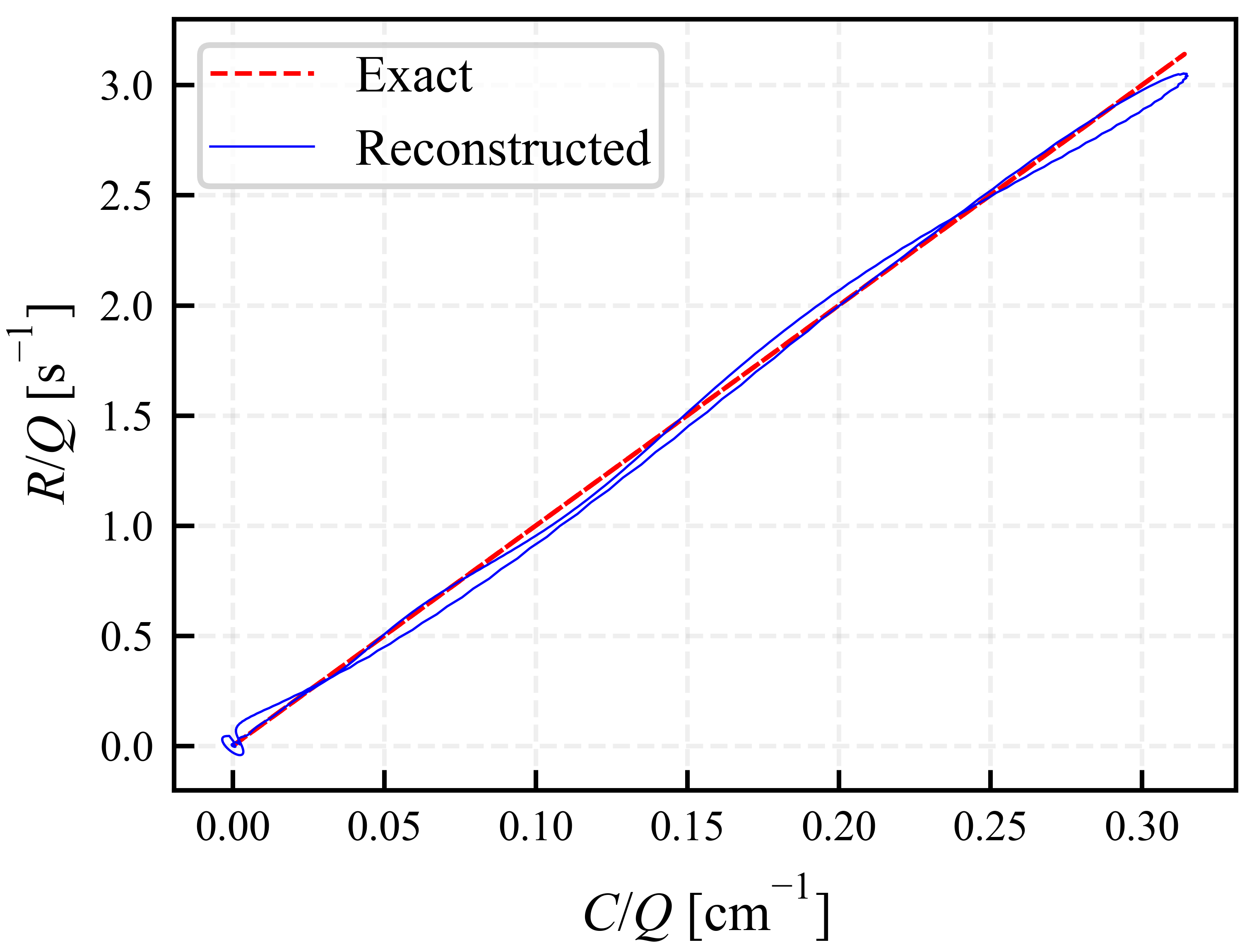} 
    \put (-15,30){\footnotesize (d)}
	\includegraphics[width=0.33\textwidth]{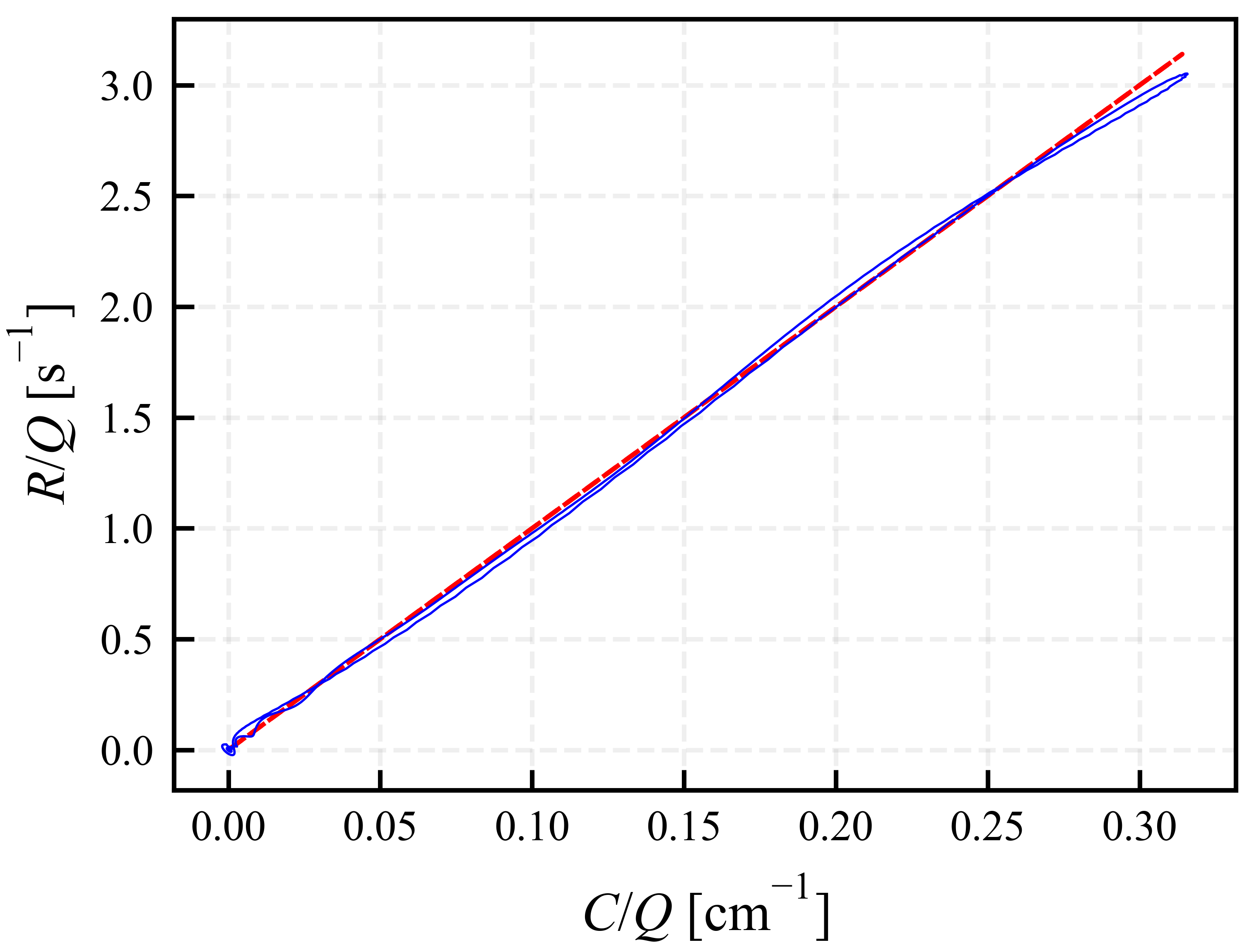}
    \put (-15,30){\footnotesize (e)}
	\includegraphics[width=0.33\textwidth]{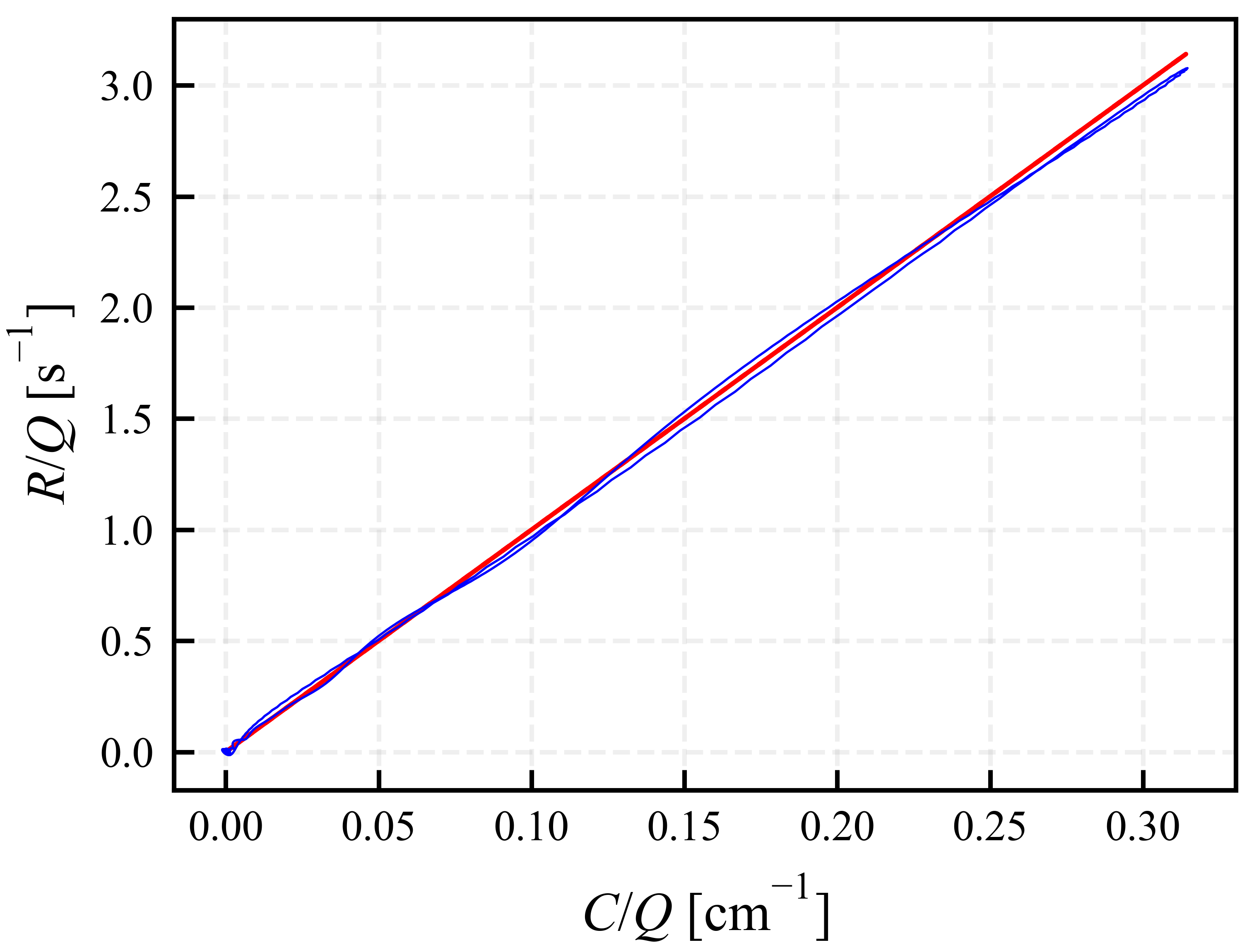}
    \put (-15,30){\footnotesize (f)}
	\caption{Y-procedure reconstruction using the inverse problem formulation and the TSVD regularized solution. The top row presents the inlet flux entering zone 3 (with the concentration in the thin reaction zone in insets), while the bottom row presents the reconstructed reaction-rate vs concentration curve. The three columns correspond to different values of the truncation mode number $m_\mathrm{cut}$: (a,d) 25, (b,e) 27 and (c,f) 30. The exact results from the forward simulation are shown for comparison along with the reconstructed results (see the legend in panel (d)).} 
	\label{fig:TSVD-linear}
\end{figure*} 

The TSVD method~\citep{Hansenbook2010} requires one to first compute the singular value decomposition (SVD) of the $M \times M$ matrix $\boldsymbol A$:

\begin{equation}
    \boldsymbol{A}=\boldsymbol{U \Sigma V}^T.
        \label{eq:SVD}
\end{equation}
Here $\boldsymbol U$ and $ \boldsymbol V$ are orthonormal matrices whose columns contain the $M$ left and right singular vectors, $\boldsymbol u_i$ and $\boldsymbol v_i$, respectively, while $\boldsymbol{\Sigma} = diag(\sigma_1,\sigma_2,..., \sigma_M)$ is a diagonal matrix containing the $M$ singular values in decreasing order of magnitude $(\sigma_1 \geq \sigma_2 \geq ...\geq \sigma_M >0)$. The two sets of $M$ vectors $\boldsymbol u_i$ and $\boldsymbol v_i$ form orthogonal bases for the range and the domain of the operator $\boldsymbol A$~\citep{roberts-linalg}. This means that $\boldsymbol{b}$ and $\boldsymbol{w}$ in Eq.~\eqref{eq:Aw_b} can be expressed as sums of these bases vectors: $\boldsymbol{b} = \sum_{i=1}^M (\boldsymbol{u}_i^T \boldsymbol{b})\,\boldsymbol{u}_i$ and $\boldsymbol{w} = \sum_{i=1}^M (\boldsymbol{v}_i^T \boldsymbol{w})\,\boldsymbol{v}_i$. 

On substituting the SVD decomposition of Eq.~\ref{eq:SVD} into Eq.~\eqref{eq:Aw_b}, one can obtain the direct solution for $\boldsymbol{w}$ (equivalent to $\boldsymbol{A}^{-1}\boldsymbol{b}$) as
\begin{equation}\label{eq:svdsol}
    \boldsymbol{w}_\mathrm{full}=\boldsymbol{V} \boldsymbol{\Sigma}^{-1} \boldsymbol{U}^T \boldsymbol{b} =\sum_{i=1}^{M} \frac{\boldsymbol{u}_i^T \boldsymbol{b}}{\sigma_i} \boldsymbol{v}_i.
\end{equation}

The measurement noise in $\boldsymbol{b}$ (arising from the noise in $F_{out}^*$, see Eq.~\eqref{eq:A_b_def}) will be present in the high-index (large $i$) high-frequency SVD modes that have small values of $\sigma_i$. This fact is illustrated by the Picard plot in Fig.~\ref{fig:Picard}(a), which depicts the variation of the singular values, along with the magnitudes of the SVD components of $\boldsymbol{b}$, $|\boldsymbol{u}_i^T \boldsymbol{b}|$, for $F_{out}^*$ in Fig.~\ref{fig:OF}. We use $M = 500$ basis functions to construct $\boldsymbol{A}$ and thus there are 500 SVD modes in total. Fig.~\ref{fig:Picard}(a) shows that the SVD components initially decrease systematically with $i$. In the absence of noise, this behaviour would have continued, but the presence of white noise stops the decrease. Indeed, those large-$i$ modes for which $|\boldsymbol{u}_i^T \boldsymbol{b}|$ fluctuates randomly without decreasing are the noise-dominated modes of $\boldsymbol{b}$. 

Eq.~\eqref{eq:svdsol} shows that these noisy components of $\boldsymbol{b}$ will be multiplied by $1/\sigma_i$ and thus get amplified in $\boldsymbol{w}$.
The Picard plot Fig.~\ref{fig:Picard}(a) also depicts the variation of the magnitude of the components of $\boldsymbol{w}$, $|\boldsymbol{u}_i^T \boldsymbol{b}|/\sigma_i$. The components decrease with $i$ initially but then start to increase, so that the magnitudes of the large-$i$ noisy components of $\boldsymbol{w}$ become comparable and even greater than that of the physically-relevant small-$i$ components. As a consequence, if all SVD modes are retained then the resulting solution $\boldsymbol{w}_\mathrm{full}$ will be dominated by noise. This observation naturally suggests a method for obtaining a meaningful solution for $\boldsymbol{w}$---simply truncate the SVD expansion and retain only the small $i$ components that are not dominated by noise. This truncated-SVD regularized solution is given by
\begin{equation}
\label{eq:tsvdsol}
\boldsymbol{w}_\mathrm{TSVD}=\sum_{i=1}^{m_\mathrm{cut}}\frac{\boldsymbol{u}_i^T \boldsymbol{b}}{\sigma_i} \boldsymbol{v}_i, \;\;\;{m_\mathrm{cut}} < M,
\end{equation}
where the truncation mode number $m_\mathrm{cut}$ is typically much smaller than $M$. 

The truncation mode-number $m_\mathrm{cut}$ is a regularization parameter like the filtration parameter $\lambda$ (Sec.~\ref{sec:filtration}) and must likewise be selected with care. Too small a value will produce a noisy recontruction while too large a value will lead to oversmoothing. Unlike the Fourier-filtration method, however, the noise dampening in the TSVD method is not done with an adhoc filter (Eq.~\eqref{eq:filter}) but is based on the SVD modes, which provide a problem specific basis since they are constructed from the matrix $\boldsymbol A$. Furthermore, the Picard plot provides guidance in choosing $m_\mathrm{cut}$ by revealing the mode-number beyond which noise begins to dominate the modes of $\boldsymbol{w}$. A good choice for $m_\mathrm{cut}$ would correspond to the transition region where the components stop decreasing and begin to increase. The corresponding SVD components contain meaningful physical information up to this index; beyond this, they mostly contain noise which would overwhelm the reconstructed $\boldsymbol{w}$.	Fig.~\ref{fig:Picard}(b) presents a zoom of the transition region and shows that the transition occurs near the mode number 26.


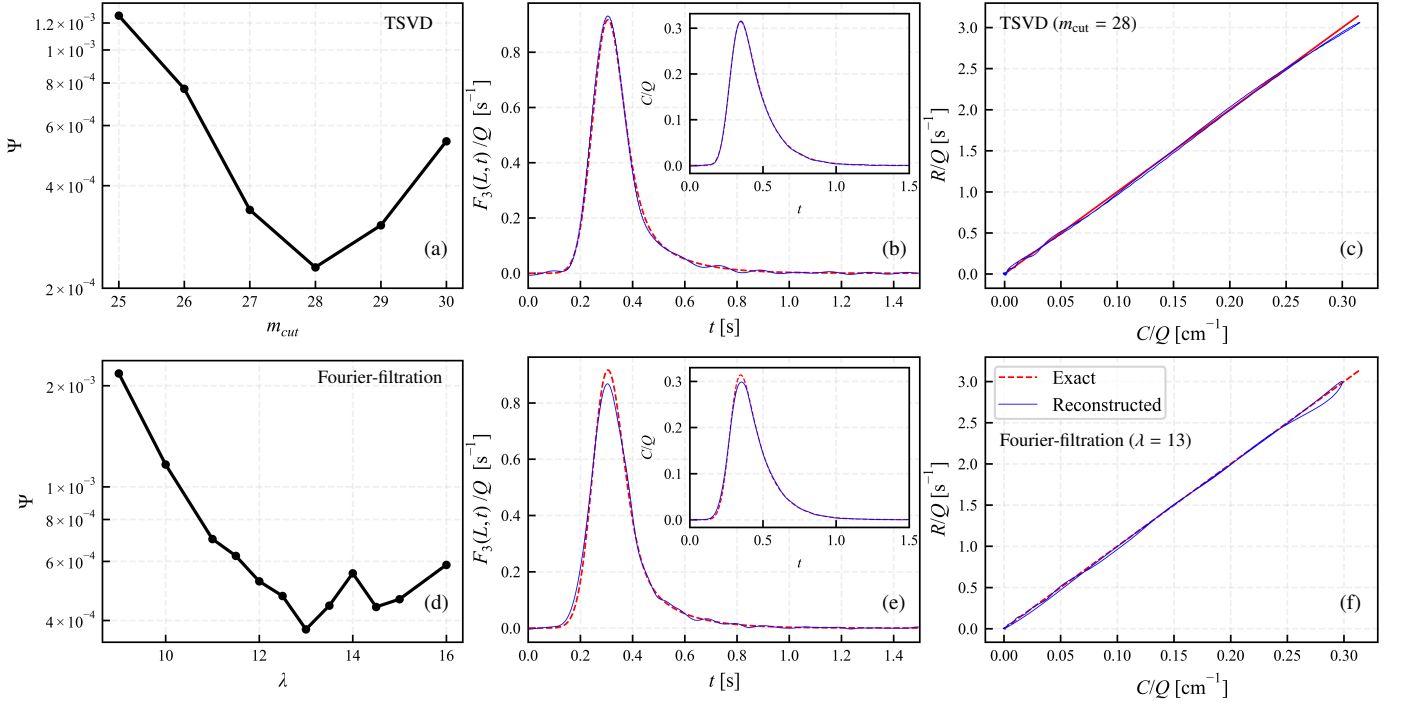
\begin{figure*}
	\centering
	\includegraphics[width=0.33\textwidth]{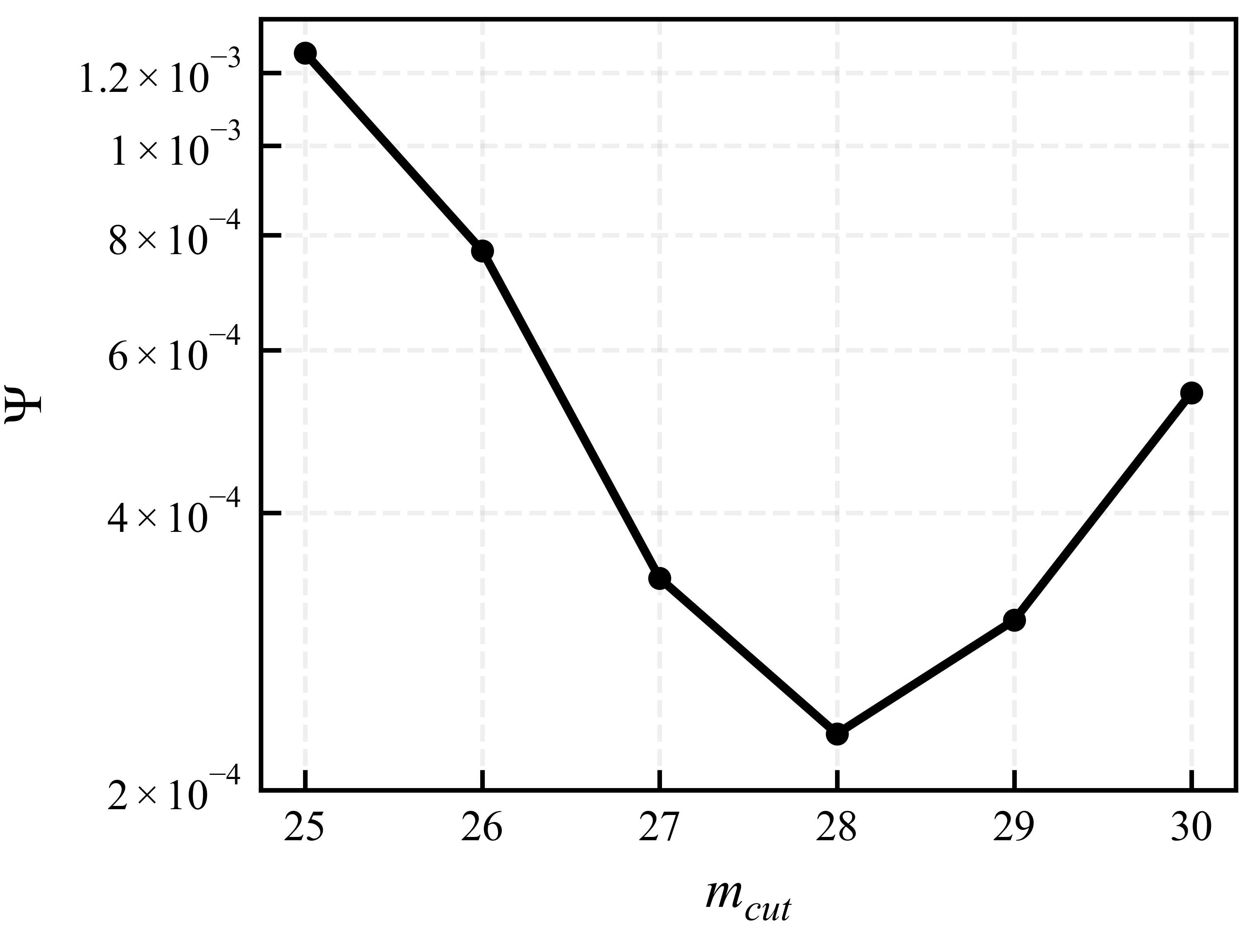}
    \put (-15,35){\footnotesize (a)}
    \put(-30,120){\scriptsize TSVD}
    \includegraphics[width=0.33\textwidth]{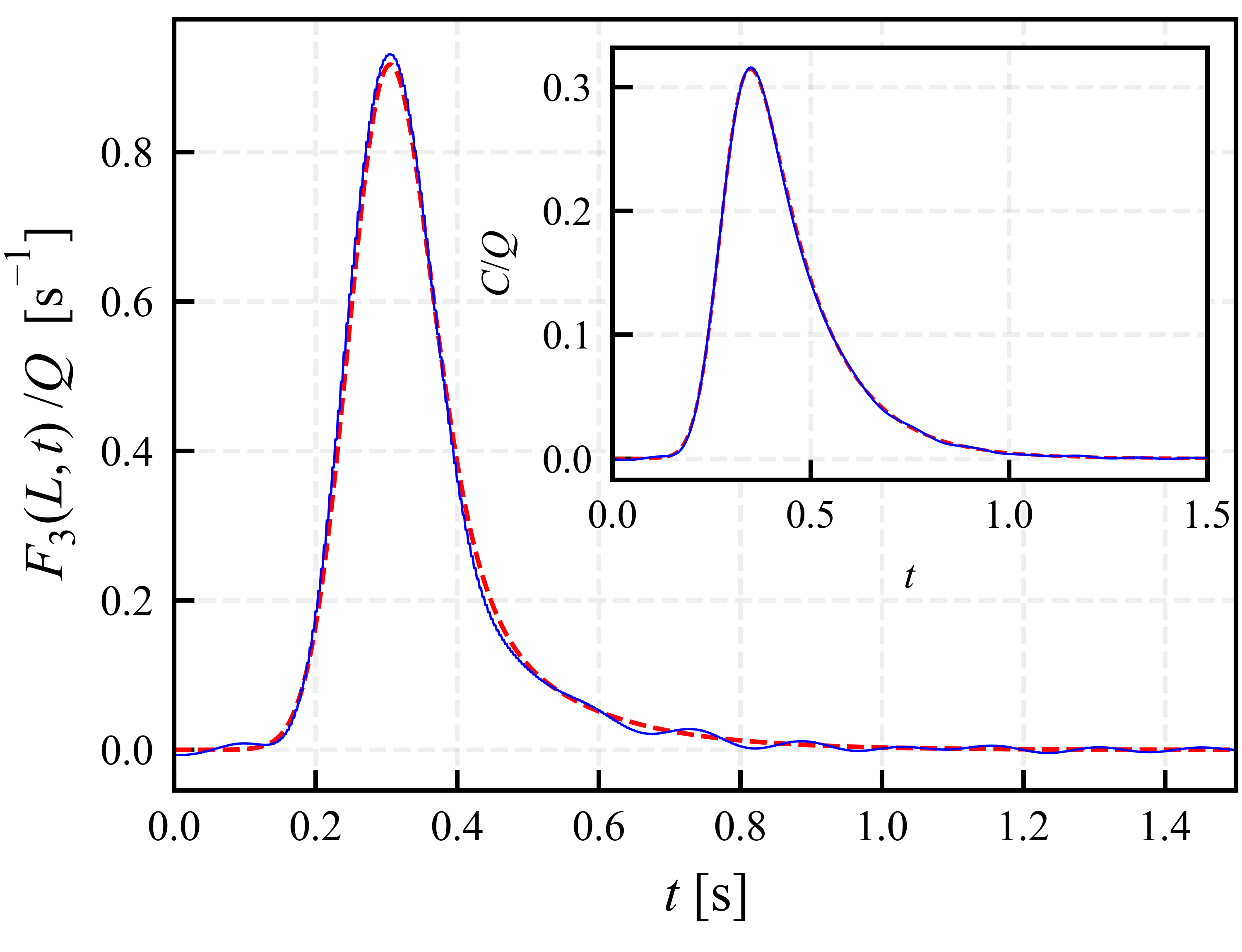}
    \put (-15,35){\footnotesize (b)}
    \includegraphics[width=0.33\textwidth]{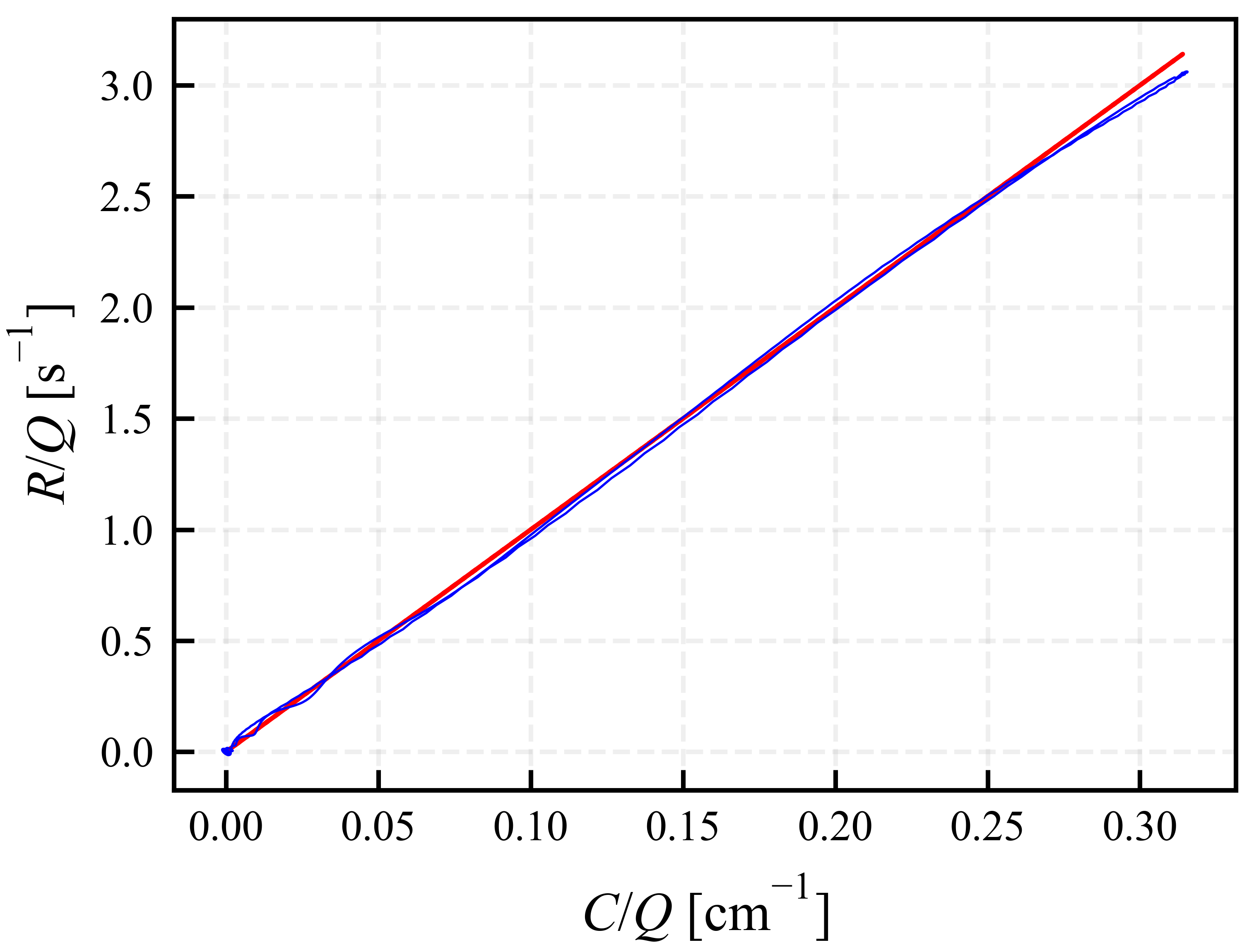}
     \put (-143,120){\scriptsize TSVD ($m_\mathrm{cut} = 28$)}
    \put (-15,35){\footnotesize (c)}
    \hspace{0.02\textwidth}
    \includegraphics[width=0.33\textwidth]{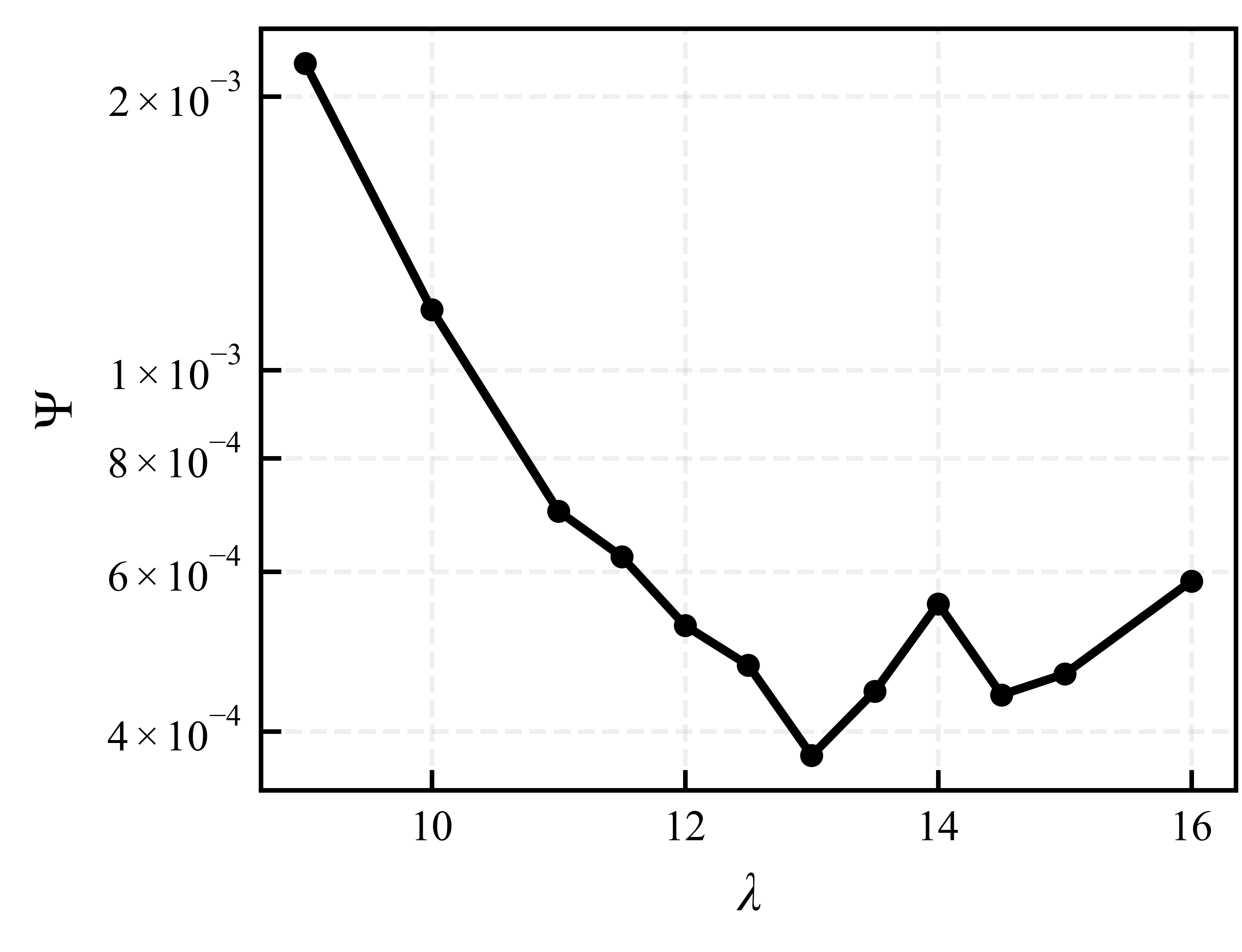}
    \put(-55,120){\scriptsize Fourier-filtration}
    \put (-15,35){\footnotesize (d)}
    \includegraphics[width=0.33\textwidth]{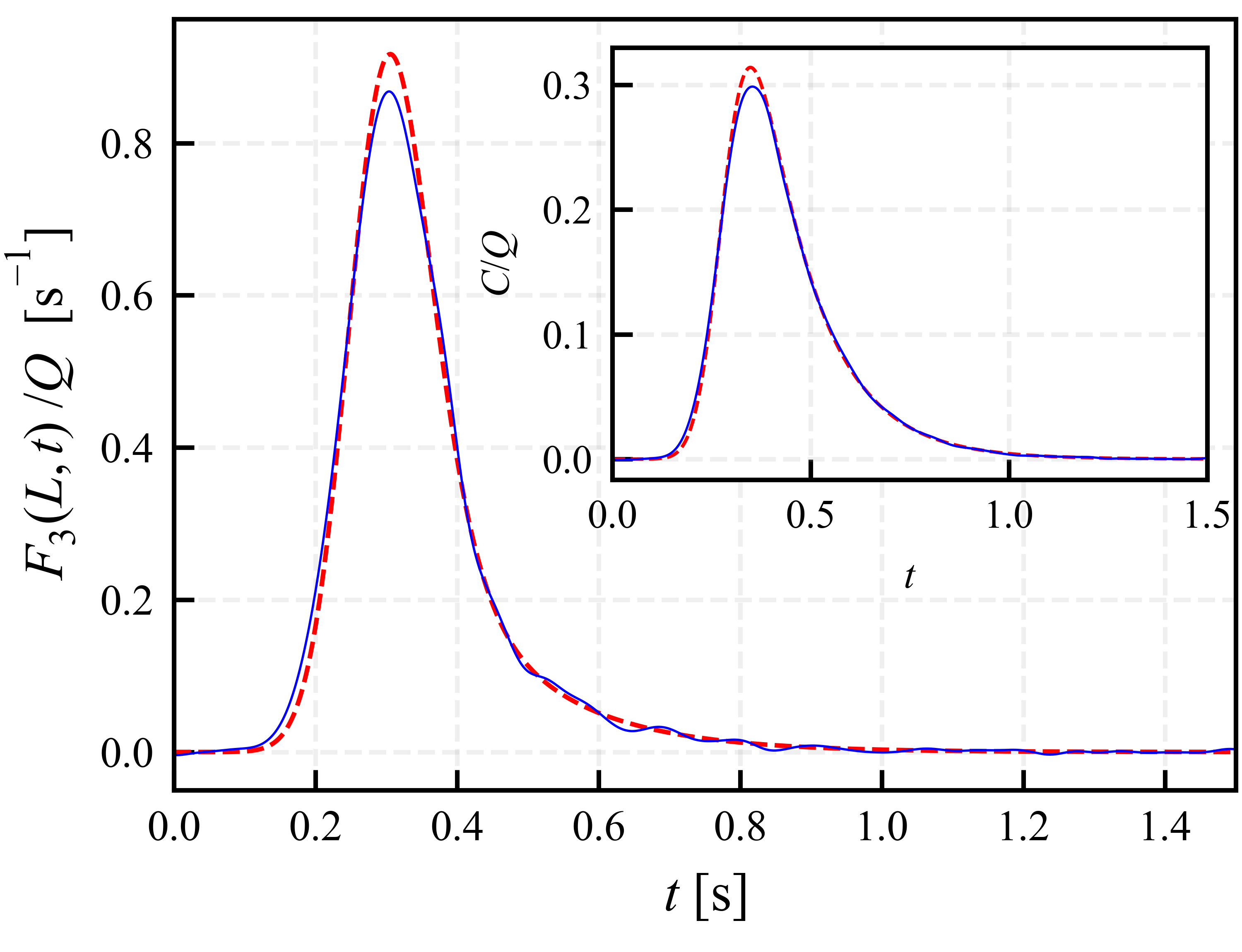}
    \put (-15,35){\footnotesize (e)}
    \includegraphics[width=0.33\textwidth]{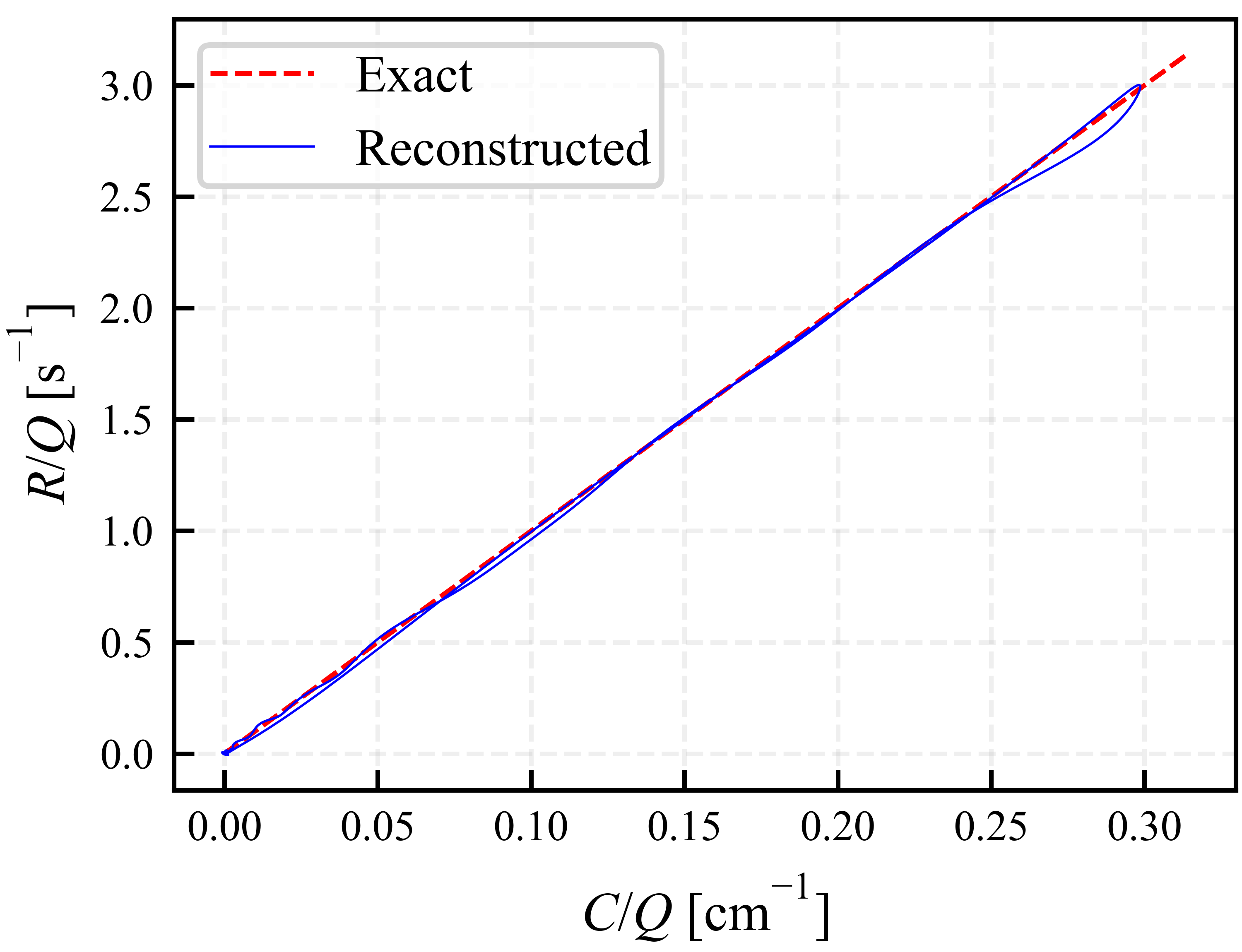}
    \put (-143,96){\scriptsize Fourier-filtration ($\lambda = 13$)}
    \put (-15,35){\footnotesize (f)}
	\caption{Selection of the regularization parameter for the TSVD method (top row) and the Fourier-filtration method (bottom row). (a) Variation of the degree of multivaluedness $\Psi$ (Eq.~\eqref{eq:psi}) with the cutoff mode number $m_\mathrm{cut}$, (b) reconstructed flux entering zone 3 (the inset shows the concentration in the thin reaction zone) for $m_\mathrm{cut} = 28$ where $\Psi$ is minimum, (c) corresponding reconstructed $R-C$ curve. (d) Variation of the degree of multivaluedness $\Psi$  with the Fourier-filtration parameter $\lambda$, (e) reconstructed flux entering zone 3 (the inset shows the concentration in the thin reaction zone) for $\lambda = 13$ where $\Psi$ is minimum, (f) corresponding reconstructed $R-C$ curve. 
    }
	\label{fig:psi}
\end{figure*}

Once a value of $m_\mathrm{cut}$ is selected and $\boldsymbol{w}_\mathrm{TSVD}$ is computed, one can reconstruct $F_3(L,t)$ and $C(t)$, and then $R(t)$, as discussed at the end of the previous section.
This TSVD based reconstruction (with $M = 500$) is shown in Fig.~\ref{fig:TSVD-linear} for three values of $m_\mathrm{cut}$ near the transition in the Picard plot: $25$ (Figs.~\ref{fig:TSVD-linear}(a,d)), $27$ (Figs.~\ref{fig:TSVD-linear}(b,e)) and $30$ (Figs.~\ref{fig:TSVD-linear}(c,f)). The top row presents the reconstructed flux entering zone 3 
while the bottom row shows the reaction rate function. The exact result from the forward simulation is shown for comparison. Since the values of $m_\mathrm{cut}$ were selected not arbitrarily but rationally on the basis of the Picard plot, we find that the TSVD results are quite good for all three values of $m_\mathrm{cut}$. While further fine-tuning could be done by adjusting $m_\mathrm{cut}$, we instead propose a new method for directly identifying the optimal value of $m_\mathrm{cut}$ in the next section.

Regarding the number of basis functions $M$, it is important to note that the direct solution $\boldsymbol{w}_\mathrm{full}$ will not converge as $M$ is increased. This is because increasing $M$ introduces higher frequency modes which will in turn amplify the noise in the solution. So the noise will keep growing as $M$ increases and the solution will diverge. In contrast, the regularized solution $\boldsymbol{w}_\mathrm{TSVD}$ converges  and increasing $M$ serves to refine the retained low-index SVD modes. Thus, with the TSVD method, the reconstructed profiles of $R(t)$ and $C(t)$ converge on increasing $M$; we find that $M = 500$ is sufficient for well-refined and converged reconstructions. 


\section{Determining the regularization parameter for state defining experiments}
\label{sec:single-valued}


In state defining experiments, where the concentration of adsorbed species on the catalyst surface does not change during the duration of a pulse, the reaction rate during the pulse is fully determined by the gas-phase concentration. As a consequence, the curve of $R$ vs $C$ will be single-valued \citep{REDEKOP20116441}. We now use this property to identify the optimal value of the regularization parameter $m_\mathrm{cut}$. When the gas pulse passes through the thin reaction zone, $R(t)$ and $C(t)$ both increase from zero, reach their maximum values simulaneously, and then decrease together to zero. In the $R-C$ plane, the passage of the pulse will trace out a curve that starts at the origin, extends outward in the first quadrant during the increasing phase of the pulse, and then returns to the origin during the decreasing phase of the pulse. In the absence of noise, or if we could perfectly reconstruct $R(t)$ and $C(t)$, the increasing and decreasing portions of the $R-C$ curve will exactly overlap to reveal the underlying single-valued apparent rate function $R(C)$. In practise, the presence of noise and the imperfect reconstruction of $R(t)$ and $C(t)$ leads to an imperfect overlap so that there are many portions of the $R-C$ curve where one has two values for $R$ at the same value of $C$ (as seen to varying extents in Figs.~\ref{fig:fourier}(d-f) and~\ref{fig:TSVD-linear}(d-f)). 

Now, in applications we will obviously not know the exact rate function or even its functional form (linear or otherwise). However, for state defining experiments, we do know that $R(C)$ should be single valued. And so we can use this knowledge to tune the regularization parameter, selecting the value which yields a $R-C$ curve that is closest to being single-valued. To implement this strategy, we define a degree of multivaluedness $\Psi$:
\begin{equation}
	\Psi=\int_0^{C_\mathrm{max}} \left[R_+(c)-R_{-}(c)\right]^2 ~ dc,
    \label{eq:psi}
\end{equation}
where $C_\mathrm{max}$ is the maximum value of $C$ in the reconstructed curve and $R_+$ and $R_-$ correspond respectively to the larger and smaller rate values at any given concentration $C$. If the reconstructed $R(C)$ were single-valued then $R_+ = R_-$ and $\Psi=0$. The larger the value of $\Psi$, therefore, the more the reconstructed curve deviates from being single-valued. 

\begin{figure*}
	\centering

    \includegraphics[width=0.33\textwidth]{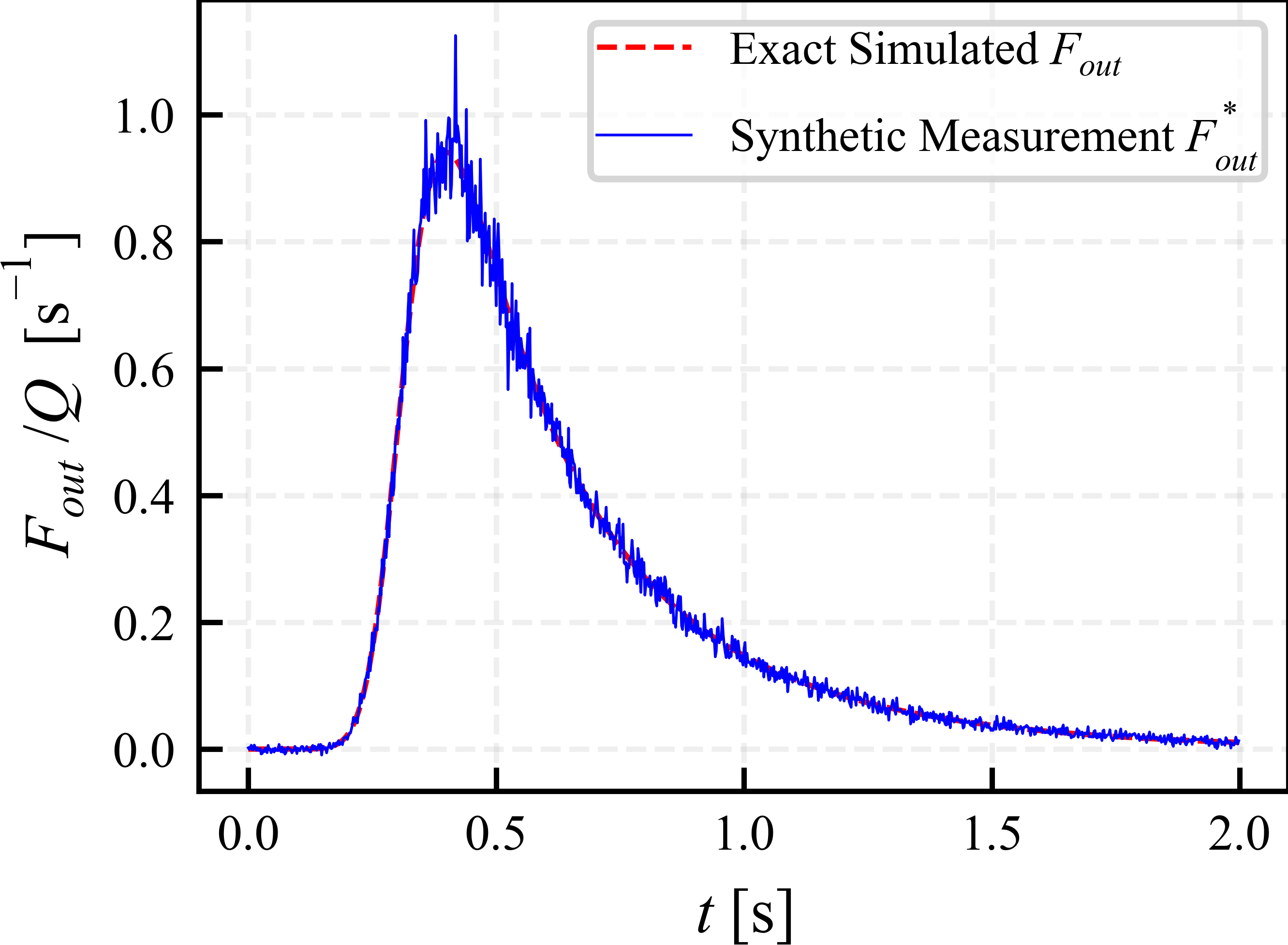}
    \put (-15,35){\footnotesize (a)}  
    \includegraphics[width=0.33\textwidth]{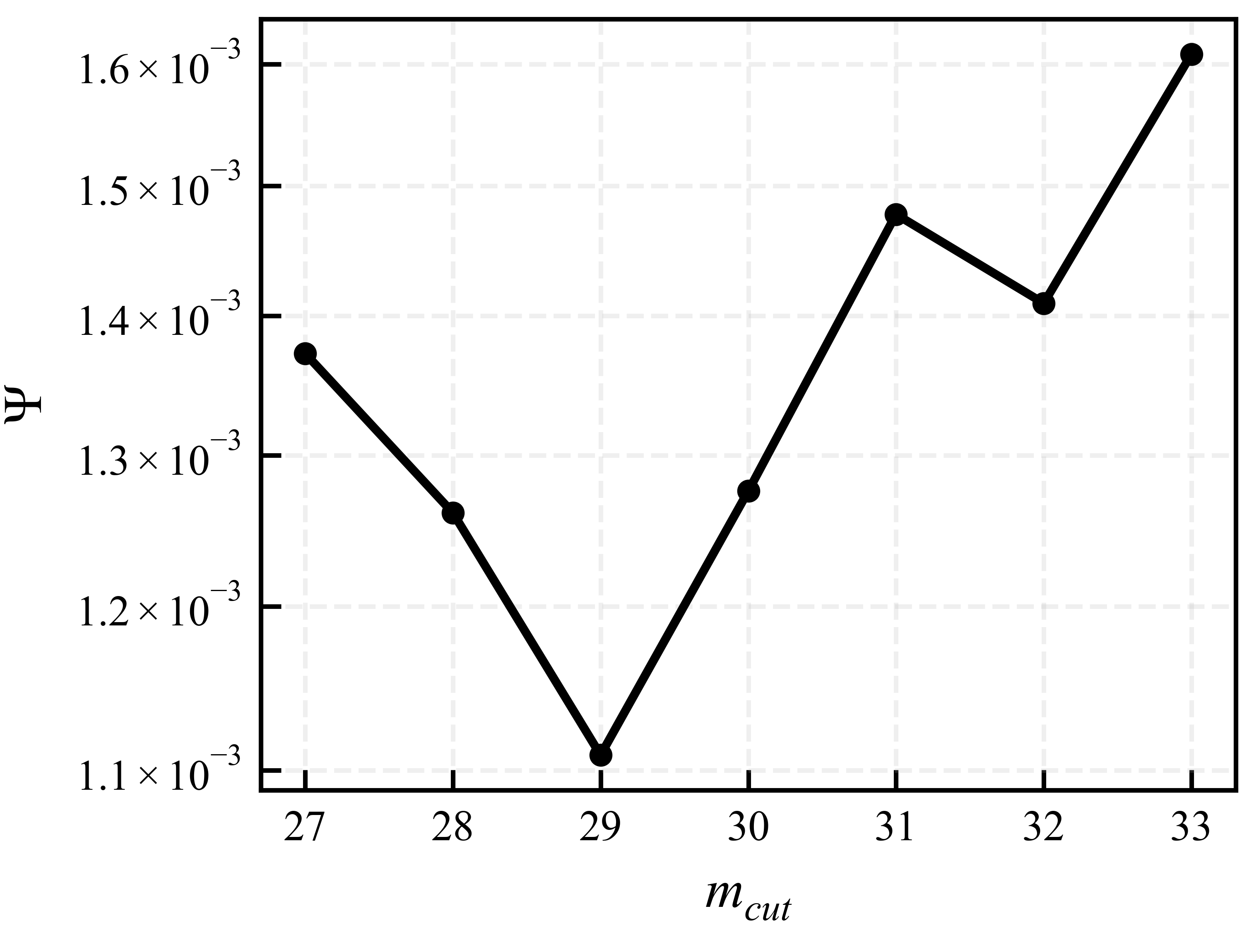}
    \put(-130,120){\scriptsize TSVD}
    \put (-15,35){\footnotesize (b)}
        \includegraphics[width=0.33\textwidth]{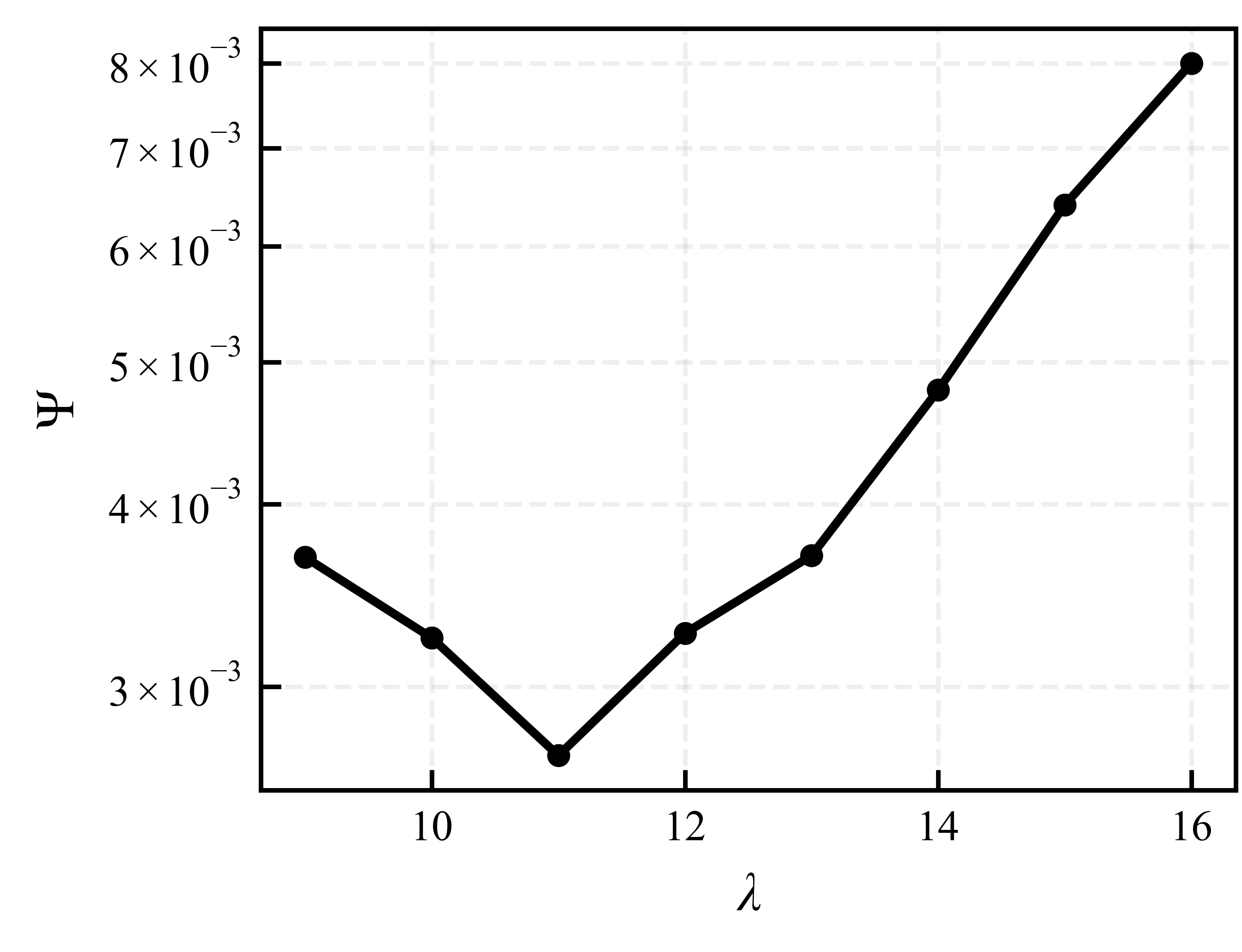}
        \put(-130,120){\scriptsize Fourier-filtration}
    \put (-15,35){\scriptsize (c)}

        \includegraphics[width=0.33\textwidth]{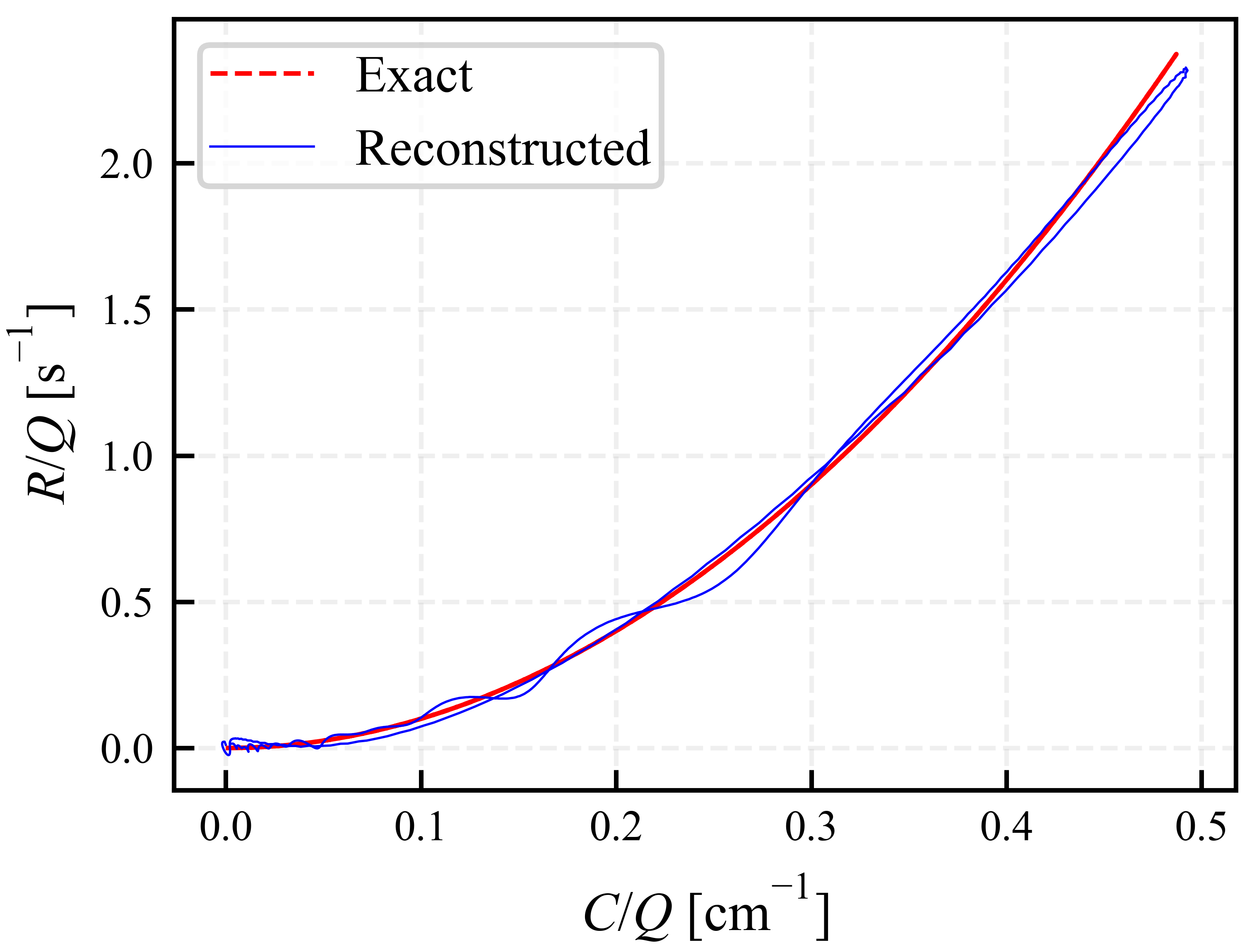}
    \put (-15,35){\footnotesize (d)}
    \put (-143,96){\scriptsize TSVD ($m_\mathrm{cut} = 29$)}
    \includegraphics[width=0.33\textwidth]{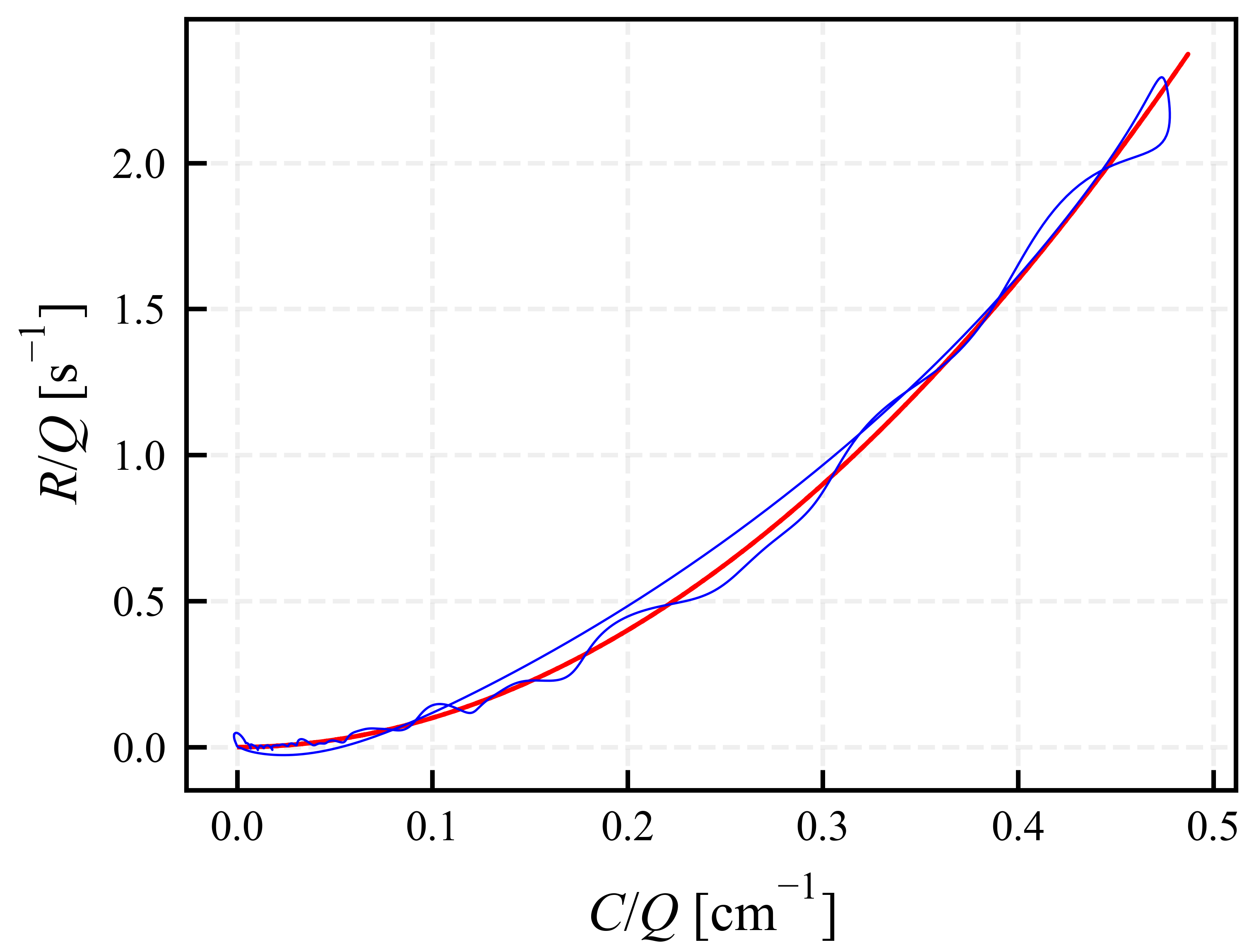}
    \put (-15,35){\footnotesize (e)}
    \put (-143,120){\scriptsize Fourier-filtration ($\lambda = 11$)}
    \includegraphics[width=0.33\textwidth]{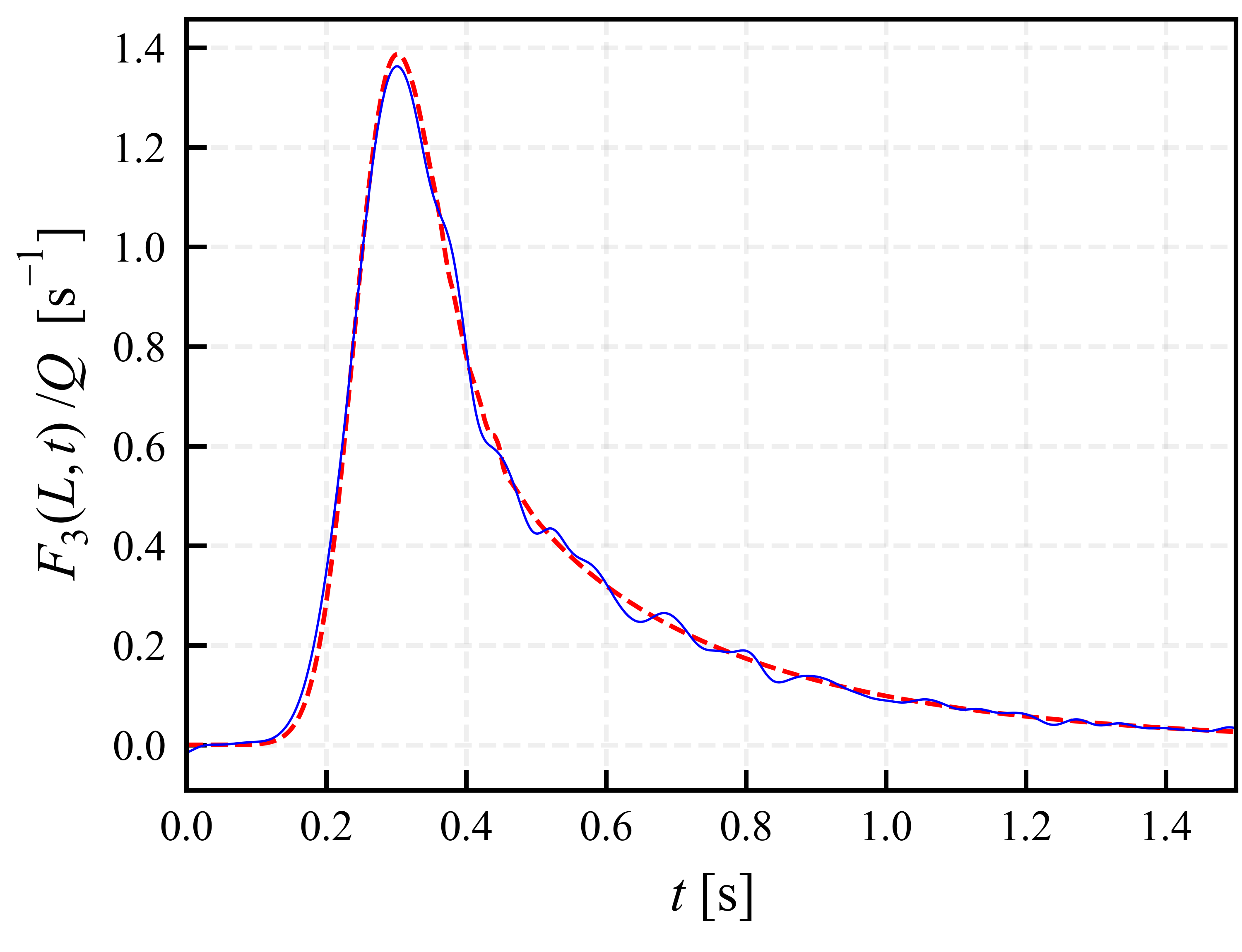}
    \put (-15,35){\footnotesize (f)}
    \put (-60,120){\scriptsize Fourier-filtration }

	\caption{Y-procedure reconstruction for a quadratic reaction using the TSVD method and Fourier-filtration, with the corresponding optimal regularization parameter values. (a) The synthetic measurement of the outlet flux, (b,c) Variation of the degree of multivaluedness $\Psi$ (Eq.~\eqref{eq:psi}) with the cutoff mode number $m_\mathrm{cut}$ and Fourier-filtration parameter $\lambda$, respectively. (d,e) Optimal reconstructed $R-C$ curves from the TSVD method and Fourier-filtration. respectively. (f) Optimal reconstructed flux entering zone 3 from the Fourier filtration method.}
	\label{fig:quadratic}
\end{figure*}

\begin{figure*}
	\centering
        \includegraphics[width=0.33\textwidth]{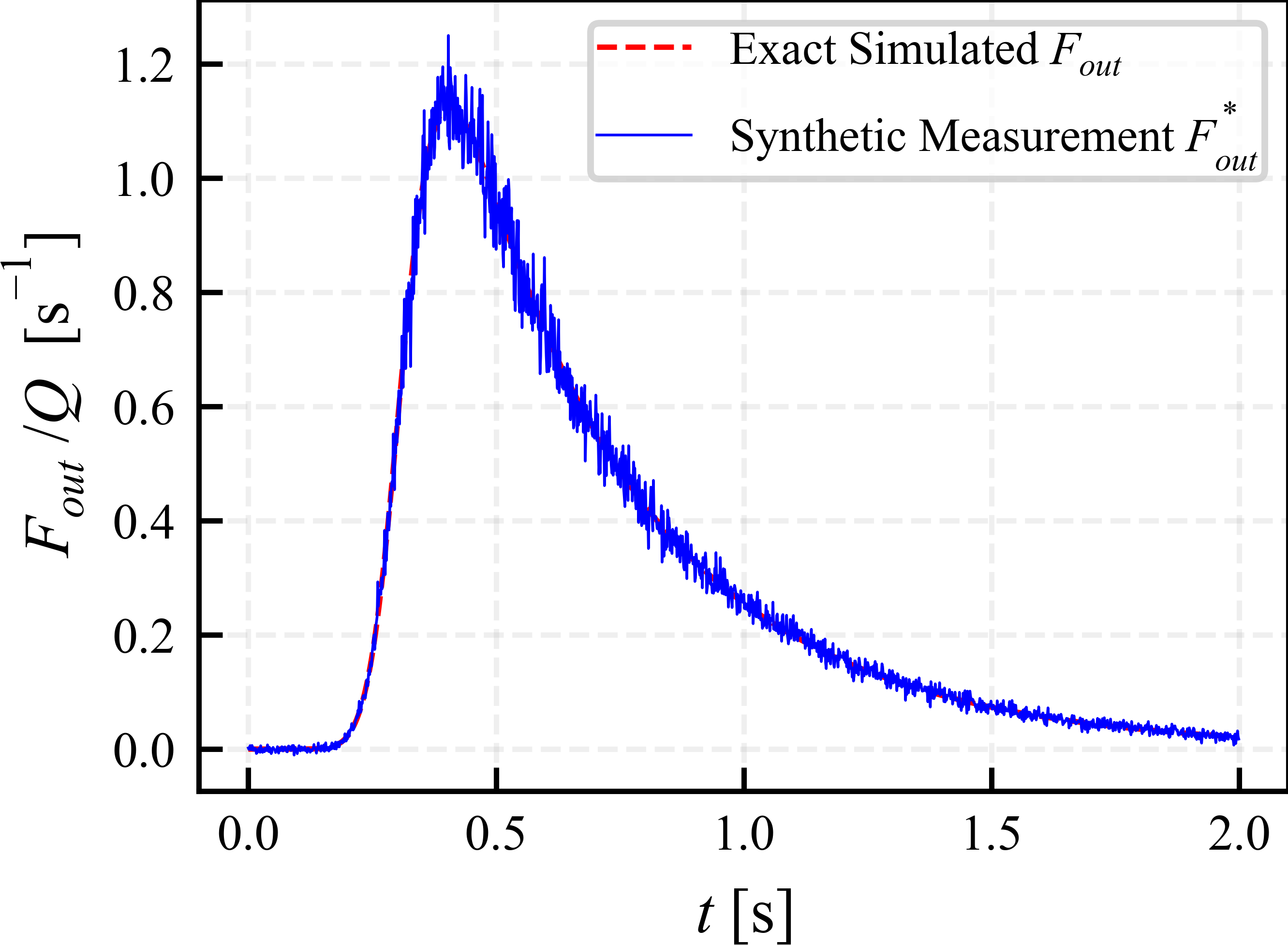}
    \put (-15,35){\footnotesize (a)}
    \includegraphics[width=0.33\textwidth]{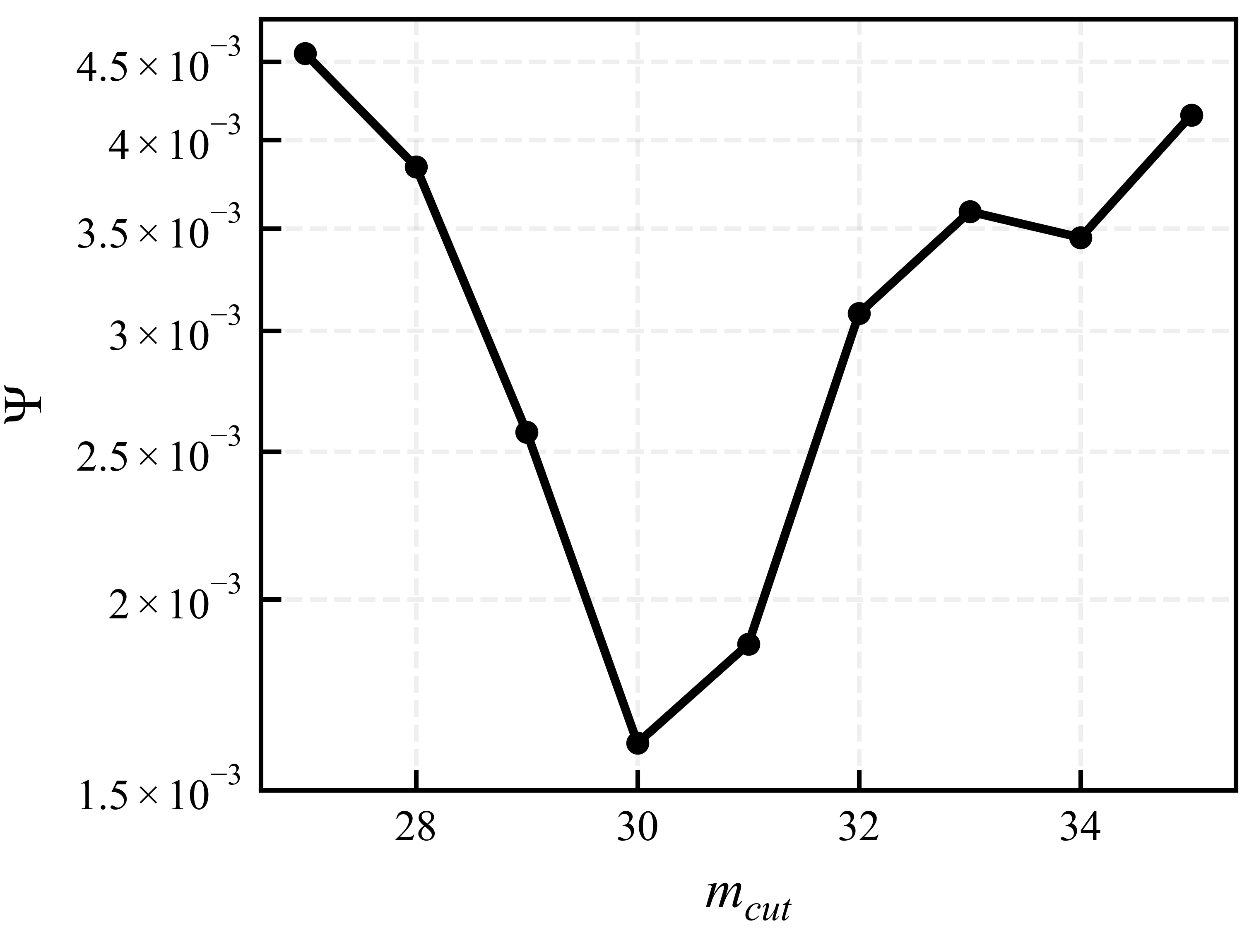}
    \put(-35,120){\scriptsize TSVD}
    \put (-15,37){\footnotesize (b)}
    \includegraphics[width=0.33\textwidth]{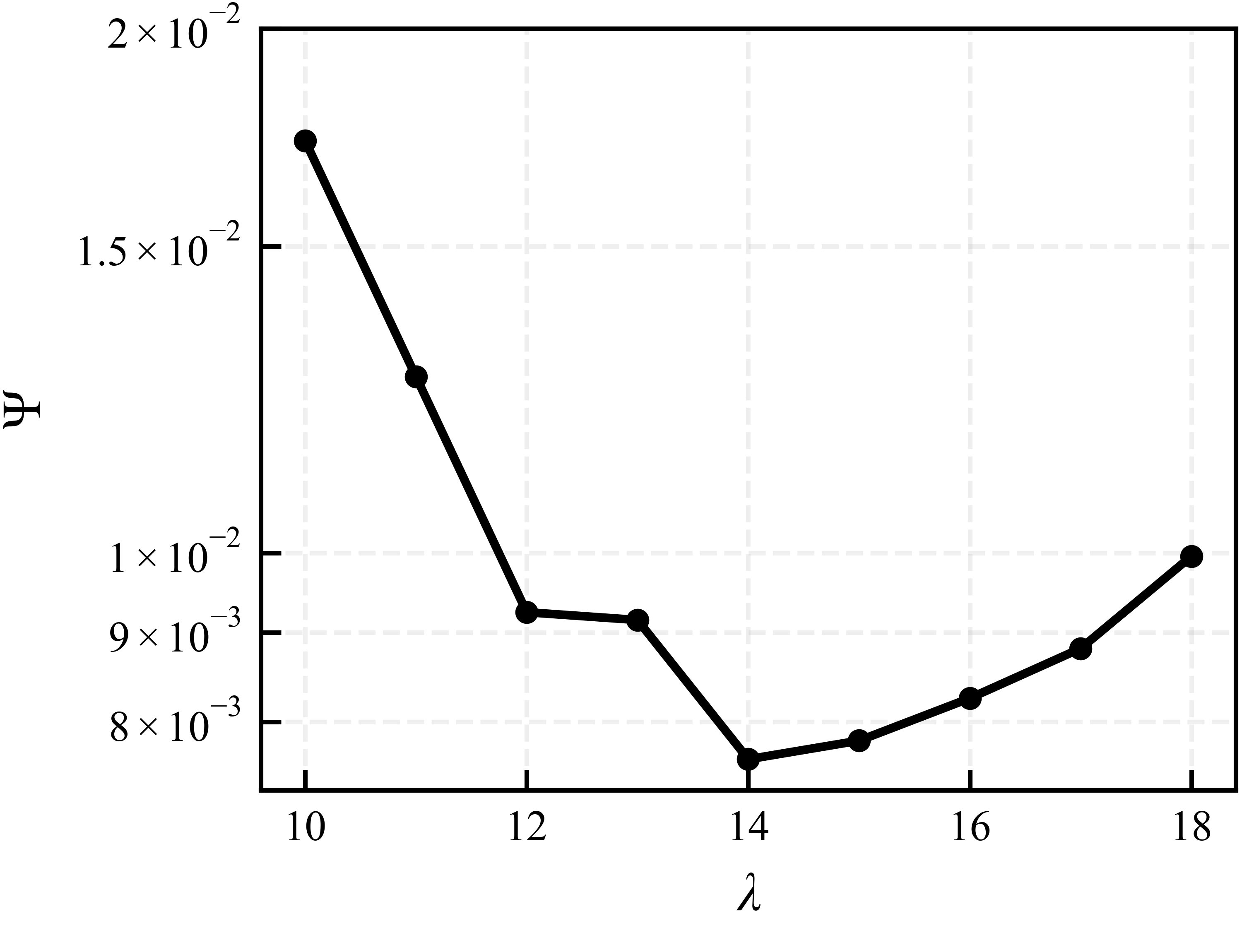}
    \put(-60,120){\scriptsize Fourier-filtration}
    \put (-15,37){\footnotesize (c)}

    \includegraphics[width=0.33\textwidth]{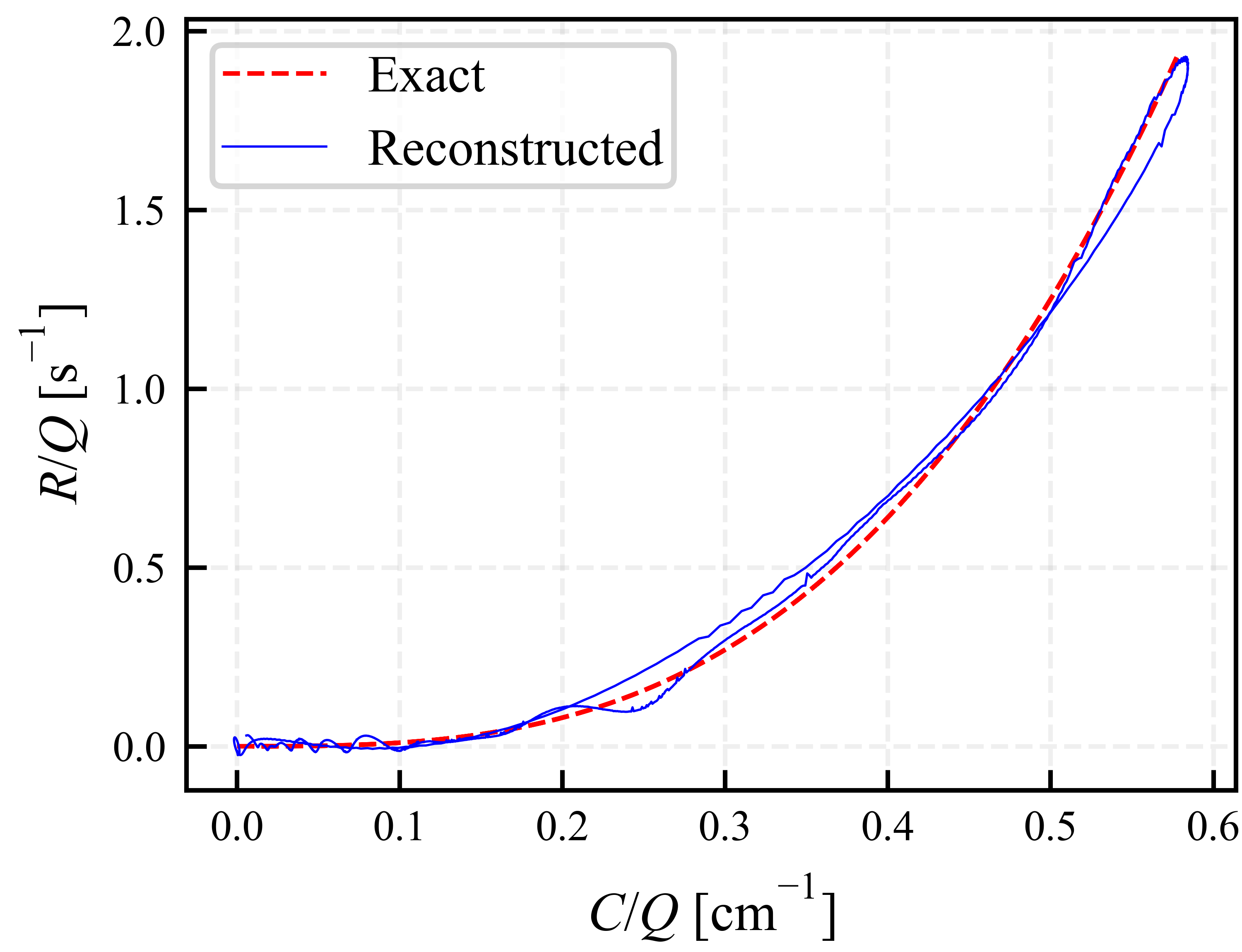}
    \put (-143,96){\scriptsize TSVD ($m_\mathrm{cut} = 30$)}
    \put (-15,37){\footnotesize (d)}
        \includegraphics[width=0.33\textwidth]{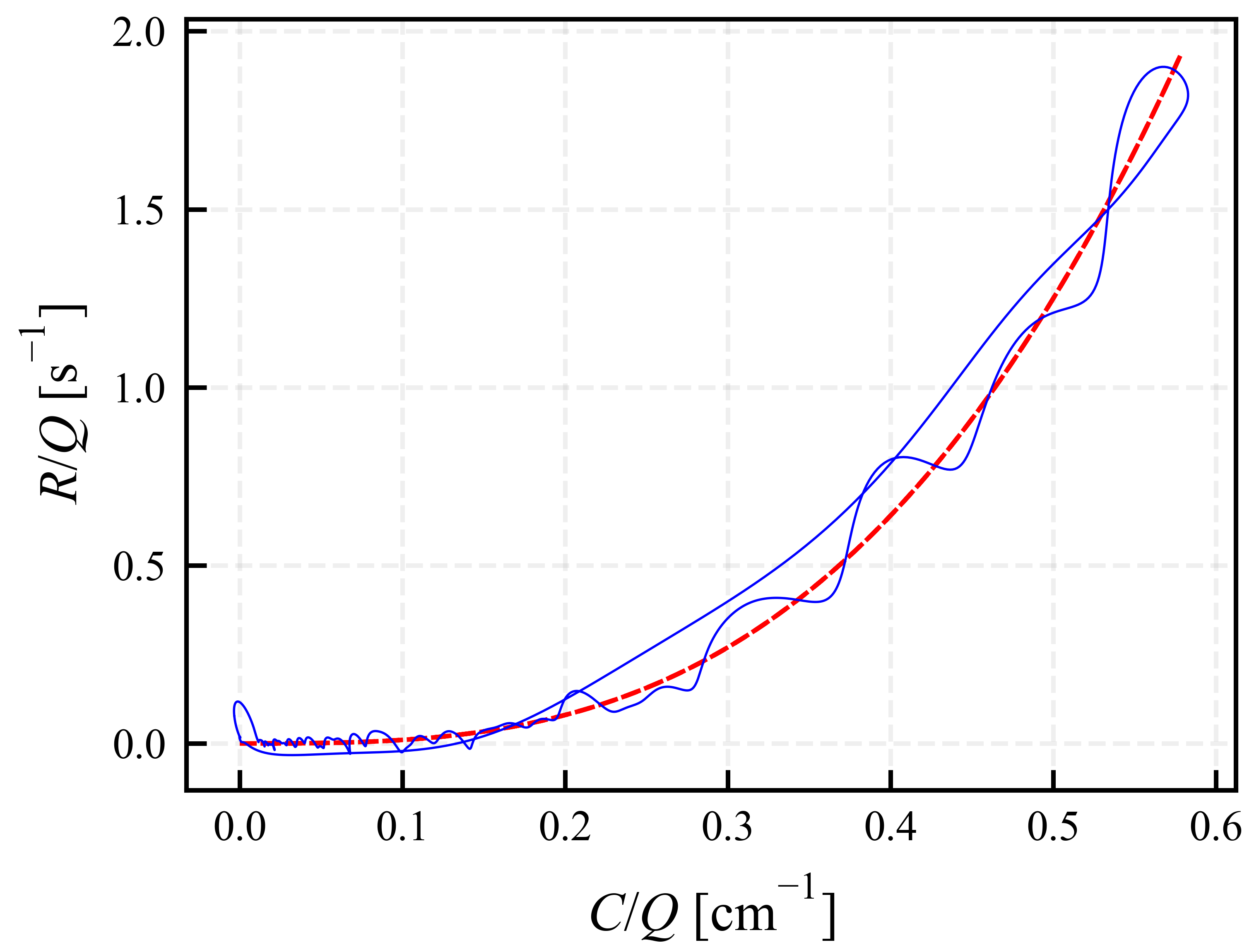}
    \put (-15,37){\footnotesize (e)}
    \put (-143,120){\scriptsize Fourier-filtration ($\lambda = 14$)}
    \includegraphics[width=0.33\textwidth]{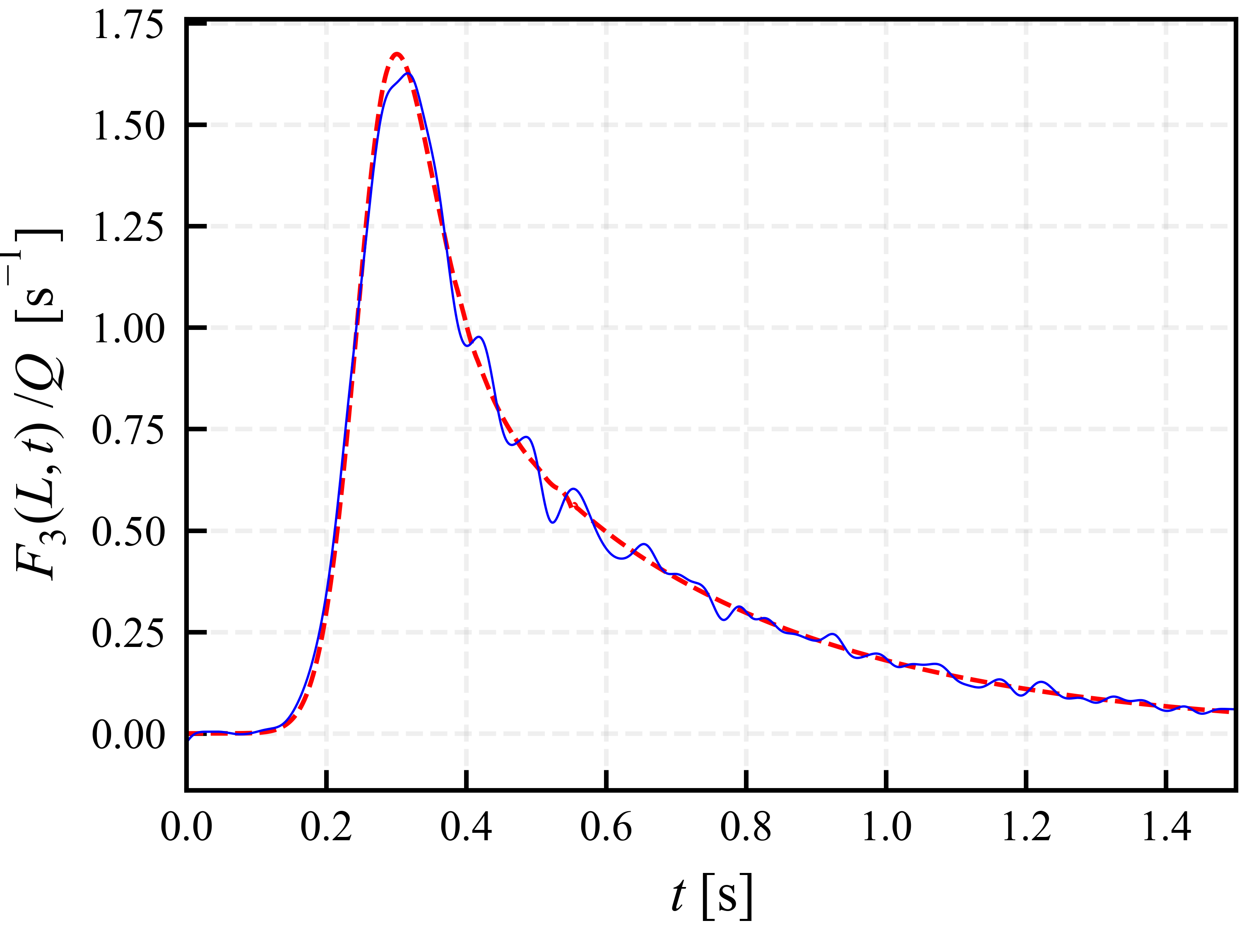}
    \put (-15,37){\footnotesize (f)}
    \put (-60,120){\scriptsize Fourier-filtration }
    
    \caption{Y-procedure reconstruction for a cubic reaction using the TSVD method and Fourier-filtration, with the corresponding optimal regularization parameter values. (a) The synthetic measurement of the outlet flux, (b,c) Variation of the degree of multivaluedness $\Psi$ (Eq.~\eqref{eq:psi}) with the cutoff mode number $m_\mathrm{cut}$ and Fourier-filtration parameter $\lambda$, respectively. (d,e) Optimal reconstructed $R-C$ curves from the TSVD method and Fourier-filtration. respectively. (f) Optimal reconstructed flux entering zone 3 from the Fourier filtration method.}
    \label{fig:cubic}
\end{figure*}

We calculate $\Psi$ for the reconstructed curves of Figs.~\ref{fig:TSVD-linear}(d-f) which correspond to $m_\mathrm{cut} = 25, 27$ and 30; we also calculate $\Psi$ for reconstructed curves obtained at intermediate and neighbouring values of $m_\mathrm{cut}$ to cover the transition region in the Picard plot (Fig.~\ref{fig:Picard}). The results are shown in Fig.~\ref{fig:psi}(a). The variation of $\Psi$ with $m_\mathrm{cut}$ exhibits a clear minimum; the value of $m_\mathrm{cut}$ at the minimum ($m_\mathrm{cut}=28$) is the choice that yields a reconstructed rate function that is closest to being single-valued. The corresponding reconstructed outlet flux and rate function are shown in Figs.~\ref{fig:psi}(b,c). The entire automated procedure---that starts with $F_{out}^*(t)$, constructs the discrete inverse problem, solves it via T-SVD using different numbers of modes, calculates $\Psi$, selects the cutoff mode, and plots the final reconstruction---takes approximately 25 $\si{s}$ of CPU time on an Intel i7 (11th generation) processor.


This strategy of selecting the regularization parameter, based on minimizing the degree of multivaluedness of the reconstructed $R(C)$, can also be applied to $\lambda$ in the Fourier-filtration method. The values of $\Psi$ for reconstructions obtained using different values of $\lambda$ (which unlike $m_\mathrm{cut}$ is not restricted to integer values) are displayed in Fig.~\ref{fig:psi}(d). We see that a minimum is present for the Fourier-filtration method as well, though it is not as well defined as in the case of the TSVD method (Fig.~\ref{fig:psi}(a)). The reconstruction corresponding to $\lambda = 13$, for which $\Psi$ is the smallest in Fig.~\ref{fig:psi}(d), is shown in Figs.~\ref{fig:psi}(e,f). The optimal Fourier-filtration reconstruction is good but inferior to the optimal TSVD reconstruction, particularly near the maximum value of concentration where the Fourier-filtration produces oversmoothing---compare the peaks of Fig.~\ref{fig:psi}(b) and Fig.~\ref{fig:psi}(e), as well as the upper end of the $R-C$ curves in Fig.~\ref{fig:psi}(c) and Fig.~\ref{fig:psi}(f). Note that since one does not have to compute the integrals required for formulating the discrete inverse problem, the Fourier-filtration method takes less CPU time to produce the final reconstruction in Fig.~\ref{fig:psi}(f) (approximately 4 $\si{s}$ as compared to 25 $\si{s}$ for producing Fig.~\ref{fig:psi}(c) via the TSVD method).

This section has shown how the single-valued property of $R(C)$, for state-defining experiments, can be used to select the regularization parameter in an objective manner. This selection procedure can be applied to any regularization method, be it the TSVD approach introduced here (Sec.~\ref{sec:inverse}), or the previously used Fourier-filtration method (Sec.~\ref{sec:filtration}). 

\section{Reconstructing nonlinear kinetics}\label{sec:nonlinear}

In this section, we test the TSVD reconstruction method for cases with nonlinear kinetics, i.e., for which the $R(C)$ curve is not a straight line. In reconstructing such cases, one does not know the shape the $R-C$ curve and so the objective strategy of selecting the regularization parameter, developed in Sec.~\ref{sec:single-valued}, becomes particularly helpful. 

We generate synthetic measurements of $F_{out}^*(t)$ by running forward simulations of the TAP reactor, starting with the inlet pulse of Eq.~\eqref{eqn:F_inlet}, and using a nonlinear kinetic rate expression in the thin reaction zone: $R = k_n C^n$ (see Eq.~\eqref{eq:R}). The synthetic data obtained after adding noise to the outlet flux from the forward simulation (as in Eq.~\eqref{eq:noise}) is shown in Fig.~\ref{fig:quadratic}(a) for a second order ($n = 2$) reaction and in Fig.~\ref{fig:cubic}(a) for a third order reaction ($n = 3$). Note that while we still present results after normalizing with the strength of the inlet pulse $Q$, the normalized results are not entirely independent of $Q$. Rather, because of the nonlinear kinetics, the results will remain the same when $Q$ is changed only if the reaction rate is also adjusted so that $k_nQ^{(n-1)}$ remains constant. For Figs.~\ref{fig:quadratic} and~\ref{fig:cubic}, we have chosen the rate constants such that $k_2 Q = k_3 Q^2 = k$ (see the end of \S~\ref{sec:model}); the corresponding reaction rates have the same order of magnitude, which facilitates a comparison of the nonlinear-reaction cases with each other and with the linear case considered in the previous sections. 

Let us consider the second-order case first. From $F_{out}^*(t)$ in Fig.~\ref{fig:quadratic}(a), we calculated the vector $\boldsymbol{b}$ using Eq.~\eqref{eq:A_b_def} and then solve the inverse problem Eq.~\eqref{eq:Aw_b} using the TSVD method. The $R-C$ curve is then reconstructed and the multivaluedness parameter $\Psi$ is computed (Eq~\ref{eq:psi}). The inverse-problem solution and reconstruction is repeated for a range of values of $m_\mathrm{cut}$ from 27 to 33, which span the transition region in the corresponding Picard plot (not shown). The variation of $\Psi$ with $m_\mathrm{cut}$ is shown in Fig.~\ref{fig:quadratic}(b); the minimum occurs at $m_\mathrm{cut} = 29$ which is then selected as the optimal value. The corresponding reconstructed $R-C$ curve is shown in Fig.~\ref{fig:quadratic}(d). For comparison, we also apply the Fourier-filteration procedure of Sec.~\ref{sec:filtration} to this second-order reaction case. Again using the minimum of $\Psi$ to select the filtration parameter as $\lambda = 11$ (see Fig.~\ref{fig:quadratic}(c)), the optimal reconstruction is obtained and plotted in Fig.~\ref{fig:quadratic}(e). The TSVD method proposed here is seen to produce a significantly better reconstruction than the Fourier-filteration method (compare Figs.~\ref{fig:quadratic}(d,e)).

The analogue of Fig.~\ref{fig:quadratic} for a third-order reaction is shown in Fig.~\ref{fig:cubic}. The improvement of the TSVD method over Fourier-filtration is even more pronounced in this case. Why does the TSVD method yield better results than Fourier-filtration? One plausible reason is that Fourier functions are a generic basis set, while the SVD modes provide a basis set that is tailored to the problem under study, by virtue of being computed from the matrix $\boldsymbol{A}$ that represents the diffusive transport process in zone~3. 

In general, reconstructing nonlinear kinetics is more challenging than linear kinetics as is apparent from comparing the nonlinear results in Figs.~\ref{fig:quadratic}(d,e) and Figs.~\ref{fig:cubic}(d,e) with the linear results in Figs.~\ref{fig:psi}(c,f). The reason is that the flux profile is more asymmetric when the kinetics are nonlinear, i.e., the profile has a steeper rising front and a more gradual decaying rear (compare Fig.~\ref{fig:cubic}(a) and Fig.~\ref{fig:OF}). Resolving the front requires one to retain high-frequency modes which however bring in noise that contaminates the reconstruction of the gradually decaying rear. This is illustrated, for the Fourier-filtration method, in Figs.~\ref{fig:quadratic}(f) and~\ref{fig:cubic}(f), where the decaying portion of the reconstructed $F_3(L,t)$ (the flux entering zone 3) is seen to exhibit spurious oscillations; these oscillations are significantly smaller in case of the TSVD method and this leads to better reconstructions of the $R-C$ curves. Spurious oscillations are present in the linear-reaction case as well, but they are much weaker since the flux profiles are less asymmetric (see Figs.~\ref{fig:psi}(b,e)).


\section{Concluding remarks}
\label{sec:conclusion}

The Y-procedure $\citep{YABLONSKY20076754}$ is a potent technique for analysing pulse responses of the thin-zone TAP reactor because it enables one to reconstruct the $R-C$ curve without having to make any assumptions on the reaction rate expression. The most challenging step of the procedure is the backtracking of diffusive transport to reconstruct the flux and concentration at the inlet of the third zone from noisy measurements of the flux exiting the third zone. This is an illposed problem and a direct solution, for example via Fourier transform, will produce a result that is entirely contaminated by amplified noise. To deal with this noise, previous implementations of the Y-procedure have applied a Fourier-based noise filter to the measured signal. This step requires human intervention to decide the degree of filtration and the result can suffer from over smoothing. 

In this work, we have developed a new method for dealing with this ill-posed step of the Y-procedure. Using a basis of localized pulses (which does not require the signal to be periodic), we formulated the problem of calculating the zone 3 inlet flux as a discrete inverse problem, $\boldsymbol {A w = b}$. A regularized solution to this problem was obtained using the truncated-SVD method. This technique uses the SVD modes of the matrix $A$ to represent the flux and then filters out the noisy components by truncating all modes beyond index $m_\mathrm{cut}$. The Picard plot reveals the transition from information rich modes to noise-dominated modes and thereby guides the choice of $m_\mathrm{cut}$. 

We have demonstrated the effectiveness of the inverse-problem formulation and the TSVD method on synthetic outlet flux data, generated considering linear and nonlinear kinetics. The TSVD reconstructions are more accurate than those obtained from the Fourier-filtration approach, with the former yielding appreciably more accurate reconstructions of nonlinear kinetics. 

We have also developed an objective method for selecting the regularization parameter in state defining experiments based on the property that the rate is a single-valued function of the gas phase concentration in such small-pulse experiments. This strategy has been shown to be effective for both the new TSVD method and the Fourier-filtration method. Importantly, with such an objective selection strategy, human judgement is no longer required to choose the regularization or filtration parameter. One can therefore automate the Y-procedure, enabling the analysis of a large number of outlet flux pulses and the rapid reconstruction of $R-C$ curves, possibly even in real time while experiments are being performed.

This method for selecting the regularization parameter does not apply, of course, to state altering experiments wherein the pulse of gas is large enough for the concentration of adsorbed species on the catalyst surface to change during the course of a pulse. In such experiments, the rate $R$ depends on the evolving adsorbed species concentration in addition to the gas phase concentration $C$, and so the $R-C$ curve traced out by a pulse takes the form of a multivalued petal plot~\citep{REDEKOP20116441,ROSSKUNZ201846}. An important task for the near future is to extend the inverse problem formulation to state altering pulses and to develop a strategy for selecting the regularization parameter when the true $R-C$ curve is multivalued. The Picard plot which accompanies the TSVD reconstruction will be especially helpful in this regard. 

Our formulation of the discrete inverse problem paves the way for further development of the Y-procedure through the use of other well-developed techniques from the field of inverse problems. Apart from the TSVD method used here, the other standard method for solving inverse problems is the Tikhonov regularization which treats the problem as a penalized optimization problem. While closely related to the TSVD method \citep{Hansenbook2010}, it has the advantage of being computationally more efficient since one need not compute the SVD of the matrix $\boldsymbol{A}$. It also offers more flexibility by allowing different choices for the penalization function. Typically, one chooses to minimize the norm of the solution vector $\boldsymbol{w}$ (in which case it is akin to the TSVD method but with a gradual dampening applied to the SVD modes instead of a hard truncation); however, other choices, such as a high-order temporal derivative of the flux, could lead to better results. Modern machine learning methods for inverse problems~\citep{MLhandbook,MLregularization,MLImage} may also produce better reconstructions by 'learning' the particularities of the profiles of the inlet and outlet pulses, to which a general purpose method like TSVD is agnostic. Furthermore, machine learning methods may be more effective in the presence of non-Gaussian noise \cite{ROELANT2007269}. A promising direction for future work, therefore, is to apply and compare different solution strategies for the discrete inverse problem formulated here, while using synthetic data generated with a non-Gaussian noise model \cite{ROELANT2007269}.

\section*{Acknowledgments}

J.R.P. is grateful to Vishal Vasan (ICTS-TIFR, Bangalore) for illuminating discussions on inverse problems during the CMI-BIRS workshop on the \textit{Mathematical and Computational Foundations of Climate Modeling} (25w5449), held at the Chennai Mathematical Institute, India in Aug 2025. J.R.P. and Anjali also thank Sonali Das and her group at IIT Bombay for helpful discussions on the TAP reactor. Anjali acknowledges financial support from IIT Bombay via her Institute Postdoctoral Fellowship.

\section*{Appendix}

\begin{figure}
    \centering    
    {\includegraphics[width=0.4\textwidth]{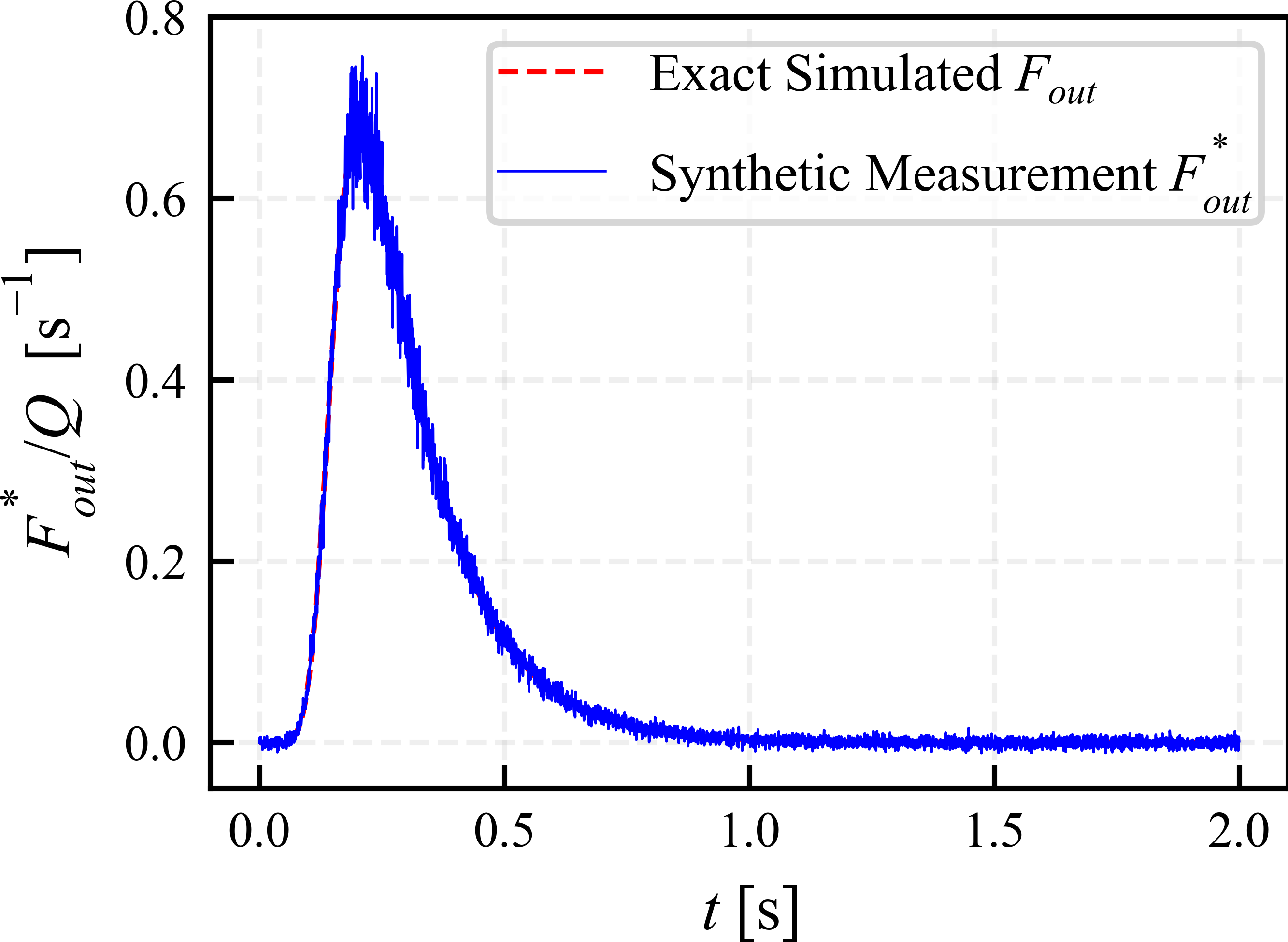}} 
	\caption{
    Outlet flux at the exit of zone 3 corresponding to a narrower inlet pulse than that which produces the outlet flux in Fig.~\ref{fig:OF}. Here too, we consider a linear reaction in zone 2. Noise is added to the exact outlet flux from the forward simulation (red-dashed line) to obtain the synthetic measured outlet flux (blue line).}
	\label{fig:OFnarrow}
\end{figure}

In this appendix, we consider an inlet pulse that is narrower than what is used in the main text. Specifically, rather than setting  $\mu = 1/4$ \si{s} and $\sigma = 1/20$ \si{s} in Eq.~\eqref{eq:Fin}, we now set $\mu = 1/12.5$ \si{s} and $\sigma = 1/40$. The resulting synthetic measurement of the outlet flux is shown in Fig.~\ref{fig:OFnarrow}, which may be compared with Fig.~\ref{fig:OF}. 

\begin{figure*}
	\centering
	\includegraphics[width=0.33\textwidth]{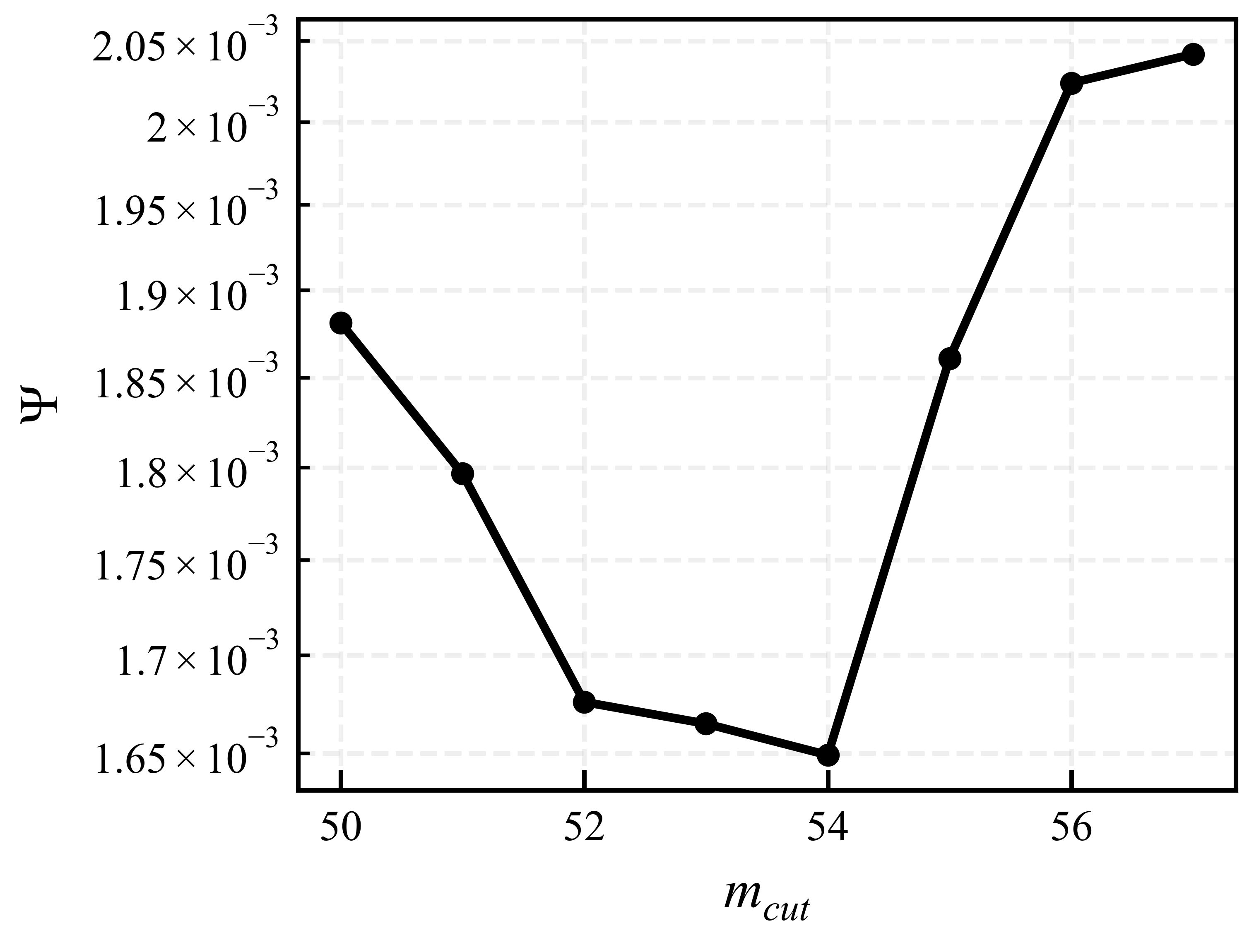}
    \put (-15,35){\footnotesize (a)}
    \put(-50,120){\scriptsize TSVD}
    \includegraphics[width=0.33\textwidth]{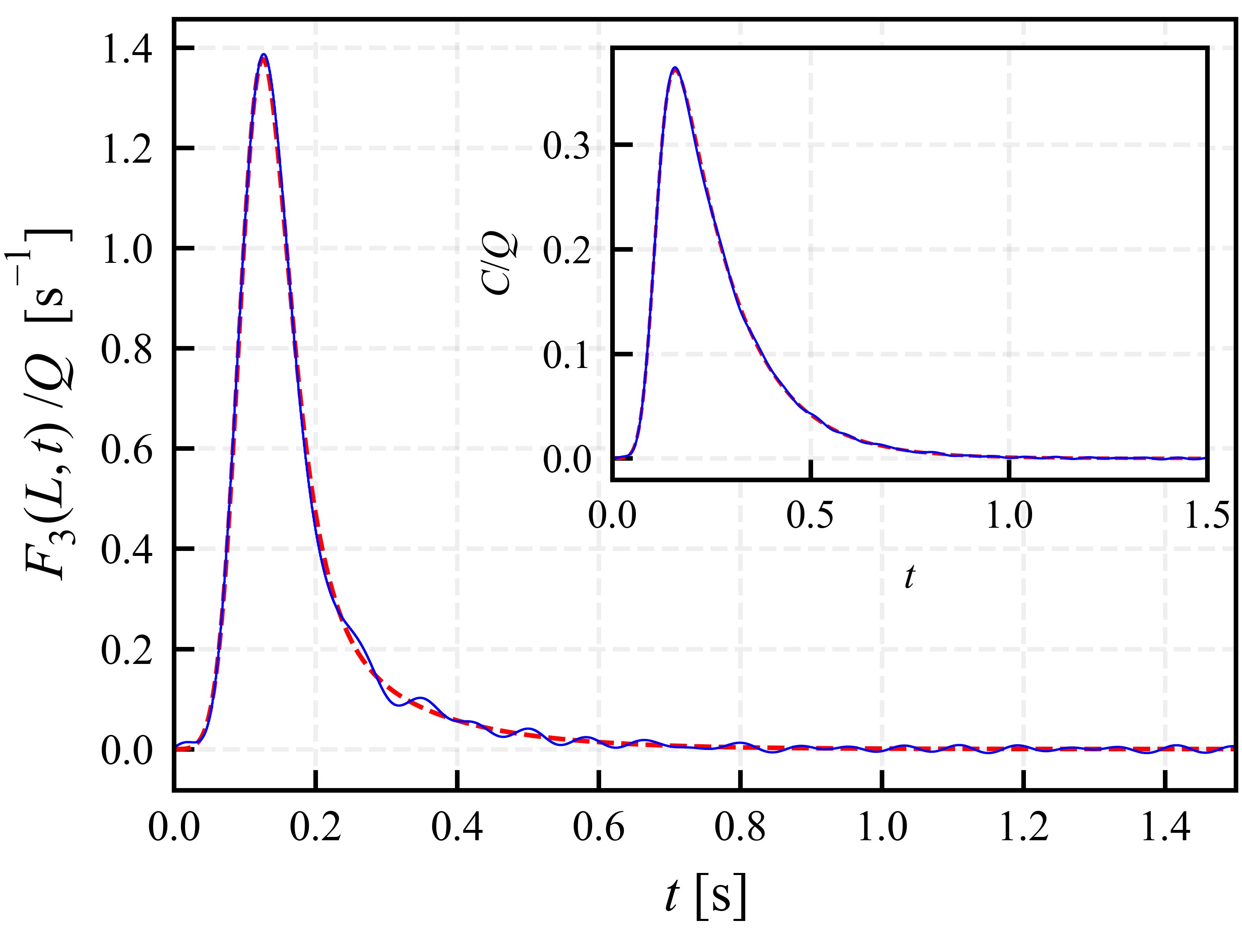}
    \put (-15,35){\footnotesize (b)}
    \includegraphics[width=0.33\textwidth]{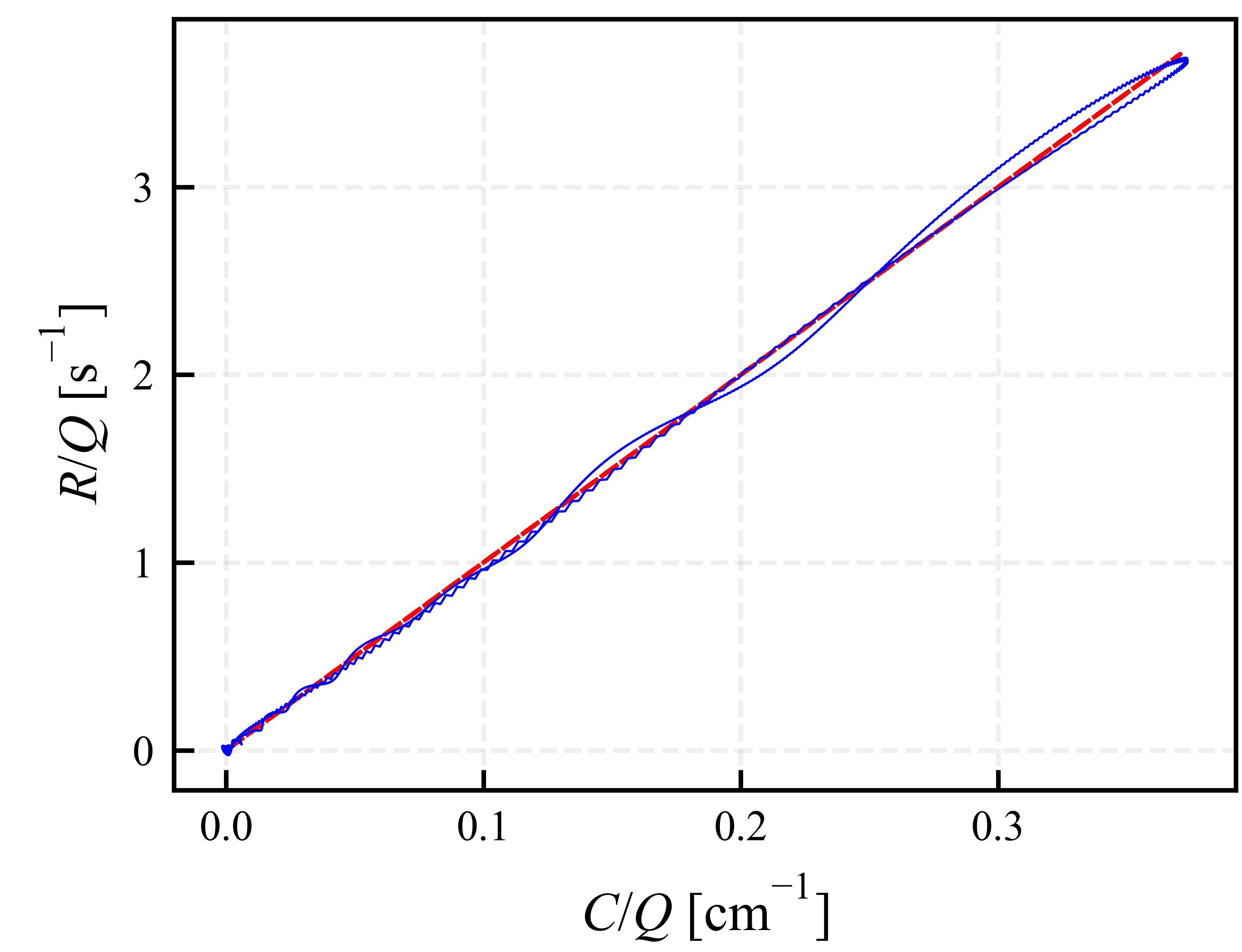}
     \put (-143,120){\scriptsize TSVD ($m_\mathrm{cut} = 54$)}
    \put (-15,35){\footnotesize (c)}
    \hspace{0.02\textwidth}
    \includegraphics[width=0.33\textwidth]{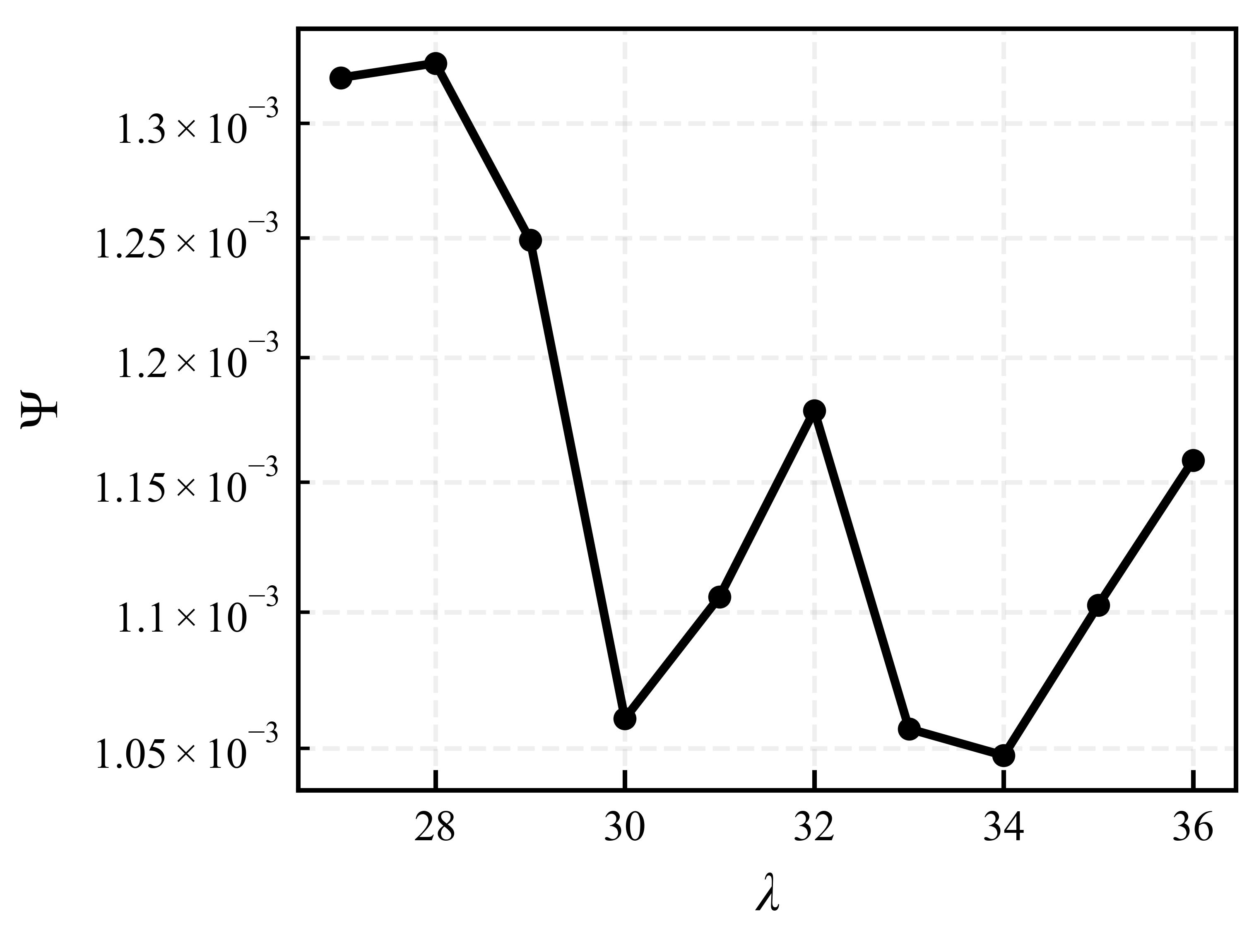}
    \put(-55,120){\scriptsize Fourier-filtration}
    \put (-15,35){\footnotesize (d)}
    \includegraphics[width=0.33\textwidth]{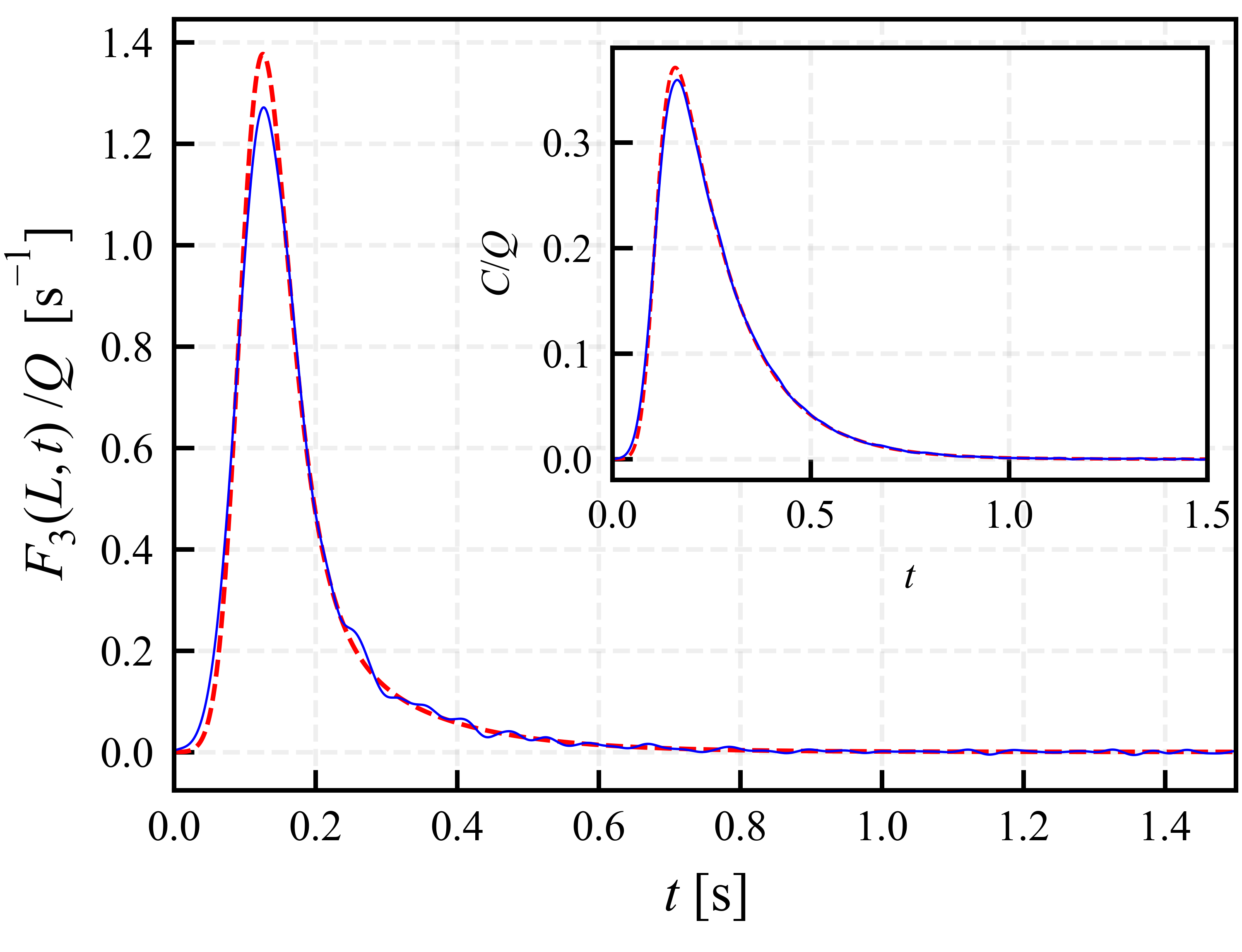}
    \put (-15,35){\footnotesize (e)}
    \includegraphics[width=0.33\textwidth]{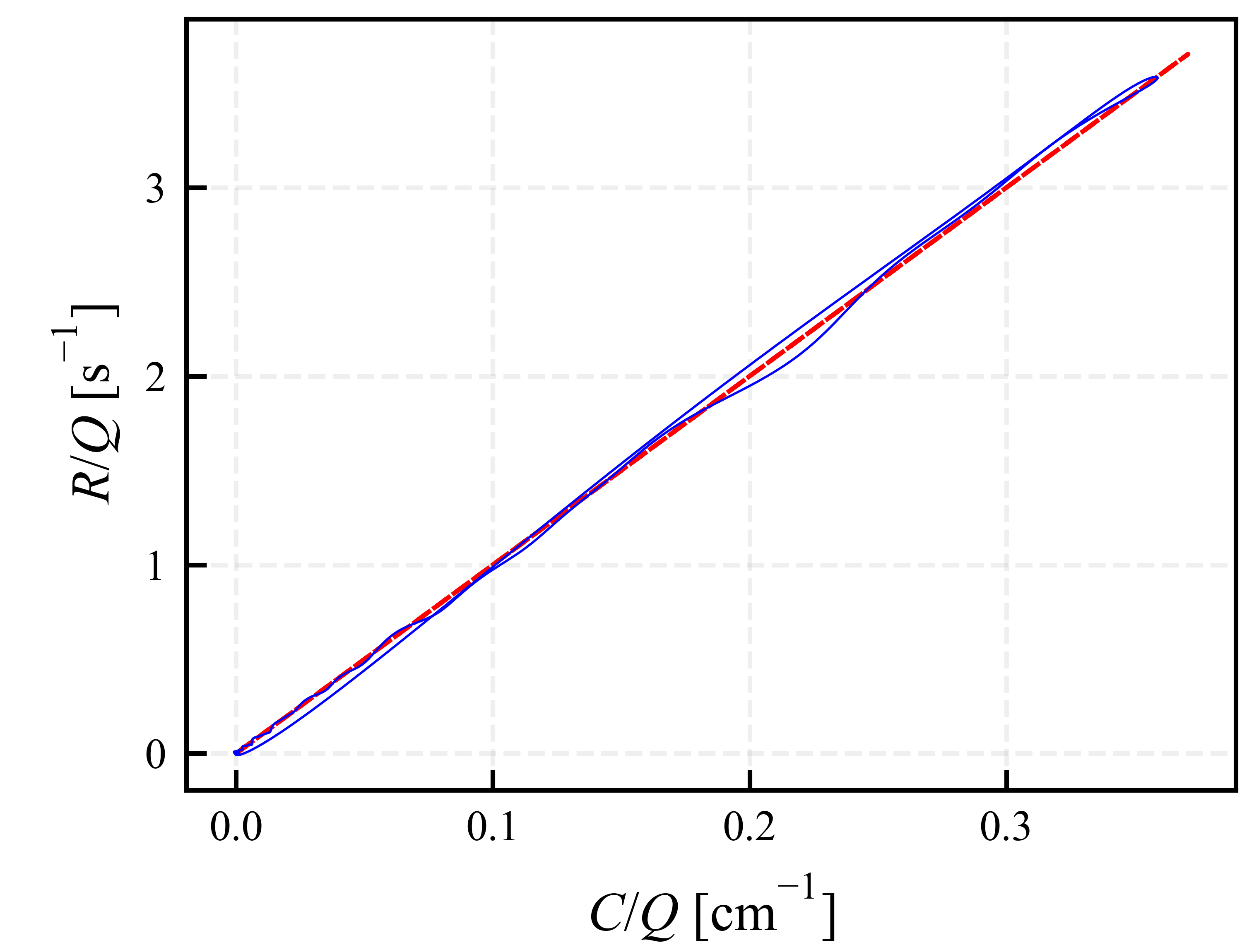}
    \put (-143,96){\scriptsize Fourier-filtration ($\lambda = 34$)}
    \put (-15,35){\footnotesize (f)}
	\caption{Analogue of Fig.~\ref{fig:psi} for a narrower pulse ($\mu = 1/12.5$ \si{s} and $\sigma=1/40$ \si{s}). (a) Variation of the degree of multivaluedness $\Psi$ (Eq.~\eqref{eq:psi}) with the cutoff SVD mode number $m_\mathrm{cut}$, (b) reconstructed flux entering zone 3 (the inset shows the concentration in the thin reaction zone) for $m_\mathrm{cut} = 54$ where $\Psi$ is minimum, (c) corresponding reconstructed $R-C$ curve. (d) Variation of the degree of multivaluedness $\Psi$  with the Fourier-filtration parameter $\lambda$, (e) reconstructed flux entering zone 3 (the inset shows the concentration in the thin reaction zone) for $\lambda = 34$ where $\Psi$ is minimum, (f) corresponding reconstructed $R-C$ curve. 
    }
	\label{fig:narrow-linear}
\end{figure*}
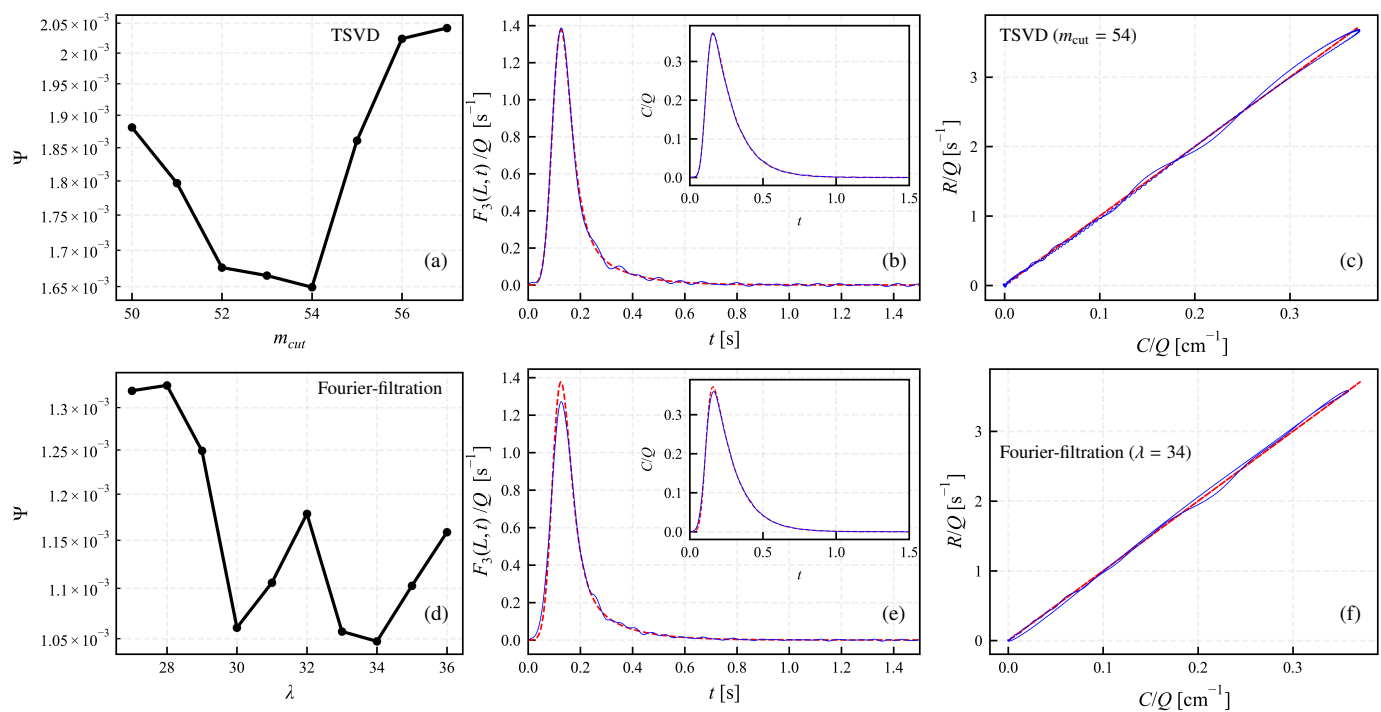

Applying the Y-procedure to $F_{out}^*$ in Fig.~\ref{fig:OFnarrow}, using both the TSVD and Fourier-filtration methods along with the $\Psi$-minimisation procedure, we obtain the reconstructions shown in Fig.~\ref{fig:narrow-linear}. This figure is the analogue of Fig.~\ref{fig:psi} but for a narrower pulse. As is the case in Fig.~\ref{fig:psi}, Fourier-filtration is seen to produce more over-smoothing than the TSVD method---compare the peaks of Fig.~\ref{fig:narrow-linear}(b) and Fig.~\ref{fig:narrow-linear}(e), as well as the upper end of the $R-C$ curves in Fig.~\ref{fig:narrow-linear}(c) and Fig.~\ref{fig:narrow-linear}(f). We also note that, for the narrower pulse, the plot of $\Psi$ against $\lambda$ has two minima with nearly the same value of $\Psi$ (the second local minimum is less prominent for the case of the broader pulse in Fig.~\ref{fig:psi}); we proceed with the global minimum for the final reconstruction. How differences in the inlet flux profile affects the variation of $\Psi$ and thereby the accuracy of the reconstruction via the Y-procedure is a question that merits further investigation in future work. 


  \bibliographystyle{elsarticle-num-names} 
  \bibliography{ref}



%
%
%
\end{document}